\documentclass[12pt,a4paper,notitlepage]{article}

\usepackage{graphics}
\usepackage[pdftex]{graphicx,color}
\usepackage{epsf}
\usepackage{amsmath}
\usepackage{amssymb}
\usepackage{amsfonts}
\usepackage{verbatim} 

\numberwithin{equation}{section}

\newcommand{\ket}[1]{\left|#1\right\rangle}

\def \mr{\mathrm}
\def \slashpartial {\partial\!\!\!\slash}
\def \slashD { D\!\!\!\!\slash\,}
\def \slashp { p\!\!\slash}

\begin{document}

\begin{titlepage}

\begin{center}\large
Vilnius university\\[-6pt]
Faculty of physics\\[-6pt]
Theoretical physics department

\vspace{60pt}
\Large
Jurgis Pa\v{s}ukonis\\

\vspace{30pt}
\Large

IMPLEMENTATION OF NON-LINEAR GAUGE-FIXING\\[-7pt]
IN FEYNARTS PACKAGE

\vspace{40pt}
\Large
Master's thesis\\
\large
(Theoretical physics and astronomy studies program)

\vspace{60pt}

\vspace{80pt}

Vilnius 2007

\end{center}

\end{titlepage}

\newpage

\pagestyle{empty}

\tableofcontents

\newpage

\pagestyle{myheadings}


\section{Introduction}

The accepted theoretical model of elementary particles and their interactions -- the Standard Model (SM) -- has been very successful at explaining the experimental data available up until now. It describes all the observed particles -- leptons and hadrons -- that interact by electromagnetic, weak and strong forces, and the predictions of the model have been tested in the particle collider experiments up to the energy scale of around 100\hspace{3pt}GeV. However, it is clear that some new effects should occur at higher energies. One of such effects is the highly anticipated Higgs scalar particle that is predicted by the SM and is yet to be found experimentally. From another point of view, it has been shown that the SM can not be valid for arbitrary high energies, therefore, new physics should be discovered at higher energies, possibly already at the scale of 1000\hspace{3pt}GeV.

Theoretical physicists have been creating models that could predict this new  physics at higher energies, one of the most realistic ones being the Minimal Supersymmetric Standard Model (MSSM). However, there is no experimental data available at higher energies to test the predictions of these new models (and to constrain the imagination of theoretical physicists), and that is probably the biggest problem today in the field of high-energy elementary particle physics. Of course, huge effort is being put into building particle colliders that would operate at higher energies: for example, the Large Hadron Collider (LHC) at CERN, which will reach energies of 14000\hspace{3pt}GeV, will start its operation within the coming year and the physics community is eagerly waiting for the results. It is clear, though, that the experimental data will be more and more difficult to attain at higher energies, and so we must look for alternatives.

One of the alternatives comes from the very nature of quantum physics: due to the quantum fluctuations of the energy of intermediate particles during a particle collision process, the physics at higher energies (e.g. 1000\hspace{3pt}GeV) affects the results of the experiments where the initial and final particles are at lower energies (e.g. 100\hspace{3pt}GeV). These effects, naturally, become weaker as the difference between the energy of the experiment and the energy of the new physics grows. But that brings up an important point: if we increase the precision of the experiments and the theoretical predictions, we can analyze these small effects, and thus analyze the physics at higher energies without actually having experiments at these energies! Such analyses are called \emph{precision measurements}, and are becoming increasingly important, as the experiments at higher energies become more difficult to perform.

The theoretical predictions, that can be compared with experiments, are obtained by calculating amplitudes for the interaction processes. The amplitude can be represented by a sum of \emph{Feynman diagrams}, while each Feynman diagram represents an algebraic expression, which is constructed using the \emph{Feynman rules} of the theory. Now, the Feynman diagrams are classified according to the number of loops they contain into tree-level, one-loop level, two-loop level and so on, and each additional loop level gives smaller corrections to the amplitude than the previous one. A typical process in the SM would contain several diagrams at the tree-level, which can be quite easily calculated by hand, however already at one-loop level, which is essential for any precision calculation, the number of diagrams grows to several hundreds! Such calculation is practically impossible to do by hand, therefore, it is necessary to develop programs that can generate and calculate Feynman diagrams \emph{automatically}. 

There are already tools developed, that are capable of automatically generating and calculating Feynman diagrams at one-loop level in the SM and MSSM -- the most notable being the FeynArts package \cite{feynarts, feynartsWWWW} and GRACE \cite{grace} -- but together with the development of such tools there is also a need for ways to check the validity of the computed results, in order to use them with any reliability. It turns out that one
powerful tool to perform such checks arises from the procedure of gauge-fixing
the theory, which is performed to deduce the Feynman rules from the Lagrangian
of the theory.

Gauge-fixing lets us freely choose the gauge in which to perform
calculations, and this gauge choice can be parameterized by the
so-called gauge-fixing parameters. The Feynman rules will then
depend on the gauge choice, however, the final results have to be
gauge-independent. What makes this interesting is that the
gauge-invariance of the final result occurs through a combination
of all the diagrams, while each one by itself is gauge-dependent,
so if some diagram is missed, or some other mistake occurs, the
final result will very likely become gauge-dependent. Therefore,
if the result does appear to be gauge-invariant, it is a strong
indication for the correctness of the result. The gauge-invariance check is also very suitable for automatic calculations,
because it by itself can be performed automatically. If the Feynman rules
with unspecified gauge parameters are used, the gauge parameters will carry over to the
expression for the final result. The result can then be evaluated
numerically with various choices for these parameters and it must not depend on
their values. Naturally, the more gauge-fixing parameters we have available, the
larger set of gauges we can explore, and the more stringent the test becomes. 

The standard derivation of the vertices in the Standard Model uses \emph{linear} gauge-fixing which results in gauges parameterized by 3 numbers $(\xi_W,\xi_Z,\xi_A)$, that appear in the particle propagators. A natural extension of this procedure is to use a \emph{non-linear} gauge-fixing which can result in a much wider set of available gauges with parameters that modify not the propagators but the vertices of the theory. The non-linear gauge-fixing of the Standard Model was presented in \cite{nonlinear}, where it was used to simplify certain tree-level calculations.  It was later recognized to be very useful for automatic calculations and it was fully implemented in GRACE \cite{grace}, which was used then to perform one-loop level calculations in the Standard Model and test the results for the full non-linear gauge invariance.

The purpose of this work is to implement the same class of non-linear gauge-fixing that was used in \cite{nonlinear, grace} in the FeynArts/FormCalc package. The advantage of this package is that it is open-source and freely distributable, which means that the implementation, once tested to be correct, can be used by the whole physics community to check their calculations for gauge invariance.

In the previous papers \cite{jp1, jp2} we have already explored the linear gauge-fixing in detail and tried to implement a subset of the non-linear gauges in FeynArts. In this final work we give a full presentation of the Standard Model, non-linear gauges, and their implementation in FeynArts. We start by discussing a Yang-Mills gauge theory with a broken symmetry, focusing on the gauge-fixing procedure. The results of this discussion are then applied to the specific case of the Standard Model with the non-linear gauge-fixing, which in the end yields a complete list of propagators and vertices of the theory. We then implement these new vertices in the FeynArts package and use it to carry out some automatic amplitude calculations and check the results for gauge invariance.


\section{Theoretical background}

In this section we present the general Yang-Mills (YM) Lagrangian with a broken symmetry, which is the basis for most of the realistic quantum field theories, including the Standard Model, presented later in this work. The goal of this section is to familiarize ourselves with the YM Lagrangian and the gauge-fixing procedure, arriving at the final form of the Lagrangian, which can then be used in the next section for describing the Standard Model.

Note that the following discussion is fairly standard and closely follows the one found in \cite{ps}, however, we present a more complete discussion of the linear gauge-fixing by considering the most general linear case and consequently finding the largest possible set of linear gauge-fixing parameters.

\subsection{Yang-Mills Lagrangian and symmetry breaking}

We consider a gauge theory for a multiplet of real scalar fields
$\phi_i$, transforming as some representation $R$ of the gauge
group $G$. The infinitesimal law of gauge transformation for the
scalar fields $\phi_i$ and the gauge fields $A_\mu^a$ is then \cite{ps}:
\begin{align}
\delta\phi_i(x) &= i \alpha^a(x) t^a_{ij} \phi_j(x), \\
\label{eq:deltaA} \delta A^a_\mu(x) &= \frac{1}{g}\partial_\mu
\alpha^a(x) + f^{abc}A^b_\mu(x) \alpha^c(x),
\end{align}
where $\alpha^a(x)$ are the infinitesimal parameters of the
transformation, $t^a_{ij}$ are the generators of the
representation $R$, and $f^{abc}$ are the structure constants of
the gauge group $G$, meaning that the generators $t^a$ (in any
representation) commute as:
\begin{equation}
[t^a,t^b] = i f^{abc} t^c.
\end{equation}
The parameter $g$ is called \emph{coupling constant} and can be
chosen independently for each simple or $U(1)$ subgroup of $G$.

If the fields $\phi_i$ are real, the generators $t^a$ will have to
be purely imaginary in order to make $\delta\phi_i$ real. In this
case it is convenient to define \emph{real} generators
\begin{equation}
T^a \equiv -i t^a,
\end{equation}
giving the transformation
\begin{equation}
\label{eq:deltaphi}
\delta\phi_i(x) = - \alpha^a(x) T^a_{ij}
\phi_j(x),
\end{equation}
and the commutation relationship
\begin{equation}
[T^a, T^b] = f^{abc} T^c.
\end{equation}

In order to construct a Lagrangian, as usual in Yang-Mills
theories, a covariant derivative
\begin{equation}
\label{eq:Dphi}
(D_\mu \phi)_i = \partial_\mu \phi_i - i g A^a_\mu t^a_{ij} \phi_j
=
\partial_\mu \phi_i + g A^a_\mu T^a_{ij} \phi_j.
\end{equation}
and a field strength tensor
\begin{equation}
F^a_{\mu\nu} = \partial_\mu A^a_\nu - \partial_\nu A^a_\mu + g
f^{abc} A^b_\mu A^c_\nu,
\end{equation}
are defined. An invariant Lagrangian for this gauge theory is:
\begin{equation}
\label{eq:l-unbroken} \mathcal{L} = -\frac{1}{4}(F^a_{\mu\nu})^2 +
\frac{1}{2}(D_\mu \phi_i)^2 - V(\phi),
\end{equation}
where $V(\phi)$ is the symmetry-breaking potential.

So far, in the Lagrangian above, the gauge fields are massless, but before proceeding to the quantization we have to include the
effects of the spontaneous symmetry breaking, which occurs if the
potential $V(\phi)$ does not have a local minimum at $\phi=0$. A
``normal" potential would have a minimum at $\phi=0$ like $V(\phi)
= m^2 |\phi|^2$, in which case the field fluctuates around 0, and
the potential term gives the mass to the field. Now if we take $V$
\emph{not} to have a minimum at 0, like $V(\phi) = a |\phi|^4 - b
|\phi|^2$, the value of the field will ``roll down" to the minimum
of the potential at some $\phi_0$, and fluctuations can only
happen around that minimum value. In other words, the field will
acquire a vacuum expectation value (VEV),
\begin{equation}
(\phi_0)_i \equiv \langle \phi_i \rangle,
\end{equation}
which is saying that the field is non-zero even in the vacuum.

In order for perturbation theory to work, we have to analyze
fields that have zero vacuum expectation value, so we redefine
$\phi_i$ as
\begin{equation}
\label{eq:chidef}
\phi_i(x) = \phi_{0i} + \chi_i(x),
\end{equation}
and from now on treat $\chi_i(x)$ as our field of interest.

Plugging in the new definition (\ref{eq:chidef}) into expression
(\ref{eq:l-unbroken}) yields the Lagrangian in terms of $\chi_i$.
Expanding it up to the \emph{quadratic terms} we get:
\begin{align}
\label{eq:L-broken} \mathcal{L}_2(A,\chi) =
&-\frac{1}{2}A^a_\mu(-g^{\mu\nu}
\partial^2 + \partial^\mu \partial^\nu) A^a_\nu +
\frac{1}{2}(\partial_\mu \chi_i)^2 \nonumber\\
&+  g \partial^\mu \chi_i A^a_\mu {F^a}_i + \frac{1}{2} g^2
{F^a}_i {F^b}_i A^a_\mu A^{\mu b} - \frac{1}{2}M_{ij}\chi_i\chi_j,
\end{align}
with constant matrices $F^a$:
\begin{equation}
\label{eq:Fdefinition}
{F^a}_i \equiv T^a_{ij} \phi_{0j},
\end{equation}
and $M_{ij}$ the coefficient of the quadratic term in the Taylor
series expansion of the potential $V(\phi)$ at $\phi_0$:
\begin{equation}
\label{eq:M}
M_{ij} \equiv \left. \frac{\partial V(\phi)}{\partial
\phi_i
\partial \phi_j}\right|_{\phi_0}.
\end{equation}
Note that in the above Lagrangian the gauge bosons acquired a mass
with the mass matrix $g^2 {F^a}_i {F^b}_i$, and the scalar fields
have the mass matrix $M_{ij}$. This is basically the famous Higgs mechanism, when the gauge bosons, initially massless, acquire a mass through a spontaneous symmetry breaking of the scalar field potential.

In addition to the quadratic terms in (\ref{eq:L-broken}), the
full Lagrangian also includes various higher-order terms, which
after quantization describe the interactions between the fields. For the remainder of this section we will not explicitly write those terms out, and instead focus on the quadratic terms and how they are affected by the gauge-fixing. The reason for that is that the quadratic terms define the propagators, and thus the particle content of the theory, and the higher-order terms is then treated as a perturbation of the quadratic Lagrangian.

\subsection{Gauge-fixing}

The Yang-Mills Lagrangian described above still can not be used for direct calculation of Feynman diagrams because it possesses a gauge symmetry, which makes the straightforward path integrals badly divergent. The solution for this problem is the gauge-fixing procedure, which we describe here. As a result of it we will get an \emph{effective} Lagrangian, which can already be used for calculating matrix elements.

\subsubsection{Faddeev-Popov procedure}

Quantization of a theory, using the path-integral method, proceeds
by analyzing the quantity
\begin{equation}
\label{eq:Z} Z = \int \mathcal{D} A \, \mathcal{D} \chi \, e^{i
\int \mathcal{L}(A,\chi)},
\end{equation}
from which the propagators and interaction vertices can be read
off. However, for a Lagrangian that is invariant under some gauge
symmetry the straightforward analysis fails. The reason for this
is basically because the integral in (\ref{eq:Z}) runs over many
field configurations that are gauge-equivalent, instead of
counting each \emph{physical} configuration just once. In order to
avoid this problem, it is required to ``fix the gauge", that is,
to subject the fields to some constraint, which would prevent
over-counting, and perform quantization under this constraint.

Gauge-fixing is usually done by the Faddeev-Popov procedure, as a
result of which $Z$ is expressed as \cite{ps, weinberg2}:
\begin{equation}
\label{eq:Z-fixed} Z = C \cdot \int \mathcal{D} A \, \mathcal{D}
\chi \, \exp \left[ i \int d^4 x \, (\mathcal{L}[A,\chi] + \mathcal{L}_{\mathrm{gf}}(G^a))
\right] \det \left( \left. \frac{\delta
G^a[A_\alpha,\chi_\alpha;x]}{\delta \alpha^b(y)}
\right|_{\alpha=0} \right).
\end{equation}
Here $G^a[A,\chi;x]$ are arbitrary constraining
functions\footnote{Technically they are functionals, because they
can (and usually do) depend also on field derivatives at $x$} -- there are as
many of them as the number of gauge fields, and they are
``constraining'' in a sense that the condition $G^a=0$ serves as a
constraint on the fields in deriving (\ref{eq:Z-fixed}). Then the new gauge-fixing Lagrangian term $\mathcal{L}_{\mathrm{gf}}$ is \emph{any} function of the $G^a$, while $C$ is an irrelevant normalization constant. Subscript $\alpha$ near a field means that the field is gauge-transformed with gauge
transformation parameters $\alpha$.

The sum $(\mathcal{L}+\mathcal{L}_{\mathrm{gf}})$ now looks as an effective Lagrangian, which we could just use in the calculation of diagrams, but there is still the strange determinant factor preventing us from that. Luckily, it can also be turned into an effective Lagrangian term as follows. It can be shown, that given any
Hermitian matrix $D^{ab}(x,y)$, its determinant can be evaluated
by the field integral:
\begin{equation}
\label{eq:determinant} \det(D^{ab}(x,y)) = \int \mathcal{D}
\bar{c} \, \mathcal{D} c \, \exp \left( i \int d^4x \, d^4y \,
\bar{c}^a(x) D^{ab}(x,y)  c^b(y) \right),
\end{equation}
where $\bar{c}^a(x)$ and $c^a(x)$ are \emph{anticommuting} fields.
Using this relation, we can express (\ref{eq:Z-fixed}) as follows:
\begin{equation}
\label{eq:Z-ghost}
 Z = C \cdot \int \mathcal{D} A \, \mathcal{D} \chi
\, \mathcal{D} \bar{c} \, \mathcal{D} c \, \exp \left[ i \int d^4
x \, (\mathcal{L}[A,\chi] + \mathcal{L}_{\mathrm{gf}}(G^a) +
\mathcal{L}_\mr{gh}[\bar{c},c,A,\chi]) \right],
\end{equation}
\begin{equation}
\label{eq:Lg}
\mathcal{L}_\mr{gh} \equiv \bar{c}^a(x) \left( \left.
\frac{\delta G^a[A_\alpha,\chi_\alpha]}{\delta \alpha^b}
\right|_{\alpha=0} \right) c^b(x).
\end{equation}
Note that in the above equations compared to
(\ref{eq:determinant}) one coordinate integral is missing, because
the functional derivative $\frac{\delta G(x)}{\delta \alpha(y)}$
is ``local", in a sense that it contains a $\delta(x-y)$ factor,
which cancels the integral.  With $Z$ expressed as
(\ref{eq:Z-ghost}) it is apparent that we can account for the
determinant factor by introducing fictitious fields $c$ and
$\bar{c}$, called \emph{ghosts}, with their Lagrangian
$\mathcal{L}_\mr{gh}$ as in (\ref{eq:Lg}) and then use the regular
quantum field theory methods for calculations. Since the fields
$c$ where introduced as anticommuting numbers, the associated
particles, when calculating Feynman, are to be treated in a way
similar to fermions.

In the end we see, that the consequence of the gauge-fixing procedure is the new
effective Lagrangian
\begin{equation}
\mathcal{L}_{\mathrm{eff}} = \mathcal{L} + \mathcal{L}_{\mathrm{gf}} + \mathcal{L}_\mr{gh}
\end{equation}
which has to be used instead of the original $\mathcal{L}$ for the calculation of propagators and interaction
vertices, also adding the new \emph{ghost} particles to the theory.

\subsubsection{Linear gauge-fixing}

Let us now turn back to the Yang-Mills Lagrangian (\ref{eq:L-broken}) and apply the gauge-fixing procedure, resulting in the effective Lagrangian. Here we will discuss \emph{linear} gauge-fixing, meaning that the constraining functions $G^a$ are linear functions of the fields, and later we will consider the non-linear case. The usual choices for $G^a$ and $\mathcal{L}_{\mathrm{gf}}$ for gauge theories is:
\begin{align}
G^a(x) &= \frac{1}{\sqrt{\xi}}[\partial_\mu A^{a\mu}(x) - \xi g
{F^a}_i \chi_i(x)], \\
\mathcal{L}_{\mathrm{gf}} &= -\frac{1}{2}G^a G^a,
\end{align}
with an arbitrary gauge-fixing parameter $\xi$ - these are the
so-called $R_\xi$ gauges. Such a choice is enough to do calculations, but it lets us explore only a one-dimensional space of gauges, while in the context of this work we are interested to have as much gauge choosing freedom as possible, that is, as many gauge-fixing parameters as possible. For that purpose   here we will generalize the $R_\xi$ gauges by considering the most
general linear constraint, constructed from gauge fields'
four-divergences and scalar fields:
\begin{equation}
\label{eq:G} G^a = \partial_\mu A^{a\mu} - {K^a}_i \chi_i,
\end{equation}
with any real matrix ${K^a}_i$. Note that any overall factors
multiplying the constraints, or any linear redefinition of them
(as $G'^a = L^{ab} G^b$) is irrelevant, because it can be absorbed
in the definition of $\mathcal{L}_{\mathrm{gf}}$. This is consistent with thinking of
$G^a$ as constraints, because such redefinitions would actually
yield the same constraints on the fields. The most general $\mathcal{L}_{\mathrm{gf}}$
we are then interested in is quadratic in $G$:
\begin{equation}
\label{eq:B}
\mathcal{L}_{\mathrm{gf}} = -\frac{1}{2}G^a L^{ab} G^b,
\end{equation}
where $L^{ab}$ is any real square matrix and $-\frac{1}{2}$ is for
later convenience. First order of $G$ should not appear in $\mathcal{L}_{\mathrm{gf}}$,
because the total derivative term $\partial_\mu A^{a\mu}$ is
irrelevant in the Lagrangian, and the first-order field terms
$\chi_i$ would shift the minimum of the potential, so we have to
redefine the fields again. Then the higher order terms of $G$ in $\mathcal{L}_{\mathrm{gf}}$
would yield higher order field terms in $\mathcal{L}_{\mathrm{eff}}$,
which will be a matter of non-linear gauge-fixing. Another
note: $L^{ab}$ can be taken to be symmetrical, because the
antisymmetric part of it cancels out in (\ref{eq:B}) and,
therefore, is irrelevant. To summarize, our gauge-fixing choice
has now two matrices of parameters $K$ and $L$ (instead of a
single number $\xi$). The goal now is to find all the relations
these parameters have to satisfy, in order for the theory to be
quantized properly, and to find the maximum number of really
independent parameters.

We now proceed with the calculation, using our definitions for
$G^a$ and $\mathcal{L}_{\mathrm{gf}}$. Plugging (\ref{eq:G}) into (\ref{eq:B}) we get:
\begin{equation}
\mathcal{L}_{\mathrm{gf}} = -\frac{1}{2}\partial_\mu A^{a\mu} (L^{ab}) \partial_\nu
A^{b\nu} - \frac{1}{2}\chi_i ({K^a}_i L^{ab} {K^b}_j) \chi_j +
\partial_\mu A^{a\mu} (L^{ab} {K^b}_i) \chi_i,
\end{equation}
which then together with (\ref{eq:L-broken}) yields:
\begin{align}
\label{eq:Leff} \mathcal{L}_{\mathrm{eff}} = \mathcal{L}_2 + \mathcal{L}_\mr{gf}
=&-\frac{1}{2}A^a_\mu(-g^{\mu\nu}\partial^2 + (1-L)\partial^\mu
\partial^\nu - g^2 F F^T g^{\mu\nu})^{ab}A^b_\nu \nonumber\\
&+\frac{1}{2}\chi_i(-\partial^2-K^T L K - M)_{ij}\chi_j \\
&-\partial_\mu A^{a\mu} {(g F-LK)^a}_i \chi_i \nonumber
\end{align}
here we switched to matrix notation: ${F^a}_i={(F)^a}_i$,
${K^a}_i={(K)^a}_i$, $L^{ab}=(L)^{ab}$, $M_{ij}=(M)_{ij}$.

We now enforce a condition on the $\mathcal{L}_{\mathrm{eff}}$ that
\emph{the resulting propagators have to be diagonal in particle
species}. This is necessary in order for the fields in the
Lagrangian to represent the physical particles (mass eigenstates)
and for the perturbation theory to be valid. This diagonalization
can be done by field redefinition, but it is much more convenient
to use gauge-fixing to achieve that, when possible. The last term
in (\ref{eq:Leff}) is a cross-term between $A$ and $\chi$ and we
have to get rid of it somehow. Obviously, we can use the
gauge-fixing to do that easily by enforcing the first relation on
our parameters:
\begin{equation}
LK = g F,
\end{equation}
or
\begin{equation}
K = g L^{-1}F,
\end{equation}
leaving only $L$ as the independent matrix. Here, taking an
inverse, we assume that $L$ is non-singular, so we will have to
take a limit, if we want to consider some of the eigenvalues of
$L$ to be zero. Using the above expression for $K$,
(\ref{eq:Leff}) becomes:
\begin{align}
\label{eq:Leff2} \mathcal{L}_{\mathrm{eff}} =
&-\frac{1}{2}A^a_\mu(-g^{\mu\nu}\partial^2 + (1-L)\partial^\mu
\partial^\nu - g^2 F F^T g^{\mu\nu})^{ab}A^b_\nu \nonumber\\
&+\frac{1}{2}\chi_i(-\partial^2-g^2 F^T L^{-1} F - M)_{ij}\chi_j.
\end{align}

One remaining ingredient for the effective Lagrangian is the ghost term $\mathcal{L}_\mr{gh}$, which we calculate now. A
variation in $G^a$ under a gauge transformation, using
(\ref{eq:G}), (\ref{eq:deltaA}) and (\ref{eq:deltaphi}) is:
\begin{align}
\delta G^a &= \partial_\mu (\delta A^{a\mu}) - {K^a}_i (\delta
\chi_i) \nonumber \\
&= \frac{1}{g} \partial^2 \alpha^a + f^{abc}\partial_\mu A^{b\mu}
\alpha^c + {K^a}_i T^b_{ij} \phi_j \alpha^b \nonumber \\
&= \frac{1}{g}
\partial^2 \alpha^a + f^{abc}\partial_\mu ( A^{b\mu} \alpha^c ) +
{K^a}_i {F^b}_i \alpha^b + {K^a}_i T^b_{ij} \chi_j \alpha^b.
\label{eq:deltaG}
\end{align}
Then the ghost Lagrangian is:
\begin{equation}
\label{eq:Lgh}
\mathcal{L}_\mr{gh} = \bar{c}^a(x)\left(-\delta^{ab} \partial^2 - g {K^a}_i
{F^b}_i - g f^{acb} \left(\partial_\mu A^{c\mu}(x) + A^{c\mu}(x) \partial_\mu\right) - g {K^a}_i T^b_{ij}
\chi_j(x) \right) c^b(x),
\end{equation}
where we multiplied (\ref{eq:deltaG}) by an overall factor $(-g)$
-- we can do that because it just changes the overall irrelevant
factor $C$ in (\ref{eq:Z-fixed}). Let's again separate the quadratic term that gives the ghost propagators:
\begin{equation}
\mathcal{L}_\mr{gh2} = \bar{c}^a(x)(-\partial^2 - g K F^T)^{ab}
c^b(x),
\end{equation}
while the other two terms in $\mathcal{L}_\mr{gh}$ give ghost couplings
to the gauge and scalar fields.

To summarize, our final quadratic gauge-fixed Lagrangian, after
enforcing the relation $LK=gF$ and including ghosts is:
\begin{align}
\label{eq:Lfull2} \mathcal{L}_{\mathrm{eff2}} =
&-\frac{1}{2}A^a_\mu(-g^{\mu\nu}\partial^2 + (1-L)\partial^\mu
\partial^\nu - g^2 F F^T g^{\mu\nu})^{ab}A^b_\nu \nonumber\\
&+\frac{1}{2}\chi_i(-\partial^2-g^2 F^T L^{-1} F - M)_{ij}\chi_j\\
&+\bar{c}^a(-\partial^2 - g^2 L^{-1} F F^T)^{ab} c^b. \nonumber
\end{align}
We can easily read off the propagators for the gauge, scalar and
ghosts fields from this Lagrangian, but they will not necessarily be diagonal
in the species space, and, therefore, the fields in this
Lagrangian will not represent the \emph{physical} particles. In
order to proceed with finding the physical fields and their
(diagonal) propagators, we have to know the $F$ matrix, appearing
in the Lagrangian, and so we will continue this analysis when we consider the Standard Model in particular.

\subsubsection{Non-linear gauge-fixing}

Linear constraints are the simplest way to perform gauge-fixing. Linear
gauge-fixing is also essential because linear $G^a$ yields quadratic
$\mathcal{L}_{\mathrm{gf}}$ and so it participates in defining physical fields and
propagators. However, according to the Faddeev-Popov procedure of gauge-fixing,
$G^a$ can be \emph{any} functions of the fields, and so in particular they can
contain non-linear combinations of fields. Of course, we don't want to
introduce non-renormalizability into the theory by the gauge-fixing procedure,
so we have to keep the total mass dimension of the fields in $\mathcal{L}_{\mathrm{gf}}$
terms up to four. Additional requirement is that we would like to keep the quadratic part of $\mathcal{L}_{\mathrm{gf}}$ unchanged, since we have already determined how it should look. 

Assuming we keep $\mathcal{L}_{\mathrm{gf}}$ the same function of $G^a$, the modification that respects the requirements above is adding a quadratic term to $G^a$:
\begin{equation}
\label{eq:Gageneral}
G^a = \partial_\mu A^{a\mu} - {K^a}_i \chi_i + G_\mr{nl}^a,
\end{equation}
where $G_\mr{nl}^a$ is the non-linear, that is, quadratic part. In this case the quadratic part of $\mathcal{L}_{\mathrm{gf}}$ will remain the same as before, but it will have terms of third and fourth order, that will give contributions to the interaction vertices. Terms of third and higher order in fields are not allowed in $G^a$, because vector and scalar fields have mass dimension 1, so that would lead to higher than four mass dimension in the Lagrangian.

Apparently there is a lot of freedom for choosing the non-linear gauge-fixing terms $G_\mr{nl}^a$, which can lead to various modifications to the interaction vertices of the theory. We will explore those choices in detail in the following sections of this work, when we discuss the Standard Model specifically.

To summarize, the full effective Lagrangian of the Yang-Mills theory is
\begin{align}
\label{eq:Lfull} \mathcal{L}_{\mathrm{eff}} =
&-\frac{1}{2}A^a_\mu(-g^{\mu\nu}\partial^2 + (1-L)\partial^\mu
\partial^\nu - g^2 F F^T g^{\mu\nu})^{ab}A^b_\nu \nonumber\\
&+\frac{1}{2}\chi_i(-\partial^2-g^2 F^T L^{-1} F - M)_{ij}\chi_j\\
&+\bar{c}^a(-\partial^2 - g^2 L^{-1} F F^T)^{ab} c^b + \nonumber \\
&+\mathcal{L}_\mr{i} + \mathcal{L}_\mr{gfi} + \mathcal{L}_\mr{ghi}, \nonumber
\end{align}
with the quadratic terms written out explicitly. Note that the constant $g$ and constant matrices $F$ and $M$ are the parameters of the specific theory, while the matrix $L$ is an arbitrary matrix of gauge-fixing parameters - we will specify those constants in the next section for the case of the Standard Model. The terms of higher than quadratic order that describe the interactions are unspecified, but grouped in three categories:
 $\mathcal{L}_\mr{i}$ is the interaction part of the original Lagrangian (\ref{eq:l-unbroken}) containing terms of higher than quadratic order, $\mathcal{L}_\mr{gfi}$ are the interaction terms of the gauge-fixing Lagrangian arising from the nonlinear constraints $G_\mr{nl}^a$, and $\mathcal{L}_\mr{ghi}$ are the interaction terms involving ghosts which we have seen in (\ref{eq:Lgh}), with some additional terms from the non-linear part of gauge-fixing. 

We will use this effective Lagrangian in the next section, applying the framework of this section to the specific case of the Standard Model.


\section{The Standard Model}

In the previous section we have described the Yang-Mills gauge theory with a broken symmetry in general, and showed how the procedure of gauge-fixing is performed, leading to a new effective Lagrangian. Here we will describe the Standard Model, also called the Glashow-Weinberg-Salam model, which is the accepted theory of electroweak interactions, and which is, in fact, a particular case of a Yang-Mills theory with a broken symmetry. 

The goal of this section is to give a theoretical background for amplitude calculations in the Standard Model and arrive at the full explicit expression for the Lagrangian. The most important section in the context of this work is the discussion of non-linear gauge-fixing, which gives us the new interaction terms, that we later implement in the FeynArts package.

\subsection{The boson content and the propagators}
\label{sec:bosonpropagators}

Here we will work out the details of the quadratic part of the full effective Lagrangian (\ref{eq:Lfull}), which will tell us the physical particles of the theory and their propagators. First we will determine the matrices $F$ and $M$ and then diagonalize the propagators, in order to find the physical fields.

The electroweak theory goes along the general framework of the previous section,
with the gauge group
\begin{equation}
G = SU(2)\times U(1),
\end{equation}
with structure constants (assuming $t^4$ is the generator of $U(1)$):
\begin{equation}
f^{abc} = \left\{ \begin{array}{ll}
	\epsilon^{abc}, & a,b,c \in \{1,2,3\}, \\
	0, & \mr{otherwise}.
\end{array} \right.
\end{equation}
The real scalar fields $\phi_i$ are taken to transform as the real
components of the usual two-dimensional complex representation of
$SU(2)$ with generators:
\begin{align}
t^a &= \frac{\sigma^a}{2}, \quad a \in \{ 1, 2, 3 \} \\
t^4 &= \frac{1}{2},
\end{align}
where $\sigma^a$ are the Pauli matrices and $t^4$ is the generator
of $U(1)$. We define the real components of the complex
two-dimensional vector $\phi$ as:
\begin{equation}
\label{eq:Phidefinition}
\phi = \frac{1}{\sqrt{2}} \left(%
\begin{array}{c}
    -i\phi^1 - \phi^2 \\
    \phi^4 + i \phi^3 \\
\end{array} %
\right),
\end{equation}
then the real representation matrices $T^a=-it^a$ in $\phi^i$
space look as:
\begin{align}
T^1 = \frac{1}{2} \left(
\begin{array}{cccc}
    0 & 0 & 0 &+1 \\
    0 & 0 &-1 & 0 \\
    0 &+1 & 0 & 0 \\
   -1 & 0 & 0 & 0 \\
\end{array}
\right), \quad T^2 = \frac{1}{2}\left(
\begin{array}{cccc}
    0 & 0 &+1 & 0 \\
    0 & 0 & 0 &+1 \\
   -1 & 0 & 0 & 0 \\
    0 &-1 & 0 & 0 \\
\end{array}
\right), \nonumber
\\
\label{eq:T}
 T^3 = \frac{1}{2}\left(
\begin{array}{cccc}
    0 &-1 & 0 & 0 \\
   +1 & 0 & 0 & 0 \\
    0 & 0 & 0 &+1 \\
    0 & 0 &-1 & 0 \\
\end{array}
\right), \quad T^4 = \frac{1}{2}\left(
\begin{array}{cccc}
    0 &-1 & 0 & 0 \\
   +1 & 0 & 0 & 0 \\
    0 & 0 & 0 &-1 \\
    0 & 0 &+1 & 0 \\
\end{array}
\right),
\end{align}
and their commutation relations are, naturally:
\begin{align}
[T^a,T^b] &= \epsilon^{abc} T^c, \quad a,b,c \in \{1,2,3\}, \\
[T^a,T^4] &= 0.
\end{align}

Now, following the previous section, we consider the symmetry
breaking of this group, which happens because the scalar fields
are subjected to some potential $V(\phi)$. A renormalizable potential that obeys the gauge symmetry and doesn't have a local minimum at $\phi=0$, as required for symmetry breaking, is:
\begin{equation}
\label{eq:Vexpression}
V(\phi) = -\frac{\mu^2}{2} (\phi^i\phi^i) + \frac{\lambda}{4} (\phi^i\phi^i)^2,
\end{equation} 
with some unknown parameters $\mu$ and $\lambda$. As a result
of it, $\phi$ acquires a vacuum expectation value (VEV) at the minimum of $V$:
\begin{equation}
\label{eq:v}
|\phi_0| = v \equiv \sqrt{\frac{\mu^2}{\lambda}}. 
\end{equation}
Note that all directions of $\phi$ are equivalent under the gauge transformation, so any choice of direction for $\phi_0$ gives the same results, so we can arbitrarily choose it as:
\begin{equation}
\phi_0 = \left( \begin{array}{c}
    0 \\
    0 \\
    0 \\
    v \\
    \end{array} \right),
\end{equation}
It is important to note that there is one unbroken
symmetry direction: the combination of generators
\begin{equation}
Q = T^3+T^4
\end{equation}
leaves $\phi_0$ invariant, and, therefore, this transformation is
still a symmetry of the theory, even after $\phi$ acquires a VEV.

Now we calculate $g{F^a}_i$ according to (\ref{eq:Fdefinition}) to be:
\begin{equation}
g{F^a}_i = g (T^a)_{ij} (\phi_0)_j = \frac{v}{2}
\left(%
\begin{array}{cccc}
  g & 0 & 0 & 0 \\
  0 & g & 0 & 0 \\
  0 & 0 & g & 0 \\
  0 & 0 & -g' & 0 \\
\end{array}%
\right),
\end{equation}
here down the matrix is the $a$-dimension, and to the right is the
$i$-dimension. Note that we introduced two different coupling
constants, $g$ and $g'$, because, as mentioned earlier, they can
be chosen independently for each semi-simple or $U(1)$ subgroup.

\subsubsection{Gauge bosons}

Let us now continue with the task of diagonalizing propagators and
first consider the gauge bosons, with their Lagrangian
\begin{equation}
\mathcal{L}_A = -\frac{1}{2}A^a_\mu(-g^{\mu\nu}\partial^2 +
(1-L)\partial^\mu
\partial^\nu - g^2 F F^T g^{\mu\nu})^{ab}A^b_\nu.
\end{equation}
Because of the different spacetime tensor coefficients of $L$ and
of the mass term $FF^T$ each one has to be diagonalized
separately. We first proceed with the mass term, because it is
already given:
\begin{equation}
g^2 F F^T = \frac{v^2}{4}
\left(%
\begin{array}{cccc}
  g^2 & 0 & 0 & 0 \\
  0 & g^2 & 0 & 0 \\
  0 & 0 & g^2 & -gg' \\
  0 & 0 & -gg' & g'^2 \\
\end{array}%
\right).
\end{equation}
The diagonalization of this term is done by field redefinition, so
we want to express the mass term as:
\begin{equation}
g^2 F F^T = U^\dag M_A U,
\end{equation}
with unitary transformation $U$, and diagonal mass matrix $M_A$.
We take:
\begin{equation}
M_A = \frac{v^2}{4}
\left(%
\begin{array}{cccc}
  g^2 & 0 & 0 & 0 \\
  0 & g^2 & 0 & 0 \\
  0 & 0 & g^2+g'^2 & 0 \\
  0 & 0 & 0 & 0 \\
\end{array}%
\right),
\end{equation}
\begin{equation}
\label{eq:U} U=
\left(%
\begin{array}{cccc}
  \frac{1}{\sqrt{2}} & -\frac{i}{\sqrt{2}} & 0 & 0 \\
  \frac{1}{\sqrt{2}} & \frac{i}{\sqrt{2}} & 0 & 0 \\
  0 & 0 & \frac{g}{\sqrt{g^2+g'^2}} & -\frac{g'}{\sqrt{g^2+g'^2}} \\
  0 & 0 & \frac{g'}{\sqrt{g^2+g'^2}} & \frac{g}{\sqrt{g^2+g'^2}} \\
\end{array}%
\right),
\end{equation}
which yields physical fields $A'=UA$ and their masses as:
\begin{align}
W^\pm_\mu \equiv A'^{1,2}_\mu  = \frac{1}{\sqrt{2}}(A^1_\mu \mp i
A^2_\mu)\quad\quad\quad &m_W = g\frac{v}{2},\nonumber\\
Z^0_\mu \equiv A'^3_\mu = \frac{1}{\sqrt{g^2+g'^2}} (g A^3_\mu -
g' A^4_\mu)\quad\quad\quad &m_Z = \sqrt{g^2+g'^2} \frac{v}{2},
\label{eq:Aphysical}\\
A_\mu \equiv A'^4_\mu = \frac{1}{\sqrt{g^2+g'^2}} (g' A^3_\mu + g
A^4_\mu)\quad\quad\quad &m_A = 0.\nonumber
\end{align}
This is the main result of the electroweak theory (independently
of the gauge-fixing details), that there are three massive gauge
bosons, mediating the weak interaction, and one massless boson,
which represents the unbroken symmetry and is nothing else but the
photon.

Before going on further let's define some new constants for later convenience. Consider replacing the two unknown coupling constants $g$ and $g'$ with other two constants: electric charge $e$ and Weinberg angle $\theta_w$ defined as
\begin{equation}
e=\frac{gg'}{\sqrt{g^2+g'^2}}, \quad
\cos\theta_w = \frac{g}{\sqrt{g^2+g'^2}}, \quad
\sin\theta_w = \frac{g'}{\sqrt{g^2+g'^2}},
\end{equation}
with the inverse transformation:
\begin{equation}
g = \frac{e}{\sin\theta_w}, \quad 
g' = \frac{e}{\cos\theta_w}.
\end{equation}
It is not apparent yet, but it will be shown briefly that $e$ as defined here is indeed the electromagnetic coupling constant. The definition of $\theta_w$ allows to rewrite $Z$ and $A$ more clearly as:
\begin{eqnarray}
Z_\mu = \cos \theta_w A^3_\mu - \sin\theta_w A^4_\mu, \\
A_\mu = \sin\theta_w A^3_\mu + \cos\theta_w A^4_\mu,
\end{eqnarray}
and the masses as:
\begin{equation}
m_W = v \frac{e}{2\sin\theta_w}, \quad 
m_Z = v \frac{e}{2\cos\theta_w\sin\theta_w}.
\end{equation}
Also note an important relationship (or, alternatively, a definition of $\theta_w$):
\begin{equation}
m_W = m_Z \cos\theta_w.
\end{equation}
One more expression that will be very useful later is the covariant derivative from (\ref{eq:Dphi})
\begin{equation}
D_\mu = \partial_\mu - ig (A^1_\mu \hat{t}^1 + A^2_\mu \hat{t}^2 + A^3_\mu \hat{t}^3) - ig'A^4_\mu\hat{t}^4
\end{equation}
expressed in terms of the physical gauge fields. Note that the $\hat{t}^i$ here are generators and depend on the representation. Substituting $A^i$ with the physical fields in the end we get:
\begin{equation}
\label{eq:Dphysical}
D_\mu = \partial_\mu 
- i\frac{g}{\sqrt{2}}(W^+_\mu \hat{t}^+ + W^-_\mu \hat{t}^-) 
- i\frac{g}{\cos\theta_w}Z_\mu(\hat{t}^3 - \sin^2\theta_w \hat{q})
- i e A_\mu \hat{q},
\end{equation}
with the new generators:
\begin{align}
\hat{t}^\pm \equiv& (\hat{t}^1 \pm i \hat{t}^2),\\
\hat{q} \equiv& (\hat{t}^3 + \hat{t}^4).
\end{align}
We can see now that we have recovered the regular coupling of the field $A_\mu$ with a coupling constant $e$ and a charge given by the value of the generator $\hat{q}$.

Let's now finish analyzing gauge bosons. Some additional explanation is needed, why the $W^\pm$ are defined exactly like they are. Because there are two identical
eigenvalues of the mass matrix, any unitary transformation of
$A^1$ and $A^2$ would lead to two energy eigenstates with the same
mass. Therefore, \emph{additionally} we want to make them
eigenstates of the remaining symmetry generator $Q$, which with
the identification of $A_\mu$ as the electromagnetic field,
happens to be the \emph{electric charge} operator. Since gauge
fields form an adjoint representation of the gauge group $G$, we
can calculate their transformation properties using commutation
relations of the group operators. Specifically, if we define the
field \emph{operator}, corresponding to field $A^a$ to be a linear
superposition of generators (here we will use hats for operators
to avoid confusion):
\begin{equation}
\hat{A} \equiv \hat{T}^a A^a,
\end{equation}
then the transformation of the field operator under a gauge transformation is given by the commutator
\begin{equation}
\delta \hat{A} = - \alpha^a [ \hat{T}^a, \hat{A}].
\end{equation}
We can use this relation to find the transformation of the fields $W^\pm$. First we need the corresponding generators $\hat{T}_{W^\pm}$. Since the full field operator
\begin{equation}
\hat{A} = \hat{T}^a A^a = \hat{T}^+ W^+ + \hat{T}^- W^- + \hat{T}^Z Z + \hat{T}^A A
\end{equation}
is constructed from multiplying the fields with generators, the generators undergo an \emph{inverse} transformation to that of the fields, therefore:
\begin{equation}
\hat{T}^\pm = \frac{1}{\sqrt{2}} (\hat{T}^1 \pm i \hat{T}^2).
\end{equation}
Now we can see that the commutators corresponding to the transformation of $W^\pm$ under the remaining symmetry:
\begin{align}
\delta \hat{T}^+ &= - \alpha [\hat{Q}, \hat{T}^+] = -
\frac{\alpha}{\sqrt{2}} [ \hat{T}^3 + \hat{T}^4, \hat{T}^1 + i
\hat{T}^2] =
- \frac{\alpha}{\sqrt{2}}(\hat{T}^2-i\hat{T}^1) = +i \alpha \hat{T}^+,  \nonumber \\
\delta \hat{T}^- &= - \alpha [\hat{Q}, \hat{T}^-] = -
\frac{\alpha}{\sqrt{2}} [ \hat{T}^3 + \hat{T}^4, \hat{T}^1 - i
\hat{T}^2] = -\frac{\alpha}{\sqrt{2}}(\hat{T}^2+i\hat{T}^1) = -i
\alpha \hat{T}^-.
\end{align}
indeed show that $W^\pm$ are eigenstates, or, to put it differently,
transform \emph{irreducibly} under the charge transformation.

We can now put final constraints on the gauge-fixing parameter matrix $L$.
With the gauge fields transformed to the physical basis as above, the Lagrangian for them becomes:
\begin{equation}
\label{eq:LAdiagonal}
\mathcal{L}_A =
-\frac{1}{2}(A'^*)^a_\mu(-g^{\mu\nu}\partial^2 + (1-U L
U^\dagger)\partial^\mu
\partial^\nu - M_Ag^{\mu\nu})^{ab}A'^b_\nu.
\end{equation}
Since we can not use field redefinition anymore, we have to insist
for the gauge-fixing term
\begin{equation}
L' \equiv U L U^\dagger
\end{equation}
to be diagonal in this basis as well. This leaves us with four
real parameters, which we write, in order to conform with the
usual notation, as:
\begin{equation}
L' =
\left(%
\begin{array}{cccc}
  1/\xi_+ & 0 & 0 & 0 \\
  0 & 1/\xi_- & 0 & 0 \\
  0 & 0 & 1/\xi_Z & 0 \\
  0 & 0 & 0 & 1/\xi_A \\
\end{array}%
\right).
\end{equation}
This gives us the original matrix:
\begin{equation}
L = U^\dagger L' U =
\left(%
\begin{array}{cccc}
  \frac{\xi_+ + \xi_-}{2\xi_+\xi_-} & i \frac{\xi_+ - \xi_-}{2\xi_+\xi_-} & 0 & 0 \\
  -i \frac{\xi_+ - \xi_-}{2\xi_+\xi_-} & \frac{\xi_+ + \xi_-}{2\xi_+\xi_-} & 0 & 0 \\
  0 & 0 & \frac{\xi_A g^2 + \xi_Z g'^2}{\xi_A\xi_Z(g^2+g'^2)} & \frac{g g'(\xi_Z - \xi_A)}{\xi_A\xi_Z(g^2+g'^2)} \\
  0 & 0 & \frac{g g'(\xi_Z - \xi_A)}{\xi_A\xi_Z(g^2+g'^2)} & \frac{\xi_Z g^2 + \xi_A g'^2}{\xi_A\xi_Z(g^2+g'^2)} \\
\end{array}%
\right).
\end{equation}
Since $L$ can neither be complex nor asymmetrical, we conclude
that
\begin{equation}
\xi_+ = \xi_- \equiv \xi_W,
\end{equation}
giving the $L$ matrices:
\begin{equation}
L' =
\left(%
\begin{array}{cccc}
  1/\xi_W & 0 & 0 & 0 \\
  0 & 1/\xi_W & 0 & 0 \\
  0 & 0 & 1/\xi_Z & 0 \\
  0 & 0 & 0 & 1/\xi_A \\
\end{array}%
\right),
\end{equation}
\begin{equation}
L =
\left(%
\begin{array}{cccc}
  1/\xi_W & 0 & 0 & 0 \\
  0 & 1/\xi_W & 0 & 0 \\
  0 & 0 & \frac{\xi_A g^2 + \xi_Z g'^2}{\xi_A\xi_Z(g^2+g'^2)} & \frac{g g'(\xi_Z - \xi_A)}{\xi_A\xi_Z(g^2+g'^2)} \\
  0 & 0 & \frac{g g'(\xi_Z - \xi_A)}{\xi_A\xi_Z(g^2+g'^2)} & \frac{\xi_Z g^2 + \xi_A g'^2}{\xi_A\xi_Z(g^2+g'^2)} \\
\end{array}%
\right),
\end{equation}
and the other gauge-fixing parameter $K$:
\begin{equation}
{K^a}_i = g L^{-1} F = \frac{v}{2}
\left(%
\begin{array}{cccc}
  g\xi_W & 0 & 0 & 0 \\
  0 & g\xi_W & 0 & 0 \\
  0 & 0 & g\xi_Z & 0 \\
  0 & 0 & -g'\xi_Z & 0 \\
\end{array}%
\right).
\end{equation}

This is our final result for the parametrization of linear gauge-fixing.
It leaves us with 3 real parameters $(\xi_W, \xi_Z, \xi_A)$,  a
slight generalization of usual $R_\xi$ gauges, which would
correspond to taking all 3 parameters to be equal
$\xi=\xi_W=\xi_Z=\xi_A$.

Finally, the Lagrangian for gauge bosons in the physical basis
(\ref{eq:LAdiagonal}) becomes explicitly diagonal:
\begin{equation}
\label{eq:LAfinal} \mathcal{L}_A = \sum_{X}
-\frac{1}{2}X^*_\mu(-g^{\mu\nu}\partial^2 + (1-\xi_X)\partial^\mu
\partial^\nu - m_X^2 g^{\mu\nu})X_\nu,
\end{equation}
with $X\in \{W^\pm,Z^0,A\}$ and masses as in (\ref{eq:Aphysical}).
The momentum-space propagators can now be easily computed by taking
the inverse of the coefficient for the quadratic term in the
Lagrangian, and multiplying it by $i$. Switching to momentum
space, we replace all $\partial_\mu$ by $-ik_\mu$, which yields for gauge bosons:
\begin{equation}
\langle X^{*\mu} X^\nu \rangle =
\frac{-i}{k^2-m_X^2}\left(g^{\mu\nu}-\frac{k^\mu k^\nu}{k^2-\xi_X
m_X^2}(1-\xi_X)\right), \quad X\in\{W^\pm,Z^0,A\}.
\end{equation}

\subsubsection{Goldstone bosons and Higgs}

Let's move on now to the quadratic scalar part of the Lagrangian (\ref{eq:Lfull}):
\begin{equation}
\label{eq:Lchi}
\mathcal{L}_\chi = \frac{1}{2}\chi_i(-\partial^2-g^2 F^T L^{-1} F - M)_{ij}\chi_j
\end{equation}
An important separation occurs in this Lagrangian as follows. It is a general theorem
(called Goldstone theorem) that the mass matrix of $\chi_i$
defined in (\ref{eq:M}) satisfies
\begin{equation}
(T^a\phi_0)_i M_{ij} = {F^a}_i M_{ij} = 0,
\end{equation}
for all $a$. It follows from the fact that $V(\phi)$ is
gauge-invariant and it basically says that $V(\phi)$ doesn't form
a quadratic potential in the directions that $\phi_0$ transforms
to. Consequently, the only non-zero term allowed in $M$ is:
\begin{equation}
M_{44} \equiv m_H^2.
\end{equation}
We can confirm the above result using the explicit expression for $V(\phi)$ from (\ref{eq:Vexpression}):
\begin{equation}
M_{ij} \equiv \left. \frac{\partial V(\phi)}
{\partial\phi_i\partial \phi_j}\right|_{\phi_0} =
\left. \left(-\mu^2 \delta_{ij} + \lambda (\phi^k\phi^k) \delta_{ij} + 
2 \lambda \phi^i \phi^j \right) \right|_{\phi=\phi_0}.
\end{equation}
With $(\phi_0)_i=v\delta_{i4}$ this becomes:
\begin{equation}
M_{ij} = (v^2 \lambda - \mu^2) \delta_{ij} + 2 v^2 \lambda \delta_{i4} \delta_{j4},
\end{equation}
and after substituting $v=\sqrt{\mu^2/\lambda}$:
\begin{equation}
M_{ij} = 2 \mu^2 \delta_{i4} \delta_{j4},
\end{equation}
confirming the argument above with identification:
\begin{equation}
m_H = \sqrt{2} \mu
\end{equation}
On the other hand, the $(g^2 F^T L^{-1} F)_{ij}$ term in
(\ref{eq:Lchi}) will be zero where $i=4$ or $j=4$, because of the
zero column in $F$. Therefore, we can write $\mathcal{L}_\chi$ as:
\begin{align}
\mathcal{L}_\chi &= \mathcal{L}_G +
\mathcal{L}_H,\\
\mathcal{L}_G &= \frac{1}{2}\chi_m(-\partial^2-g^2 F^T L^{-1}
F)_{mn}\chi_n, \\
\mathcal{L}_H &= \frac{1}{2}h(-\partial^2-m_H^2) h,
\end{align}
where $m,n \in \{1,2,3\}$ and we have renamed $\chi_4\equiv h$. The Lagrangian separates and we get
three \emph{massless} fields $\chi_m$ called \emph{Goldstone
bosons} (the mass-like term is proportional to $L$, which indicates that they have no physical mass), and one \emph{massive} field $h$ with an unknown mass $m_H$, which is called the \emph{Higgs boson}. We see that the massive field is unaffected by the gauge-fixing procedure, and it already has a diagonal propagator
\begin{equation}
\langle hh \rangle = \frac{i}{k^2-m^2_H}.
\end{equation}

It's worth noting here that the Higgs boson is a physical particle, which is an essential prediction of the Standard Model and which yet has to be discovered experimentally in particle colliders. On the other hand the three Goldstone bosons are not actually physical, that is, they can not appear in the initial or final state in amplitude calculations. Instead, they represent the missing degrees of freedom needed to make the three gauge bosons $W^\pm, Z$ massive from the initially massless vector bosons. This is in general the way that the Goldstone bosons function in theories with a broken gauge symmetry, and there will always be as many of them, as the number of massive gauge bosons.

Of course, we still have to include the Gauge bosons as the intermediate particles in amplitude calculations, so we continue now with diagonalizing their remaining Lagrangian:
\begin{equation}
\mathcal{L}_G = \frac{1}{2}\chi_m(-\partial^2-g^2 F^T L^{-1}
F)_{mn}\chi_n,
\end{equation}
with the mass matrix:
\begin{equation}
M_\chi = g^2 F^T L^{-1} F =
\frac{v^2}{4}
\left(%
\begin{array}{ccc}
  g^2 \xi_W & 0 & 0  \\
  0 & g^2 \xi_W & 0 \\
  0 & 0 & (g^2+g'^2) \xi_Z  \\
\end{array}%
\right).
\end{equation}
We see that it is already diagonal in the current basis and the
masses are $\sqrt{\xi_X} m_X$, with first two Goldstone bosons
corresponding to $W$ and the third to $Z^0$. However, like with
gauge bosons, since two mass eigenvalues are equal, the states
should be chosen as eigenstates of charge operator, which, using
(\ref{eq:T}), in the representation of $\chi_m$ is:
\begin{equation}
Q=T^3 + T^4 = \left(
\begin{array}{ccc}
    0 &-1 & 0 \\
   +1 & 0 & 0 \\
    0 & 0 & 0 \\
\end{array}
\right).
\end{equation}
We use the same redefinition for Goldstone boson fields as for gauge bosons:
\begin{equation}
\label{eq:chiphdefinition}
\chi^\pm = \frac{1}{\sqrt{2}}(\chi_1 \mp i \chi_2),
\end{equation}
which means that the states, similarly to generators of gauge bosons, undergo an \emph{inverse} transformation:
\begin{equation}
\ket{\chi^\pm} = \frac{1}{\sqrt{2}} \left( \ket{\chi_1} \pm i \ket{\chi_2} \right).
\end{equation}
We can check now that these are indeed the charge eigenstates:
\begin{equation}
\delta\ket{\chi^\pm} =
-\alpha\hat{Q}\ket{\chi^\pm} = 
-\frac{\alpha}{\sqrt{2}}(\ket{\chi_2} \mp i \ket{\chi_1}) = 
\pm i \frac{\alpha}{\sqrt{2}}(\ket{\chi_1} \pm i \ket{\chi_2}) = 
\pm i \alpha \ket{\chi^\pm},
\end{equation}
so the Lagrangian in the diagonal form looks as:
\begin{equation}
\label{eq:Lchifinal} \mathcal{L}_G = \sum_X
\frac{1}{2}\chi_X(-\partial^2-\xi_X m^2_X)\chi_X,
\end{equation}
with $X\in\{W^\pm,Z^0\}$. That gives the following propagators for Goldstone bosons:
\begin{equation}
\langle \chi_X \chi_X \rangle = \frac{i}{k^2-\xi_X m^2_X}, \quad
X\in\{W^\pm,Z^0\}.
\end{equation}

\subsubsection{Ghosts}

Finally, let's look at the quadratic ghost Lagrangian piece in (\ref{eq:Lfull}):
\begin{equation}
\mathcal{L}_{g} = \bar{c}^a(-\partial^2 - g^2 L^{-1} F F^T)^{ab}
c^b.
\end{equation}
It has the mass matrix:
\begin{equation}
M_g = g^2 L^{-1} F F^T = \frac{v^2}{4}
\left(%
\begin{array}{cccc}
  g^2 \xi_W & 0 & 0 & 0 \\
  0 & g^2 \xi_W & 0 & 0 \\
  0 & 0 & g^2 \xi_Z & -gg' \xi_Z \\
  0 & 0 & -gg' \xi_Z & g'^2 \xi_Z \\
\end{array}%
\right),
\end{equation}
which, like the matrix for gauge bosons, is not diagonal. Notice,
however, that it looks very similar to the gauge boson mass
matrix, except for the $\xi$ factors. Indeed, if we use the
\emph{same} unitary transformation (\ref{eq:U}) for the ghost
fields, as we did for gauge bosons:
\begin{align}
\label{eq:physicalghosts}
c'^a &= U^{ab} c^b,  \nonumber \\
\bar{c}'^a &= \bar{c}^b (U^\dagger)^{ba},
\end{align} then the
transformed mass matrix becomes diagonal:
\begin{equation}
M'_g = U M_g U^\dagger = \frac{v^2}{4}
\left(%
\begin{array}{cccc}
  g^2 \xi_W & 0 & 0 & 0 \\
  0 & g^2 \xi_W & 0 & 0 \\
  0 & 0 & (g^2+g'^2)\xi_Z & 0 \\
  0 & 0 & 0 & 0 \\
\end{array}%
\right).
\end{equation}
The masses for the ghost fields are then, similar as for Goldstone
bosons, $\sqrt{\xi_X} m_X$, where $m_X$ is the mass of the
corresponding gauge boson. However, unlike the case with the
Goldstone bosons, there is also a massless ghost corresponding to
the massless vector field $A_\mu$. And so the diagonal Lagrangian
for ghosts is then:
\begin{equation}
\label{eq:Lgfinal} \mathcal{L}_{g} = \sum_X \bar{c}_X(-\partial^2
- \xi_X m_X^2) c_X,
\end{equation}
with $X\in\{W^\pm,Z^0,A\}$, and the propagators are
\begin{equation}
\langle \bar{c}_X c_X \rangle = \frac{i}{k^2-\xi_X m^2_X}, \quad
X\in\{W^\pm,Z^0,A\}.
\end{equation}

\subsection{Fermions}
\label{sec:fermions}

Up to now we have only included scalar and gauge vector fields in our theory, because they are the crucial ingredients of a gauge theory with Higgs mechanism. In addition to that the Standard Model, of course, contains fermions which are the matter constituents, and so here we will discuss their Lagrangian.

Let's start with a free massless two-component \emph{Weyl spinor} $\psi$, which can be either left-handed or right-handed. It has the Lagrangian:
\begin{equation}
\mathcal{L} = i\bar{\psi} \slashpartial  \psi.
\end{equation}
In a gauge theory that extends naturally to
\begin{equation}
\mathcal{L} = i\bar{\psi} \slashD \psi
\end{equation}
with a covariant derivative as in (\ref{eq:Dphi}):
\begin{equation}
D_\mu = \partial_\mu - i g A^a_\mu t^a,
\end{equation}
with generators $t^a$ dependent on the representation that a particular fermion belongs to. The only question then is how to group the fermions into representations of the $SU(2)$ and what charges of the $U(1)$ subgroup to assign to those groups. Note that the ``charge'' of $U(1)$ is just the value of the generator $t^4$, also conventionally called $Y$.

We have the following observed fermions to fit into the Standard Model: electron $e^-$, neutrino $\nu_e$, up-quark $u$, down-quark $d$, and two copies of these particles with exactly the same properties but different masses called the second generation $(\mu^-, \nu_\mu, c, s)$ and the third generation $(\tau^-, \nu_\tau, t, b)$. Note that all the particles here except for neutrinos have both left-handed and right-handed components, but it appears that these components have to be treated as basically two distinct particles, for example $e^-_L$ and $e^-_R$, only connected by a mass term, as will be shown briefly. Neutrinos have only the left-handed component, and we will skip the subscript $L$ for them.

The particles enumerated above arrange into representations as follows: the left-handed components form \emph{doublets} of $SU(2)$
\begin{equation}
E_L = 
\begin{pmatrix}
\nu_{e} \\ e^-_L
\end{pmatrix},
\quad\quad\quad
Q_L =
\begin{pmatrix}
u_L \\ d_L
\end{pmatrix},
\end{equation}
called the lepton doublet and the quark doublet with $U(1)$ charges
\begin{equation}
Y_E = -1/2, \quad\quad\quad Y_Q = 1/6
\end{equation}
respectively. The right-handed components are \emph{singlets} of $SU(2)$:
\begin{equation}
e^-_R, \quad u_R, \quad d_R,
\end{equation}
with charges
\begin{equation}
Y_e=-1, \quad Y_u=2/3, \quad Y_d=-1/3.
\end{equation}
Remember that the components of the doublets have $t^3=\pm 1/2$, while for singlets $t^3=0$, and that the electric charge $q=t^3+Y$, which gives the right charges $q_\nu=0$, $q_e=-1$, $q_u=2/3$, $q_d=-1/3$.
The other two generations look exactly the same, so we will not repeat them here, just keep in mind that all the calculations here apply equally to all three generations.

We can now write the massless Lagrangian for all fermions in the first generation:
\begin{equation}
\mathcal{L} = i \bar{E}_L \slashD E_L + i \bar{e}_R \slashD e_R +
i \bar{Q}_L \slashD Q_L + i \bar{u}_R \slashD u_R + i \bar{d}_R \slashD d_R.
\end{equation}
Using the expression for $D_\mu$ given in (\ref{eq:Dphysical}) and the values for the generators ($t^i=\sigma^i/2$ for doublets) we get explicitly:
\begin{align}
\label{eq:Lfgauge}
\mathcal{L} =& i \bar{E}_L \slashpartial E_L + i \bar{e}_R \slashpartial e_R +
i \bar{Q}_L \slashpartial Q_L + i \bar{u}_R \slashpartial u_R + i \bar{d}_R \slashpartial d_R\\
&+ g(W^+_\mu J^{\mu+}_W + W^-_\mu J^{\mu-}_W + Z_\mu J^\mu_Z) + e A_\mu J^\mu_A,\nonumber\\
J^{\mu+}_W \equiv& \frac{1}{\sqrt{2}}(\bar{\nu_e} \gamma^\mu e_L 
+ \bar{u}_L \gamma^\mu d_L),\\
J^{\mu-}_W \equiv& \frac{1}{\sqrt{2}}(\bar{e}_L \gamma^\mu \nu_e 
+ \bar{d}_L \gamma^\mu u_L),\\
J^\mu_A \equiv& 
- \bar{e}\gamma^\mu e
+\tfrac23 \bar{u}\gamma^\mu u
-\tfrac13 \bar{d}\gamma^\mu d,
\\
J^\mu_Z \equiv& \frac{1}{\cos\theta_w}\left[
\tfrac12\bar{\nu_e}\gamma^\mu\nu_e
-\tfrac12\bar{e}_L\gamma^\mu e_L
+\tfrac12\bar{u}_L\gamma^\mu u_L
-\tfrac12\bar{d}_L\gamma^\mu d_L
- \sin^2\theta_w J^\mu_A \right] 
.
\end{align}
The different fermion currents $J^{\mu+}_W$, $J^{\mu-}_W$, $J^\mu_Z$, $J^\mu_A$ that were defined here each couple to a different gauge boson. Also note that the fields $e$\footnote{not to be confused with the coupling constant $e$!}, $u$, $d$ without a subscript $L$ or $R$ denote the sum of both left (e.g. $e_L$) and right ($e_R$) components, and occur if the coupling is non-chiral. We can see that the electromagnetic current $J^\mu_A$, as expected, is completely non-chiral, while the $J^\mu_Z$ has a non-chiral part proportional to $J^\mu_A$ and a pure left-handed part.

We have now the correct coupling of fermions to gauge bosons, but the fermions are still massless, which in reality is not true. The mass term for a fermion (e.g. electron) looks like
\begin{equation}
\mathcal{L} = -m_e \bar{e} e = -m_e (\bar{e}_L e_R + \bar{e}_R e_L),
\end{equation}
but such term is not allowed in the gauge-invariant Lagrangian, because the left and right components belong to different representations of $SU(2)$, thus their product is not invariant. In order then for fermions to acquire mass, we again use the Higgs mechanism -- that is, we couple the fermions to the scalar fields $\phi$ and, after the symmetry breaking, the non-zero expectation value will provide the effective mass term. The allowed gauge-invariant coupling Lagrangian is \cite{ps}:
\begin{equation}
\mathcal{L} = -\lambda_e \bar{E}_L \phi e_R
-\lambda_d \bar{Q}_L \phi d_R
-\lambda_u \epsilon^{ab} \bar{Q}_{La}\phi^\dagger_b u_R + \mr{h.c.},
\end{equation}
with unknown parameters $\lambda_e, \lambda_d, \lambda_u$. The scalar field $\phi$ here is the same two-component complex doublet under $SU(2)$ as in (\ref{eq:Phidefinition}):
\begin{equation}
\phi = 
\frac{1}{\sqrt{2}} 
\begin{pmatrix}
    -i\phi^1 - \phi^2 \\
    \phi^4 + i \phi^3 
\end{pmatrix}
=
\frac{1}{\sqrt{2}} 
\begin{pmatrix}
    -i\sqrt{2}\chi^+ \\
    v + h + i\chi_3
\end{pmatrix}
\end{equation}
  In general, since such terms are not forbidden by any symmetry, there is no reason to expect that they shouldn't be included, so their addition to the Lagrangian is fully justified.

After expanding the Lagrangian we get:
\begin{align}
\label{eq:Lfscalar}
\mathcal{L}=&
-\frac{1}{\sqrt{2}}
	\left(\lambda_e\bar{e}e + \lambda_d\bar{d}d + \lambda_u\bar{u}u \right)
	(v + h) 
\\
&+i \left(
	\lambda_e\bar{\nu}_L e_R +\lambda_d\bar{u}_L d_R - \lambda_u\bar{u}_R d_L 
\right) \chi^+ \nonumber
\\
&+i \left(
	-\lambda_e\bar{e}_R \nu_L +\lambda_u\bar{d}_L u_R - \lambda_d\bar{d}_R u_L 
\right) \chi^- \nonumber
\\
&-\frac{i}{\sqrt{2}} \left(
	\lambda_e\bar{e}_L e_R - \lambda_e\bar{e}_R e_L
	+\lambda_d\bar{d}_L d_R - \lambda_d\bar{d}_R d_L
	-\lambda_u\bar{u}_L u_R + \lambda_u\bar{u}_R u_L
\right) \chi_3, \nonumber
\end{align}
and the terms multiplied by $v$ on the first line give masses:
\begin{equation}
m_e = \frac{\lambda_{e}v}{\sqrt{2}}, \quad
m_u = \frac{\lambda_{u}v}{\sqrt{2}}, \quad
m_d = \frac{\lambda_{d}v}{\sqrt{2}}.
\end{equation}
We can now freely take the masses $m_e$, $m_u$, $m_d$ as the unknown parameters instead of the couplings, and express the couplings as
\begin{equation}
\lambda_i = \frac{\sqrt{2}m_i}{v} = \frac{g}{\sqrt{2}}\frac{m_i}{m_W}.
\end{equation}
Combining (\ref{eq:Lfgauge}) with (\ref{eq:Lfscalar}) gives the full interacting Lagrangian for the fermions of the first generation:
\begin{align}
\mathcal{L}_{f1} =& 
	\bar{e} (i\slashpartial - m_e) e 
	+ \bar{\nu_e} (i \slashpartial) \nu_e 
	+ \bar{u} (i\slashpartial - m_u) u 
	+ \bar{d} (i\slashpartial - m_d) d 
\\
&+ g(W^+_\mu J^{\mu+}_W + W^-_\mu J^{\mu-}_W + Z_\mu J^\mu_Z) 
+ e A_\mu J^\mu_A,
\nonumber\\
& +\frac{ig}{\sqrt{2}} \left( \chi^+ J^+_\chi + \chi^- J^-_\chi \right) 
-\frac{ig}{2}\chi_3 J^3_\chi -\frac{g}{2}h J_h
\nonumber
\end{align}
with fermion currents that couple to the Goldstone bosons and Higgs:
\begin{align}
J^+_\chi \equiv&
	\tfrac{m_e}{m_W}\bar{\nu}_L e_R 
	+\tfrac{m_d}{m_W}\bar{u}_L d_R 
	-\tfrac{m_u}{m_W}\bar{u}_R d_L,
\\
J^-_\chi \equiv&
	-\tfrac{m_e}{m_W}\bar{e}_R \nu_L
	+\tfrac{m_u}{m_W}\bar{d}_L u_R 
	-\tfrac{m_d}{m_W}\bar{d}_R u_L,
\\
J^3_\chi \equiv&
	\tfrac{m_e}{m_W}\bar{e} \gamma^5 e
	+\tfrac{m_d}{m_W}\bar{d} \gamma^5 d
	-\tfrac{m_u}{m_W}\bar{u} \gamma^5 u,
\\
J_h \equiv&
	\tfrac{m_e}{m_W}\bar{e} e
	+\tfrac{m_d}{m_W}\bar{d} d
	+\tfrac{m_u}{m_W}\bar{u} u.
\end{align}
For the full fermion Lagrangian we then just need to sum over all three generations:
\begin{equation}
\mathcal{L}_f = \sum_{i=1}^3 \mathcal{L}_{fi}.
\end{equation}
The only difference between the different generations are masses, which arises from the fact that the fermion-scalar coupling constants $\lambda$ can be different for each generation.

There is actually an additional complication related to the inclusion of other generations: the coupling to the scalars can actually \emph{mix} generations, that is, there is a term $\lambda_d^{ij} \bar{Q}_{Li} \phi d_{Rj}$, where $i,j$ are generation indices and $\lambda_d$ is now a \emph{matrix}. That effectively yields flavor mixing of quarks in the charge-changing weak interactions (through the so-called \emph{Cabibbo-Kobayashi-Maskawa} mixing matrix), and allows a weak $CP$ violation. Those effects are not, however, of our interest here, and we will assume $\lambda$'s are diagonal, which is a pretty good approximation.

Another note: the quarks, of course, come in $SU(3)$ triplets, which is an unbroken gauge symmetry of the Standard Model, and so everywhere in the Lagrangian where there is a quark, it's actually a sum of \emph{three} different color quarks. They have to be summed-over if a quark appears as an intermediate particle in amplitude calculations. There are then also gauge-bosons associated with the $SU(3)$ called gluons and quark coupling, but we will not discuss that Lagrangian, as we will not be doing QCD calculations.

\subsection{Interactions}

Now we move on to the task of writing out all the remaining interactions in the Standard Model. Let's look back at the full effective Lagrangian for a Yang-Mills theory (\ref{eq:Lfull}) and work on the three interaction terms $\mathcal{L}_\mr{i}, \mathcal{L}_\mr{gfi}, \mathcal{L}_\mr{ghi}$ in the following sections. The goal here is to have in the end a full explicit Lagrangian in terms of the physical fields, from which the interaction vertices can be read off directly.

\subsubsection{$\mathcal{L}_\mr{i}$ - original interactions}
\label{sec:Li}

The term $\mathcal{L}_\mr{i}$ contains vector-vector interactions arising from the dynamic terms of the gauge fields, vector-scalar interactions from the gauge coupling through the covariant derivative, and, finally, scalar-scalar interactions from the Higgs potential. To find the expressions for those terms we have to go back to the very original Lagrangian (\ref{eq:l-unbroken}):
\begin{equation}
\mathcal{L} = -\frac{1}{4}(F^a_{\mu\nu})^2 +
\frac{1}{2}(D_\mu \phi_i)^2 - V(\phi).
\end{equation}
Writing it out more explicitly after the symmetry breaking it becomes:
\begin{align}
\mathcal{L} = & 
-\frac{1}{4}(\partial_\mu A^a_\nu - \partial_\nu A^a_\mu 
	+ gf^{abc} A^b_\mu A^c_\nu )^2 
\nonumber \\
&+\frac{1}{2}(\partial_\mu \chi_i + g A^a_\mu {F^a}_i 
	+g A^a_\mu T^a_{ij} \chi_j )^2
\\
&+ \frac{\mu^2}{2} (\phi_{0i}+\chi_i)^2 
	-\frac{\lambda}{4}(\phi_{0i}+\chi_i)^4.
\nonumber
\end{align}
We can now collect the higher-than-quadratic terms, grouping them by the particle species that are involved in the interaction (e.g. $\mathcal{L}_{vvv}$ for 3-vector interactions and $\mathcal{L}_{vvss}$ for vector-vector-scalar-scalar interactions):
\begin{align}
&
\mathcal{L}_\mr{i} =  \mathcal{L}_{vvvv} + \mathcal{L}_{vvv} +
	\mathcal{L}_{vvss} + \mathcal{L}_{vvs} + \mathcal{L}_{vss} +
	\mathcal{L}_{ssss} + \mathcal{L}_{sss},
\\&
\mathcal{L}_{vvvv} = -\frac{g^2}{4} f^{abc}f^{ade} 
	A^b_\mu A^c_\nu A^{d\mu} A^{e\nu},
\\&
\mathcal{L}_{vvv} = -g f^{abc} (\partial_\mu A^a_\nu) A^{b\mu} A^{c\nu},
\\&
\mathcal{L}_{vvss} =  \frac{1}{2}
	g^2 T^a_{ij} T^{b}_{ik} A^a_\mu A^{b\mu} \chi_j \chi_k,
\\&
\mathcal{L}_{vvs} =  g^2 A^a_\mu A^{b\mu} {F^a}_i T^b_{ij} \chi_j,
\\&
\mathcal{L}_{vss} = g A^{a\mu} T^a_{ij}(\partial_\mu \chi_i)\chi_j,
\\&
\mathcal{L}_{ssss} = -\frac{\lambda}{4} (\chi_i\chi_i)^2,
\\&
\mathcal{L}_{sss} = -\lambda (\phi_{0i}\chi_i)(\chi_i\chi_i).
\end{align}
All the symbols used here (structure constants $f$, generators of the scalar fields $T$, vacuum expectation value $\phi_0$, matrix $F=T\phi_0$) are defined in the earlier sections, so the only work that needs to be done here is to substitute all the definitions and the physical fields, and work out the final expressions.
The computations are quite cumbersome and not very illuminating, so we just give the final results here term-by-term. We abbreviate:
\begin{equation}
s_W\equiv\sin\theta_w, \quad c_W\equiv\cos\theta_w.
\end{equation}
%
\begin{align}
\mathcal{L}_{vvvv} = -\frac{e^2}{s_W^2} & \left\{ 
\tfrac{1}{2}\left(W^-_\mu W^{+\mu} W^-_\nu W^{+\nu} -
	W^-_\mu W^{-\mu} W^+_\nu W^{+\nu}\right) \right.
\\ \nonumber
&+c_W^2\left(W^-_\mu W^{+\mu} Z_\nu Z^{\nu} -
	W^-_\mu Z^{\mu} W^+_\nu Z^{\nu}\right)
\\ \nonumber
&+s_W^2\left(W^-_\mu W^{+\mu} A_\nu A^{\nu} -
	W^-_\mu A^{\mu} W^+_\nu A^{\nu}\right)
\\ \nonumber
&\left.+s_Wc_W\left(2W^-_\mu W^{+\mu} A_\nu Z^{\nu} 
	- W^-_\mu A^{\mu} W^+_\nu Z^{\nu}
	- W^-_\mu Z^{\mu} W^+_\nu A^{\nu}\right)\right\}.
\end{align}
%
\begin{align}
\mathcal{L}_{vvv}(x) = -i\frac{e}{s_W}&\left[\left(
	\left(\partial_{1\rho}-\partial_{2\rho}\right) g^{\mu\nu}
	+\left(\partial_{2\mu}-\partial_{3\mu}\right) g^{\nu\rho}
	+\left(\partial_{3\nu}-\partial_{1\nu}\right) g^{\rho\mu}
\right) \right.
\\ \nonumber
&\left.\times W^{+\mu}(x_1) W^{-\nu}(x_2) 
	(\cos\theta_w Z^{\rho}(x_3) + \sin\theta_w A^{\rho}(x_3))
\right]_{x_1=x_2=x_3=x},
\end{align}
the coordinates ($x_1,x_2,x_3$) are written out explicitly here only to notate that the derivatives ($\partial_1,\partial_2,\partial_3$) in parentheses act on different fields.
%
\begin{align}
\mathcal{L}_{vvss} = &
\frac{e^2}{2s_W}\left(
	i A_\mu W^{+\mu} h \chi^-
	-i A_\mu W^{-\mu} h \chi^+
	-A_\mu W^{+\mu} \chi_3 \chi^-
	-A_\mu W^{-\mu} \chi_3 \chi^+
\right) 
\\ \nonumber &
+\frac{e^2}{2c_W}\left(
	-i Z_\mu W^{+\mu} h \chi^-
	+i Z_\mu W^{-\mu} h \chi^+
	+Z_\mu W^{+\mu} \chi_3 \chi^-
	+Z_\mu W^{-\mu} \chi_3 \chi^+
\right) 
\\ \nonumber &
+ e^2 A_\mu A^\mu \chi^+ \chi^-
+ \frac{e^2(c_W^2-s_W^2)}{s_W c_W} A_\mu Z^\mu \chi^+ \chi^-
+ \frac{e^2(c_W^2-s_W^2)^2}{4 s_W^2 c_W^2} Z_\mu Z^\mu \chi^+ \chi^-
\\ \nonumber &
+ \frac{e^2}{4s_W^2} \left(
	W^+_\mu W^{-\mu} h h
	+ W^+_\mu W^{-\mu} \chi_3 \chi_3
	+ 2 W^+_\mu W^{-\mu} \chi^+ \chi^-
\right)
\\ \nonumber &
+ \frac{e^2}{8s_W^2c_W^2} \left(
	Z_\mu Z^{\mu} h h
	+ Z_\mu Z^{\mu} \chi_3 \chi_3
\right).
\end{align}
%
\begin{align}
\mathcal{L}_{vvs} = &
e m_W \left( i W^+_\mu A^\mu \chi^- - i W^-_\mu A^\mu \chi^+ \right)
+ \frac{e m_W s_W}{c_W} 
\left( -i W^+_\mu Z^\mu \chi^- + i W^-_\mu Z^\mu \chi^+ \right)
\nonumber \\ &
+ \frac{e m_W}{s_W} W^+_\mu W^{-\mu} h
+ \frac{e m_W}{2s_Wc_W^2} Z_\mu Z^{\mu} h
\end{align}
%
\begin{align}
\mathcal{L}_{vss} = &
\frac{e}{2s_W} \left[
	W^{+\mu}(h\partial_\mu\chi^- - \chi^-\partial_\mu h)
	+W^{-\mu}(h\partial_\mu\chi^+ - \chi^+\partial_\mu h)
\right] 
\\ \nonumber
+& \frac{ie}{2s_W} \left[
	W^{+\mu}(\chi_3\partial_\mu\chi^- - \chi^-\partial_\mu \chi_3)
	-W^{-\mu}(\chi_3\partial_\mu\chi^+ - \chi^+\partial_\mu \chi_3)
\right]
\\ \nonumber
	+& i e A^{\mu}(\chi^-\partial_\mu\chi^+ - \chi^+\partial_\mu \chi^-)
	+ i e\frac{c_W^2-s_W^2}{2s_Wc_W} 
	Z^{\mu}(\chi^-\partial_\mu\chi^+ - \chi^+\partial_\mu \chi^-)
\\ \nonumber
+ &\frac{e}{2s_Wc_W} Z^\mu (h\partial_\mu\chi_3 - \chi_3\partial_\mu h).
\end{align}
%
\begin{align}
\mathcal{L}_{ssss} = -\frac{e^2m_H^2}{32s_W^2m_W^2} (&
hhhh + \chi_3\chi_3\chi_3\chi_3 + 4\chi^+\chi^+\chi^-\chi^-
\\ \nonumber
& + 2 hh\chi_3\chi_3 + 4hh\chi^+\chi^- + 4\chi^+\chi^-\chi_3\chi_3
)
\end{align}
%
\begin{align}
\mathcal{L}_{sss} = -\frac{em_H^2}{4s_Wm_W} \left(
hhh + h\chi_3\chi_3 + 2h\chi^+\chi^- \right).
\end{align}
Note that we expressed the vacuum expectation value $v$ and the constant of the Higgs potential $\lambda$ in terms of the physical masses: 
\begin{equation}
v=\frac{2m_W\sin\theta_w}{e}, \quad \lambda = \frac{\mu^2}{v^2} = 
\left(\frac{m_H e}{2\sqrt{2}m_W\sin\theta_w}\right)^2.
\end{equation}

\subsubsection{$\mathcal{L}_\mr{gfi}$ - non-linear gauge-fixing}
\label{sec:Lgfi}

We now switch to the analysis of the interaction terms arising from the non-linear gauge-fixing $\mathcal{L}_\mr{gfi}$. As we discussed when analyzing the gauge-fixing in general, the constraint functions can be of the form (\ref{eq:Gageneral}):
\begin{equation}
G^a = \partial_\mu A^{a\mu} - {K^a}_i \chi_i + G_\mr{nl}^a,
\end{equation}
with $G_\mr{nl}^a$ being \emph{any} quadratic functions of the fields, and the gauge-fixing Lagrangian is
\begin{equation}
\mathcal{L}_\mr{gf} = -\frac{1}{2} G^a L^{ab} G^b.
\end{equation}
The quadratic additions $G_\mr{nl}^a$ obviously give then the interaction terms. We have actually already solved for matrices $L$ and $K$ that give nice propagators in earlier sections, and so we can rewrite $\mathcal{L}_\mr{gf}$ after changing to the basis of physical fields as (keeping  $G_\mr{nl}^a$ undefined):
\begin{align}
\label{eq:Lgfphysical}
\mathcal{L}_\mr{gf} =& 
-\frac{1}{\xi_W}G^-G^+ - \frac{1}{\xi_Z}(G^Z)^2 - \frac{1}{\xi_A}(G^A)^2 
\\
=&-\frac{1}{\xi_W} \left| \partial_\mu W^{+\mu} - \xi_W
m_W \chi^+ + G_{nl}^+ \right|^2 \nonumber \\
&- \frac{1}{2 \xi_Z} \left( \partial_\mu Z^\mu - \xi_Z m_Z \chi^0 + G_{nl}^Z
\right)^2 \nonumber \\
&- \frac{1}{2 \xi_A} \left( \partial_\mu A^\mu + G_{nl}^A \right)^2, \nonumber
\end{align}
where the non-linear additions were transformed  into $G_\mr{nl}^+$ (and its hermitian conjugate $G_\mr{nl}^-$), which can be a \emph{complex} quadratic function of the fields, and $G_\mr{nl}^Z$, $G_\mr{nl}^A$ that are \emph{real} quadratic modifications. It has to be kept in mind when choosing the quadratic additions, that, in order not to introduce unnecessary complications, the resulting $\mathcal{L}_\mr{gf}$ should respect the unbroken symmetries of the Lagrangian, which are the electromagnetic $U(1)$ and the $SU(3)$ of QCD. That means that $G_\mr{nl}^+$ should transform as a positively charged field, $G_\mr{nl}^Z$ and 
 $G_\mr{nl}^A$ should be neutral, and they all should, of course, be neutral under $SU(3)$.

In this work we will assume a particular class of non-linear gauges given by the choice, in accordance to \cite{grace}, \cite{nonlinear}:
\begin{align}
\label{eq:Gnl}
&
G_\mr{nl}^+ =  
- i \tilde{\alpha} e  (A_\mu W^{\mu+}) 
- i \tilde{\beta} \frac{ec_W}{s_W}(Z_\mu W^{\mu+})
- \tilde{\delta} \frac{e\xi_W}{2s_W}(h\chi^+)
+ i \tilde{\kappa} \frac{e\xi_W}{2s_W}(\chi_3\chi^+)
\\ &
G_\mr{nl}^Z = - \tilde{\varepsilon} \frac{e\xi_Z}{2s_Wc_W}(h\chi_3) 
\\ &
G_\mr{nl}^A = 0.
\end{align}
We have introduced here \emph{five} new gauge-fixing parameters
$$\tilde{\alpha}, \tilde{\beta}, \tilde{\delta}, \tilde{\kappa}, \tilde{\varepsilon}$$
that specify the concrete choice of the gauge. The case when they all are equal to 0 corresponds, of course, to the usual linear gauge, and their variation allows to explore the different non-linear gauges, and check the results of various calculations for more general gauge-invariance.

The remaining goal of this section is now just to plug in the definitions for $G_\mr{nl}^+$ and $G_\mr{nl}^Z$ into (\ref{eq:Lgfphysical}) and to collect the terms of higher than quadratic order, which will give us $\mathcal{L}_\mr{gfi}$. Remember that the quadratic terms in $\mathcal{L}_\mr{gf}$ are independent of $G_\mr{nl}^a$ and were already taken into account for calculation of the propagators. Again, we split $\mathcal{L}_\mr{gfi}$ into pieces describing the interactions of different particle species:
\begin{equation}
\mathcal{L}_\mr{gfi} =  \mathcal{L}_{\mr{gf}vvvv} + \mathcal{L}_{\mr{gf}vvv} +
	\mathcal{L}_{\mr{gf}vvss} + \mathcal{L}_{\mr{gf}vvs} + \mathcal{L}_{\mr{gf}vss} +
	\mathcal{L}_{\mr{gf}ssss} + \mathcal{L}_{\mr{gf}sss},
\end{equation}
and give the results term-by-term. There are no peculiarities involved in this calculation -- just the straightforward squaring of parentheses in (\ref{eq:Lgfphysical}):
\begin{align}
\mathcal{L}_{\mr{gf}vvvv} = & 
-\frac{\tilde{\alpha}^2e^2}{\xi_W} (W_\mu^-A^\mu W_\nu^+ A^\nu)
-\frac{\tilde{\beta}^2e^2c_W^2}{\xi_Ws_W^2}(W_\mu^-Z^\mu W_\nu^+ Z^\nu)
\\ \nonumber & 
-\frac{\tilde{\alpha}\tilde{\beta}e^2c_W}{\xi_Ws_W}(
	W_\mu^-A^\mu W_\nu^+ Z^\nu
	+ W_\mu^-Z^\mu W_\nu^+ A^\nu),
\\
\mathcal{L}_{\mr{gf}vvv} = &
\frac{i\tilde{\alpha}e}{\xi_W} (
	W_\mu^+ A^\mu \partial_\nu W^{-\nu} 
	- W_\mu^- A^\mu \partial_\nu W^{+\nu} )
+ \frac{i\tilde{\beta}ec_W}{\xi_Ws_W} (
	W_\mu^+ Z^\mu \partial_\nu W^{-\nu} 
	- W_\mu^- Z^\mu \partial_\nu W^{+\nu} ),
\end{align}
\begin{align}
\mathcal{L}_{\mr{gf}vvss} = &
-\frac{i\tilde{\alpha}\tilde{\delta}e^2}{2s_W} 
	( A_\mu W^{+\mu} h\chi^- - A_\mu W^{-\mu} h\chi^+ )
+\frac{\tilde{\alpha}\tilde{\kappa}e^2}{2s_W}
	( A_\mu W^{+\mu} \chi_3\chi^- + A_\mu W^{-\mu} \chi_3\chi^+ )
\nonumber \\ &
-\frac{i\tilde{\beta}\tilde{\delta}e^2c_W}{2s_W^2} 
	( Z_\mu W^{+\mu} h\chi^- - Z_\mu W^{-\mu} h\chi^+ )
+\frac{\tilde{\beta}\tilde{\kappa}e^2c_W}{2s_W^2}
	( Z_\mu W^{+\mu} \chi_3\chi^- + Z_\mu W^{-\mu} \chi_3\chi^+ ),
\\
\mathcal{L}_{\mr{gf}vvs} = &
-i\tilde{\alpha}em_W (W^+_\mu A^\mu\chi^- - W^-_\mu A^\mu\chi^+)
-\frac{i\tilde{\beta}e m_Wc_W}{s_W} (W^+_\mu Z^\mu\chi^- - W^-_\mu Z^\mu\chi^+),
\end{align}
\begin{align}
\mathcal{L}_{\mr{gf}vss} = &
\frac{\tilde{\delta}e}{2s_W}
	( h\chi^-\partial_\mu W^{+\mu} + h\chi^+\partial_\mu W^{-\mu} )
+ \frac{i\tilde{\kappa}e}{2s_W}
	( \chi_3\chi^-\partial_\mu W^{+\mu} - \chi_3\chi^+\partial_\mu W^{-\mu} )
\\ \nonumber &
+ \frac{\tilde{\varepsilon}e}{2s_Wc_W} (h\chi_3\partial_\mu Z^\mu)
\\
\mathcal{L}_{\mr{gf}ssss} = &
-\frac{\tilde{\delta}^2e^2\xi_W}{4s_W^2}(hh\chi^+\chi^-)
-\frac{\tilde{\kappa}^2e^2\xi_W}{4s_W^2}(\chi_3\chi_3\chi^+\chi^-)
-\frac{\tilde{\varepsilon}^2e^2\xi_Z}{8s_W^2c_W^2}(hh\chi_3\chi_3)
\\
\mathcal{L}_{\mr{gf}sss} = &
-\frac{\tilde{\delta}e\xi_Wm_W}{s_W}(h\chi^+\chi^-)
-\frac{\tilde{\varepsilon}e\xi_Zm_Z}{2s_Wc_W}(h\chi_3\chi_3).
\end{align}

All these gauge-fixing terms combine with the regular terms from $\mathcal{L}_\mr{i}$ to effectively alter the strength or the structure of the coupling, depending on the gauge parameters. We will show that when we write out the full Lagrangian with all its terms.

\subsubsection{$\mathcal{L}_\mr{ghi}$ - ghost interactions}
\label{sec:Lghi}

We now move on to the last part of the interaction Lagrangian $\mathcal{L}_\mr{ghi}$ - the ghost interaction terms. Remember from the description of the gauge-fixing procedure that the ghost Lagrangian is given by (\ref{eq:Lg}):
\begin{equation}
\mathcal{L}_\mr{gh} \equiv \bar{c}^a(x) \left( \left.
\frac{\delta G^a[A_\alpha,\chi_\alpha]}{\delta \alpha^b}
\right|_{\alpha=0} \right) c^b(x),
\end{equation}
and that we have already diagonalized the ghost propagators to get the ghost energy eigenstates as (\ref{eq:physicalghosts}):
\begin{equation}
c'^a \equiv 
\begin{pmatrix}
c^+ \\ c^- \\ c^Z \\ c^A
\end{pmatrix}
=
U^{ab} c^b
=
\begin{pmatrix}
\tfrac{1}{\sqrt{2}}(c^1 - ic^2) \\
\tfrac{1}{\sqrt{2}}(c^1 + ic^2) \\
c_W c^3 - s_W c^4 \\
s_W c^3 + c_W c^4
\end{pmatrix}
\end{equation}
\begin{equation}
\bar{c}'^a \equiv 
\begin{pmatrix}
\bar{c}^- \\ \bar{c}^+ \\ \bar{c}^Z \\ \bar{c}^A
\end{pmatrix}
=
\bar{c}^b (U^\dagger)^{ba}
=
\begin{pmatrix}
\tfrac{1}{\sqrt{2}}(\bar{c}^1 + i\bar{c}^2) \\
\tfrac{1}{\sqrt{2}}(\bar{c}^1 - i\bar{c}^2) \\
c_W \bar{c}^3 - s_W \bar{c}^4 \\
s_W \bar{c}^3 + c_W \bar{c}^4
\end{pmatrix}.
\end{equation}
Note also that we had to multiply the variation $\delta G/\delta \alpha$ by $(-g)$ to get (\ref{eq:Lgh}) and correctly normalized ghost propagators
\footnote{It should be kept in mind that $(g)$ is not really a single constant, but a diagonal matrix $g^{ab}$ with $g^{11}=g^{22}=g^{33}=g$ and $g^{44}=g'$.}.
We can then express the ghost Lagrangian directly in terms of the physical fields as:
\begin{equation}
\label{eq:Lghphysical2}
\mathcal{L}_\mr{gh} \equiv 
\bar{c}^a \left( \frac{\delta G^a_\alpha}{\delta \alpha^b} \right) (-g) c^b
= -
\bar{c}'^a U^{ac} \left( \frac{\delta G^c_\alpha}{\delta \alpha^d}
 \right) (gU^\dagger)^{db} c'^b
= -
\bar{c}'^a \left( \frac{\delta G'^a_\alpha}{\delta \alpha'^b} \right) c'^b
\end{equation}
with transformed $G^a$ and $\alpha^a$:
\begin{align}
G'^a =& U^{ab} G^b, \\
\alpha'^a =& (Ug^{-1})^{ab} \alpha^b.
\end{align}
The transformed constraints $G'^a$ are, of course, just the physical constraints as in (\ref{eq:Lgfphysical}): 
\begin{equation}
\label{eq:Gphysical}
G'^a =
\begin{pmatrix}
G^+ \\ G^- \\ G^Z \\ G^A
\end{pmatrix}
=
\begin{pmatrix}
 \partial_\mu W^{+\mu} - \xi_W m_W \chi^+ + G_{nl}^+ \\ 
 \partial_\mu W^{-\mu} - \xi_W m_W \chi^- + G_{nl}^- \\ 
\partial_\mu Z^\mu - \xi_Z m_Z \chi^0 + G_{nl}^Z \\ 
\partial_\mu A^\mu + G_{nl}^A
\end{pmatrix},
\end{equation}
and the transformed $\alpha'^a$ are the physical gauge transformation coefficients in terms of which we will have to find variations of the fields. So, actually, all we have to do now is to calculate the variations of the physical fields in terms of $\alpha'^a$, and then we can simply read off the ghost Lagrangian using (\ref{eq:Lghphysical2}) and the definitions for $G^\pm, G^Z, G^A$. 

The variation in the physical gauge fields is:
\begin{align}
\delta A_\mu^{'a} =& U^{ab}\delta A_\mu^b 
= U^{aa'} \left(g^{-1}\partial_\mu\alpha^{a'} + f^{a'bc} A^b_\mu \alpha^c\right)
\nonumber \\
=& \partial_\mu\alpha'^a + \left( U^{aa'} f^{a'b'c'} (U^\dagger)^{b'b}
	(gU^\dagger)^{c'c}\right) A'^b_\mu \alpha'^c.
\end{align}
After transforming the structure constants we get:
\begin{align}
\delta W^\pm_\mu =& \partial_\mu \alpha^\pm \pm ie \left(
	W^\pm_\mu\alpha^A + \frac{c_W}{s_W} W^\pm_\mu \alpha^Z 
	-A_\mu \alpha^\pm - \frac{c_W}{s_W} Z_\mu \alpha^\pm \right),
\\
\delta Z_\mu =& \partial_\mu \alpha^Z - ie\frac{c_W}{s_W}
\left( W^+_\mu \alpha^- - W^-_\mu \alpha^+ \right),
\\
\delta A_\mu =& \partial_\mu \alpha^A - ie
\left( W^+_\mu \alpha^- - W^-_\mu \alpha^+ \right).
\end{align}
Note that the new gauge transformation coefficients $\alpha^\pm,\alpha^Z,\alpha^A$ are indeed \emph{physical} in the sense that each of them yields a transformation of the corresponding physical gauge field by $\partial_\mu\alpha$. The non-Abelian part of the gauge transformation naturally
doesn't look symmetric anymore when written out in terms of the physical fields, because the symmetry is actually broken.

Now, to get the variations of the scalar fields we similarly need to transform everything to the physical basis. Here it's convenient to use a notation similar to that for gauge bosons, combining Goldstones and Higgs into one vector:
\begin{equation}
\chi'_i=
\begin{pmatrix}
\chi^+ \\ \chi^- \\ \chi_3 \\ h
\end{pmatrix}
=
U^\chi_{ij} \chi_j,
\end{equation}
with (looking back at (\ref{eq:chiphdefinition})):
\begin{equation}
U^\chi=
\left(%
\begin{array}{cccc}
  \frac{1}{\sqrt{2}} & -\frac{i}{\sqrt{2}} & 0 & 0 \\
  \frac{1}{\sqrt{2}} & \frac{i}{\sqrt{2}} & 0 & 0 \\
  0 & 0 & 1 & 0 \\
  0 & 0 & 0 & 1 \\
\end{array}%
\right).
\end{equation}
The variation is then:
\begin{align}
\delta\chi'_i =& U^\chi_{ii'} \delta\chi_{i'} = U^\chi_{ii'}\left(
	-\alpha^a {F^a}_{i'} + \alpha^a T^a_{i'j}\chi_j \right)
\nonumber \\
=&
- \left( U^\chi_{ii'} {F^{a'}}_{i'} (gU^\dagger)^{a'a} \right) \alpha'^a
- \left( U^\chi_{ii'} T^{a'}_{i'j'} (U^{\chi\dagger})_{j'j} 
	(gU^\dagger)^{a'a} \right) \alpha'^a \chi'_j,
\end{align}
which results in
\begin{align}
\delta \chi^\pm =& -m_W\alpha^\pm \pm \frac{ie}{2s_W} \left(
	\frac{c_W^2-s_W^2}{c_W}\chi^\pm\alpha^Z
	+2s_W\chi^\pm\alpha^A - \chi_3\alpha^\pm 
	\pm i h\alpha^\pm \right)
\\
\delta\chi_3 =& -m_Z\alpha^Z -\frac{ie}{2s_W}
	\left( \chi^+\alpha^- - \chi^-\alpha^+ 
	-\frac{i}{c_W}h\alpha^Z \right)
\\
\delta h =& \frac{e}{2s_W}\left(
	\chi^+\alpha^- + \chi^-\alpha^+ + \frac{1}{c_W}\chi_3\alpha^Z \right).
\end{align}
Note that again the terms just proportional to $\alpha$ end up being diagonal in the physical basis, which is just a check that we did the transformations  correctly: their diagonality is equivalent to the ghost propagators being diagonal, and that's what we started with in this section.

We can now use these expressions for variations to write out the ghost Lagrangian. First consider the part of $\mathcal{L}_\mr{ghi}$ which is present even without the non-linear gauge-fixing terms. Plugging in (\ref{eq:Gphysical}) into (\ref{eq:Lghphysical2}) with $G_\mr{nl}=0$ yields:
\begin{align}
\label{eq:Lgh0}
\mathcal{L}_{\mr{gh}0} =&
-\bar{c}^- \frac{\partial_\mu(\delta W^{+\mu})}{\delta\alpha'^a} c'^a
-\bar{c}^+ \frac{\partial_\mu(\delta W^{-\mu})}{\delta\alpha'^a} c'^a
-\bar{c}^Z \frac{\partial_\mu(\delta Z^{\mu})}{\delta\alpha'^a} c'^a
-\bar{c}^A \frac{\partial_\mu(\delta A^{\mu})}{\delta\alpha'^a} c'^a
\\ \nonumber &
+\xi_W m_W \bar{c}^- \frac{\delta\chi^+}{\delta\alpha'^a} c'^a
+\xi_W m_W \bar{c}^+ \frac{\delta\chi^-}{\delta\alpha'^a} c'^a
+\xi_Z m_Z \bar{c}^Z \frac{\delta\chi_3}{\delta\alpha'^a} c'^a.
\end{align}
Note that the operator
$$c'^a\frac{\delta}{\delta\alpha'^a}$$
just replaces the $\alpha'^a$ in the variations with $c'^a$. We can then easily write out (\ref{eq:Lgh0}) by just substituting the variations of all the fields. Also, since we are looking for the \emph{interaction} terms, we drop the quadratic ghost-antighost terms, yielding:
\begin{align}
\mathcal{L}_{\mr{ghi}0} =&
+ie (\partial_\mu \bar{c}^-) \left(
	W^{+\mu} c^A + \frac{c_W}{s_W} W^{+\mu} c^Z 
	-A^\mu c^+ - \frac{c_W}{s_W} Z^\mu c^+ \right)
\\ \nonumber &
-ie (\partial_\mu \bar{c}^+) \left(
	W^{-\mu} c^A + \frac{c_W}{s_W} W^{-\mu} c^Z 
	-A^\mu c^- - \frac{c_W}{s_W} Z^\mu c^- \right)
\\ \nonumber &
-ie\frac{c_W}{s_W}(\partial_\mu \bar{c}^Z)
\left(W^{+\mu}c^- - W^{-\mu}c^+\right)
\\ \nonumber &
-ie(\partial_\mu \bar{c}^A)
\left(W^{+\mu}c^- - W^{-\mu}c^+\right)
\\ \nonumber &
+\frac{ie\xi_W m_W}{2s_W}\, \bar{c}^- \left(
	\frac{c_W^2-s_W^2}{c_W}\chi^+c^Z
	+2s_W\chi^+c^A - \chi_3 c^+ 
	+ i h c^+ \right)
\\ \nonumber &
-\frac{ie\xi_W m_W}{2s_W}\, \bar{c}^+ \left(
	\frac{c_W^2-s_W^2}{c_W}\chi^-c^Z
	+2s_W\chi^-c^A - \chi_3 c^- 
	- i h c^- \right)
\\ \nonumber &
-\frac{ie\xi_Z m_Z}{2s_W}\, \bar{c}^Z \left(
	\chi^+ c^- - \chi^- c^+ 
	-\frac{i}{c_W}h c^Z \right).
\end{align}
This Lagrangian describes the ghost-ghost-vector interactions (first 4 lines) and ghost-ghost-scalar interactions (last 3 lines) and it would be the final result in the case of linear gauge-fixing.

With non-linear gauge-fixing the ghost interaction Lagrangian will also contain terms arising from the non-zero $G^+_\mr{nl}, G^Z_\mr{nl}$:
\begin{equation}
\mathcal{L}_\mr{ghinl} = 
-\bar{c}^- \frac{\delta G^+_\mr{nl}}{\delta\alpha'^a}c'^a
-\bar{c}^+ \frac{\delta G^-_\mr{nl}}{\delta\alpha'^a}c'^a
-\bar{c}^Z \frac{\delta G^Z_\mr{nl}}{\delta\alpha'^a}c'^a.
\end{equation}
These terms can also be calculated straightforwardly using the definitions (\ref{eq:Gnl}). It's convenient to work out the variation of the constraints term-by-term, which gives the Lagrangian split up in terms, each proportional to a different  gauge-fixing parameter:
\begin{equation}
\mathcal{L}_\mr{ghinl} = \mathcal{L}_{\mr{ghi}\alpha} + \mathcal{L}_{\mr{ghi}\beta} + \mathcal{L}_{\mr{ghi}\delta} + \mathcal{L}_{\mr{ghi}\kappa} + \mathcal{L}_{\mr{ghi}\varepsilon}.
\end{equation}
We list here the final resulting expressions:
\begin{align}
\mathcal{L}_{\mr{ghi}\alpha} =& +i\tilde{\alpha}e\bar{c}^- \left(
	A_\mu\partial^\mu c^+ + W^+_\mu\partial^\mu c^A \right)
- \tilde{\alpha} e^2 \bar{c}^- \left(
	A_\mu W^{+\mu}c^A + \tfrac{c_W}{s_W}A_\mu W^{+\mu}c^Z - \right.
\nonumber \\  &
\qquad  \left. 
	-A_\mu A^\mu c^+ - \tfrac{c_W}{s_W}A_\mu Z^\mu c^+
	-W_\mu^+W^{+\mu}c^- + W^+_\mu W^{-\mu}c^+
	\right) + \mr{h.c.}
\\ 
\mathcal{L}_{\mr{ghi}\beta} =& +i\tilde{\beta}e\tfrac{c_W}{s_W}\,\bar{c}^- \left(
	Z_\mu\partial^\mu c^+ + W^+_\mu\partial^\mu c^Z \right)
- \tilde{\beta} e^2 \tfrac{c_W}{s_W}\, \bar{c}^- \left(
	Z_\mu W^{+\mu}c^A + \tfrac{c_W}{s_W}Z_\mu W^{+\mu}c^Z -  \right.
\nonumber \\  &
\quad  \left. 
	-Z_\mu A^\mu c^+ - \tfrac{c_W}{s_W}Z_\mu Z^\mu c^+
	-\tfrac{c_W}{s_W} W_\mu^+W^{+\mu}c^- + \tfrac{c_W}{s_W} W^+_\mu W^{-\mu}c^+
	\right) + \mr{h.c.}
\end{align}
\begin{align}
\mathcal{L}_{\mr{ghi}\delta} =&
-\tilde{\delta}\frac{e\xi_Wm_W}{2s_W}\,\bar{c}^- h c^+
+\tilde{\delta}\frac{e^2\xi_W}{4s_W^2}\,\bar{c}^- \left(
	\chi^+\chi^+c^- + \chi^+\chi^-c^+  + \right.
\\ \nonumber &
\quad \left.
	+ \frac{1}{c_W}\chi^+\chi_3c^Z +i\frac{c_W^2-s_W^2}{c_W}\chi^+hc^Z + 2is_W \chi^+hc^A - i\chi_3hc^+
- hhc^+ \right) + \mr{h.c.}
\\ 
\mathcal{L}_{\mr{ghi}\kappa} =&
i\tilde{\kappa}\frac{e\xi_W}{2s_W}\, \bar{c}^-
	\left(m_W \chi_3 c^+ + m_Z \chi^+ c^Z \right)
+\tilde{\kappa}\frac{e^2\xi_W}{4s_W^2}\,\bar{c}^- \left(
	-\chi^+\chi^+c^- + \chi^+\chi^-c^+ + \right.
\\ \nonumber &
\quad \left.
	+\frac{i}{c_W}\chi^+hc^Z + \frac{c_W^2-s_W^2}{c_W}\chi^+\chi_3c^Z
	+2s_W\chi^+\chi_3c^A + i\chi_3 h c^+ - \chi_3\chi_3c^+ \right) + \mr{h.c.}
\\
\mathcal{L}_{\mr{ghi}\varepsilon} =&
-\tilde{\varepsilon}\frac{e\xi_Zm_Z}{2s_Wc_W}\bar{c}^Z h c^Z
+\tilde{\varepsilon}\frac{e^2\xi_Z}{4s_W^2c_W} \bar{c}^Z \left(
	-i\chi^+hc^- + i\chi^-hc^+  + \chi^+\chi_3c^- + \chi^-\chi_3c^+ + \right.
\nonumber \\ &
\quad \left.
+ \frac{1}{c_W}\chi_3\chi_3c^Z - \frac{1}{c_W}hhc^Z \right).
\end{align}
The abbreviation ``h.c.'' above means that a hermitian conjugate of the whole expression has to be added as well.
Among the terms above we see some ghost-ghost-vector and ghost-ghost-scalar interactions, which provide modifications to the terms in $\mathcal{L}_\mr{ghi0}$, but also we have lots of new terms describing ghost-ghost-vector-vector and ghost-ghost-scalar-scalar interactions that are only present in the case of the non-linear gauge-fixing. 

For the final ghost interaction Lagrangian we just need to collect all the parts listed in this section:
\begin{equation}
\mathcal{L}_\mr{ghi} = \mathcal{L}_\mr{ghi0} + \mathcal{L}_{\mr{ghi}\alpha} + \mathcal{L}_{\mr{ghi}\beta} + \mathcal{L}_{\mr{ghi}\delta} + \mathcal{L}_{\mr{ghi}\kappa} + \mathcal{L}_{\mr{ghi}\varepsilon}.
\end{equation}

\subsection{Full Lagrangian}

In the previous sections we have worked out all the terms needed to construct the full Lagrangian of the Standard Model in the case of non-linear gauge-fixing. We provide here a summary of the contents of the Lagrangian and it is written out explicitly in Appendix A. The full list of interaction vertices arising from the Lagrangian is given in Appendix B.

The Lagrangian contains the following pieces:
\begin{align}
\mathcal{L} =& \mathcal{L}_v + \mathcal{L}_s + \mathcal{L}_f + \mathcal{L}_g +
\\ \nonumber
&  + \mathcal{L}_{vvvv} + \mathcal{L}_{vvv} +
	\mathcal{L}_{vvss} + \mathcal{L}_{vvs} + \mathcal{L}_{vss} +
	\mathcal{L}_{ssss} + \mathcal{L}_{sss} +
\\ \nonumber
& + \mathcal{L}_{ffv} + \mathcal{L}_{ffs} + 
\\ \nonumber
& + \mathcal{L}_{ggv} + \mathcal{L}_{ggs} + \mathcal{L}_{ggvv} + \mathcal{L}_{ggss}.
\end{align}
The terms $\mathcal{L}_v, \mathcal{L}_s, \mathcal{L}_f, \mathcal{L}_g$ hold the quadratic terms for vector, scalar, fermion and ghost  fields respectively, describing their propagators -- fermion propagator terms can be found in Section \ref{sec:fermions} and the others in Section \ref{sec:bosonpropagators}. The rest of the Lagrangian describes the interactions: the terms \emph{not} involving fermions or ghosts are found by combining the results of Sections \ref{sec:Li} and \ref{sec:Lgfi}, the fermion interactions $\mathcal{L}_{ffv},\mathcal{L}_{ffs}$ are described in Section \ref{sec:fermions} and, finally, the ghost interactions were listed in the previous Section \ref{sec:Lghi}.

We have now derived the full Standard Model Lagrangian, which we will apply now in calculations. First, in the following section, we will give an example of a tree-level process amplitude calculation, which explicitly demonstrates the use of the derived Lagrangian and the effects of the non-linear gauge-fixing. Then, finally, we will describe the implementation of the Lagrangian in FeynArts and the calculations we did using the package, which is the real purpose of this work.


\section{Explicit tree-level $\gamma\gamma \rightarrow WW $ calculation}
\label{sec:explicitcalc}

In this section we will perform a tree-level calculation for the tree-level process of $W^+W^-$
production in a photon-photon interaction, using the non-linear gauge-fixing Lagrangian. The goal of this calculation is to see, that all the gauge-fixing
parameters actually cancel out in the final result as expected.
This particular process was chosen as an example because it illustrates well the non-trivial cancellation of the gauge-fixing parameters.

The tree-level diagrams for this process are the following:
\\
\\
$i\mathcal{M}_{1t}=$
\begin{minipage}{0.4\textwidth}
\includegraphics[scale=0.5]{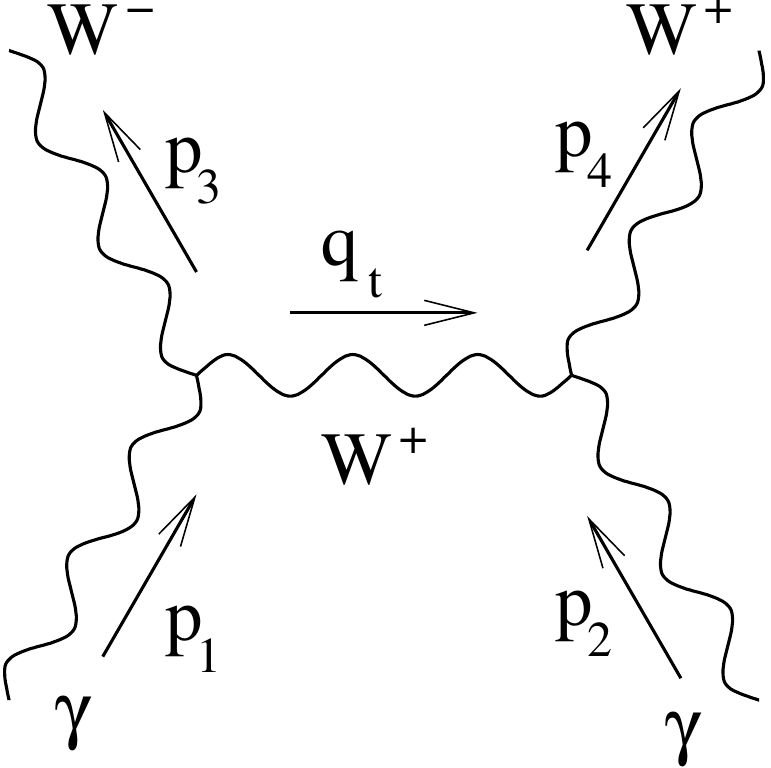}
\end{minipage}
$i\mathcal{M}_{1u}=$
\begin{minipage}{0.3\textwidth}
\includegraphics[scale=0.5]{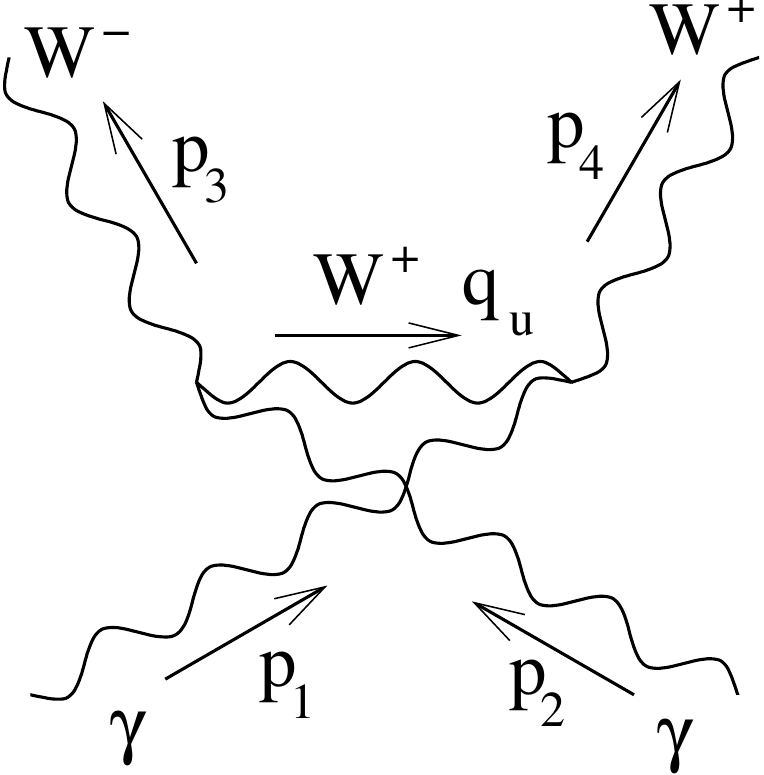}
\end{minipage}
\\
\\
\\
$i\mathcal{M}_{2t}=$
\begin{minipage}{0.4\textwidth}
\includegraphics[scale=0.5]{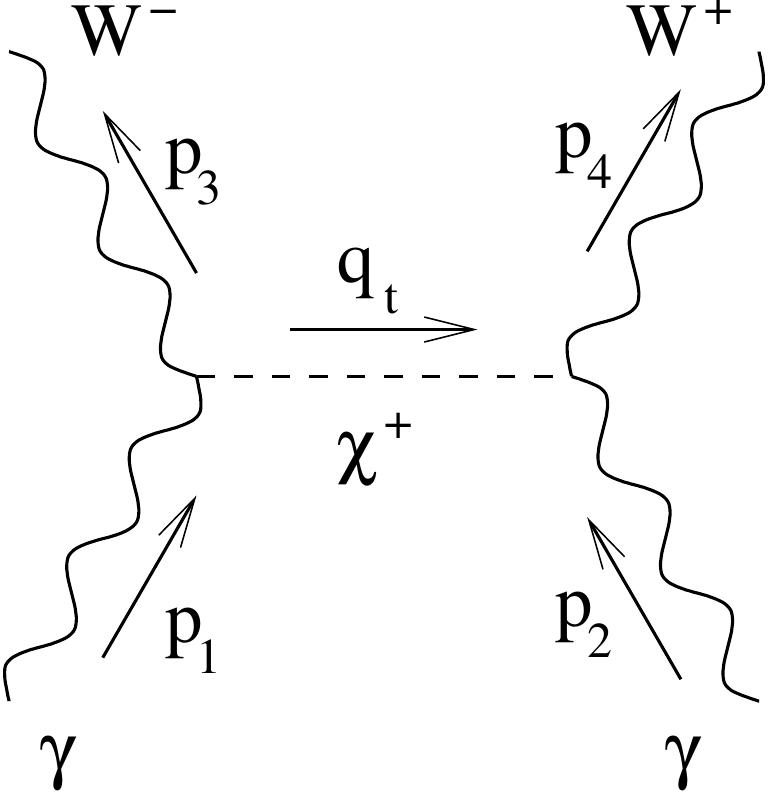}
\end{minipage}
$i\mathcal{M}_{2u}=$
\begin{minipage}{0.3\textwidth}
\includegraphics[scale=0.5]{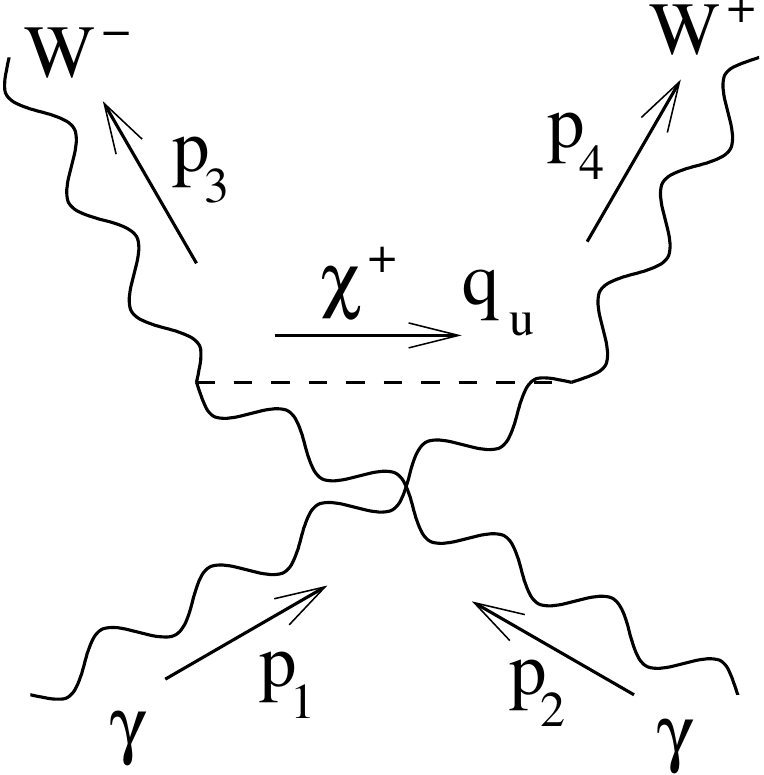}
\end{minipage}
\\
\\
\\
$i\mathcal{M}_{3}=$
\begin{minipage}{0.4\textwidth}
\includegraphics[scale=0.5]{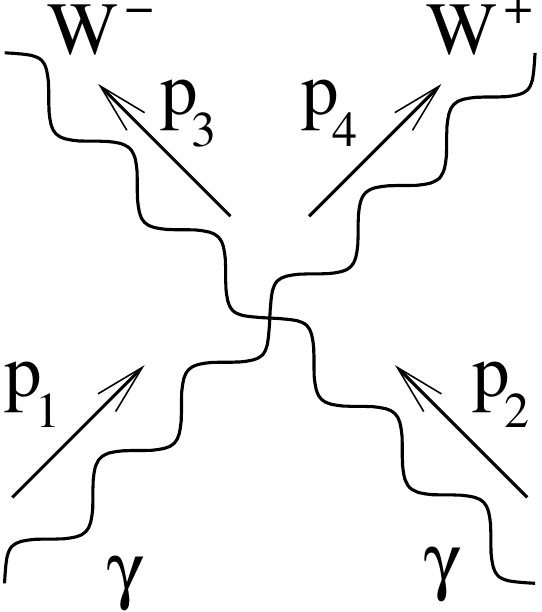}
\end{minipage}
\\
\\
\\
and the full amplitude is:
\begin{equation}
\mathcal{M} = \mathcal{M}_{1t} + \mathcal{M}_{1u} + \mathcal{M}_{2t} +
\mathcal{M}_{2u} + \mathcal{M}_{3}.
\end{equation}
The interesting fact we will see is that each diagram separately is \emph{not}
independent of the gauge-fixing parameters, corresponding to the fact that each
diagram separately is not gauge-invariant, and only the sum of all 5 diagrams
gives a gauge-invariant (and thus independent of the parameters) result.

We now proceed to the calculation of the diagrams -- all the needed Feynman rules for our Lagrangian can be found in the Appendix B. Note that all the interaction vertices involved in the process depend on the non-linear gauge-fixing parameter $\tilde{\alpha}$ and our interest will be how it cancels out in the final amplitude.

Let's define:
\begin{align}
q^\mu_t&=p^\mu_1-p^\mu_3=p^\mu_4-p^\mu_2, \quad\quad t \equiv q_t^2 = m_W^2 - 2p_1 \cdot p_3 = m_W^2
- 2p_2 \cdot p_4,\\
q^\mu_u&=p^\mu_4-p^\mu_1=p^\mu_2-p^\mu_3, \quad\quad u \equiv q_u^2 = m_W^2 - 2p_1 \cdot p_4 = m_W^2
- 2p_2 \cdot p_3,
\end{align}
where $t$, $u$ are Mandelstam variables. Also note that
\begin{equation}
p_1 \cdot p_3 = p_2 \cdot p_4, \quad\quad p_1 \cdot p_4 = p_2 \cdot p_3,
\end{equation}
and clearly
\begin{equation}
p^2_1 = p^2_2 = 0, \quad\quad p^2_3 = p^2_4 = m_W^2.
\end{equation}
We will use $\varepsilon_{i\mu}$ for the polarization of the external vector
particle with momentum $p_i$, and it is -- as usual -- orthogonal to the momentum of the particle
(no sum over $i$ here):
\begin{equation}
\varepsilon_{i\mu} \cdot p_i^\mu = 0.
\end{equation}
Throughout the calculation the following abbreviations will be used:
\begin{align}
\varepsilon_{ij}\equiv& \, \varepsilon_i \cdot \varepsilon_j^*,
\quad{\rm or} \nonumber \\
\varepsilon_{ij}\equiv& \, \varepsilon_i \cdot \varepsilon_j,
\quad{\rm or} \\
\varepsilon_{ij}\equiv& \, \varepsilon_i^* \cdot \varepsilon_j^*,
\nonumber
\end{align}
there is no potential confusion here, because each polarization vector appears
either as $\varepsilon_i$  for incoming particles, or as $\varepsilon_i^*$ 
for outgoing, but not both.

One final note on how the calculation will be organized. In order
to analyze the dependence of the final result on $\tilde\alpha$ it is convenient
to separate terms in the amplitude containing different orders of $\tilde\alpha$.
There will be terms of first and second order in $\tilde\alpha$, which we will
notate as:
\begin{align}
\mathcal{M} =& \mathcal{M}_0 + \mathcal{M}_\alpha + \mathcal{M}_{\alpha 2},\\
\mathcal{M}_0 =& \mathcal{M}_{1t0} + \mathcal{M}_{1u0} + \mathcal{M}_{2t0} +
\mathcal{M}_{2u0} + \mathcal{M}_{30},\\
\mathcal{M}_\alpha =& \mathcal{M}_{1t\alpha} + \mathcal{M}_{1u\alpha} +
\mathcal{M}_{2t\alpha} +
\mathcal{M}_{2u\alpha} + \mathcal{M}_{3\alpha},\\
\mathcal{M}_{\alpha 2} =& \mathcal{M}_{1t\alpha 2} + \mathcal{M}_{1u\alpha 2} +
\mathcal{M}_{2t\alpha 2} +
\mathcal{M}_{2u\alpha 2} + \mathcal{M}_{3\alpha 2}.
\end{align}
We will proceed by calculating each diagram in a row and splitting
the result into three terms by the order of $\tilde\alpha$ for each
diagram.

The first diagram is now:
\begin{align}
\mathcal{M}_{1t} =& -i
\varepsilon_{1\lambda}\varepsilon_{2\sigma}\varepsilon^*_{3\nu}\varepsilon^*_{
4\kappa } C^ { \mu\nu\lambda } _ { \rm { WWA } } (-q_t , -p_3 , p_1)
C^{\kappa\rho\sigma}_{\rm{WWA}}(-p_4,q_t,p_2) \nonumber \\
& \times \frac{-i}{t - m_W^2}\left( g_{\mu\rho} - (1-\xi_W)\frac{q_{t\mu}
q_{t\rho}}{t - \xi_W m_W^2} \right).
\end{align}
We use a shorthand $C^{\mu\nu\lambda}_{\rm{WWA}}(p_1,p_2,p_3)$ for the
$WWA$ interaction vertex (see Appendix B), where $(\mu,\nu,\lambda)$ are contracted with incoming
$W^-$, $W^+$, $A$ lines respectively, and $(p_1,p_2,p_3)$ are their respective
momenta. The interaction vertex $C_{\rm{WWA}}$ consists of two terms:
$C_{0\rm{WWA}}$, the part coming from the original Lagrangian, and the gauge-fixing part $C_{g\rm{WWA}}$ which is proportional to $\tilde\alpha$. Also note that
\begin{equation}
C^{\mu\nu\lambda}_{\rm{WWA}}(p_1,p_2,p_3) =
-C^{\nu\mu\lambda}_{\rm{WWA}}(p_2,p_1,p_3).
\end{equation}
Let's rewrite $\mathcal{M}_{1t}$ then as:
\begin{align}
\mathcal{M}_{1t} =&
\varepsilon_{1\lambda}\varepsilon_{2\sigma}\varepsilon^*_{3\nu}\varepsilon^*_{
4\kappa }  \frac{1}{t - m_W^2}\left( g_{\mu\rho} -
(1-\xi_W)\frac{q_{t\mu}
q_{t\rho}}{t - \xi_W m_W^2} \right) \nonumber \\
&\times (C^ { \mu\nu\lambda } _ { 0\rm { WWA } }
(-q_t , -p_3 , p_1) + C^ { \mu\nu\lambda } _ {g \rm { WWA } } (-q_t , -p_3 , p_1))
\label{eq:M1texpanded}
\\
&\times (C^{\rho\kappa\sigma}_{0\rm{WWA}}(q_t,-p_4,p_2) +
C^{\rho\kappa\sigma}_{g\rm{WWA}}(q_t,-p_4,p_2) )  \nonumber,
\end{align}
and start with the term quadratic in $\tilde\alpha$, which will come from taking
$C_g$ for both interaction vertices. It is:
\begin{align}
\mathcal{M}_{1t\alpha 2} =&
\varepsilon_{1\lambda}\varepsilon_{2\sigma}\varepsilon^*_{3\nu}\varepsilon^*_{
4\kappa } C^ { \mu\nu\lambda } _ {g \rm { WWA } } (-q_t , -p_3 , p_1)
C^{\rho\kappa\sigma}_{g\rm{WWA}}(q_t,-p_4,p_2) \nonumber \\
& \times \frac{1}{t - m_W^2}\left( g_{\mu\rho} - (1-\xi_W)\frac{q_{t\mu}
q_{t\rho}}{t - \xi_W m_W^2} \right) \nonumber \\
=&
\varepsilon_{1\lambda}\varepsilon_{2\sigma}\varepsilon^*_{3\nu}\varepsilon^*_{
4\kappa } \left(\frac{ie\tilde\alpha}{\xi_W}\right)^2 \frac{1}{t - m_W^2}
(-q_t^\mu g^{\nu\lambda} + p_3^\nu g^{\mu\lambda})
(q_t^\rho g^{\kappa\sigma} + p_4^\kappa g^{\rho\sigma}) \nonumber \\
& \times \left( g_{\mu\rho} - (1-\xi_W)\frac{q_{t\mu}
q_{t\rho}}{t - \xi_W m_W^2} \right) \nonumber \\
=&
\varepsilon_{1\lambda}\varepsilon_{2\sigma}\varepsilon^*_{3\nu}\varepsilon^*_{
4\kappa } \left(\frac{e\tilde\alpha}{\xi_W}\right)^2 \frac{q_t^\mu q_t^\rho}{t -
m_W^2}
g^{\nu\lambda} g^{\kappa\sigma} \left( g_{\mu\rho} - (1-\xi_W)\frac{q_{t\mu}
q_{t\rho}}{t - \xi_W m_W^2} \right) \nonumber \\
=& (\varepsilon_{1}\cdot \varepsilon^*_{3})
  (\varepsilon_{2}\cdot\varepsilon^*_{4})
\left(\frac{e\tilde\alpha}{\xi_W}\right)^2 \frac{1}{t - m_W^2}
\left( t - (1-\xi_W)\frac{t^2}{t - \xi_W m_W^2} \right) \nonumber \\
=&  e^2 \tilde\alpha^2 \varepsilon_{13}
  \varepsilon_{24}
\frac{t}{\xi_W(t - \xi_W m_W^2)}.
\label{eq:M1ta2}
\end{align}
Here the third equality is from $\epsilon\cdot p=0$, the fourth one from
$q_t^2=t$, and the fifth one is just some simplifying algebra.

Now we move to the term $\mathcal{M}_{1t\alpha}$ proportional to $\tilde\alpha$,
which comes from (\ref{eq:M1texpanded}) when choosing $C_g$ for one interaction
vertex and $C_0$ for the other:
\begin{align}
\mathcal{M}_{1t\alpha} =&
\varepsilon_{1\lambda}\varepsilon_{2\sigma}\varepsilon^*_{3\nu}\varepsilon^*_{
4\kappa }  \frac{1}{t - m_W^2}\left( g_{\mu\rho} -
(1-\xi_W)\frac{q_{t\mu}
q_{t\rho}}{t - \xi_W m_W^2} \right) \nonumber \\
\times& (C^ { \mu\nu\lambda } _ { 0\rm { WWA } }
(-q_t , -p_3 , p_1) C^{\rho\kappa\sigma}_{g\rm{WWA}}(q_t,-p_4,p_2) \\
&+C^{\rho\kappa\sigma}_{0\rm{WWA}}(q_t,-p_4,p_2) C^ { \mu\nu\lambda } _ {g \rm
{WWA } } (-q_t , -p_3 ,p_1) ). \nonumber
\end{align}
Vertex $C_g$ again reduces to just a constant times $q_t$ because of
orthogonal polarizations:
\begin{align}
\mathcal{M}_{1t\alpha} =&
\varepsilon_{1\lambda}\varepsilon_{2\sigma}\varepsilon^*_{3\nu}\varepsilon^*_{
4\kappa }  \frac{1}{t - m_W^2}
\left( \frac{ie\tilde\alpha}{\xi_W} \right)
\left( g_{\mu\rho} -
(1-\xi_W)\frac{q_{t\mu}
q_{t\rho}}{t - \xi_W m_W^2} \right)  \\
\times& (C^ { \mu\nu\lambda } _ { 0\rm { WWA } }
(-q_t , -p_3 , p_1) q_t^\rho g^{\kappa\sigma}
- C^{\rho\kappa\sigma}_{0\rm{WWA}}(q_t,-p_4,p_2)
q_t^\mu g^{\nu\lambda} ). \nonumber
\end{align}
In order to evaluate this expression we need to calculate the piece:
\begin{align}
K_{13} \equiv&
q_{t\mu}\varepsilon_{1\lambda}\varepsilon^*_{3\nu}
C^{\mu\nu\lambda}_{0\rm{WWA}}(-q_t,-p_3,p_1) \nonumber \\
=&
(p_1-p_3)_{\mu}\varepsilon_{1\lambda}\varepsilon^*_{3\nu}
C^{\mu\nu\lambda}_{0\rm{WWA}}(p_3-p_1,-p_3,p_1) \nonumber \\
=&
-ie
(p_1-p_3)_{\mu}\varepsilon_{1\lambda}\varepsilon^*_{3\nu}
\left( (-2p_3+p_1)^\lambda g^{\mu\nu} +
(p_1+p_3)^\mu g^{\nu\lambda} +
(-2p_1+p_3)^\nu g^{\lambda\mu} \right) \nonumber \\
=&
-ie \left(
-2(p_1 \cdot \varepsilon^*_3)(p_3 \cdot \varepsilon_1) -
(p_3^2-p_1^2)(\varepsilon_1 \cdot \varepsilon^*_3) +
2(p_3 \cdot \varepsilon_1)(p_1 \cdot \varepsilon^*_3)
\right) \nonumber \\
=&
i e m_W^2 \varepsilon_{13}.
\label{eq:K13}
\end{align}
There is also an analogous contraction:
\begin{align}
K_{24} \equiv&
-q_{t\mu}\varepsilon_{2\lambda}\varepsilon^*_{4\nu}
C^{\mu\nu\lambda}_{0\rm{WWA}}(q_t,-p_4,p_2) \nonumber \\
=&
(p_2-p_4)_{\mu}\varepsilon_{2\lambda}\varepsilon^*_{4\nu}
C^{\mu\nu\lambda}_{0\rm{WWA}}(p_4-p_2,-p_4,p_2) \nonumber,
\end{align}
which, comparing to the second line of (\ref{eq:K13}), is clearly the same as
$K_{13}$ but with all indexes $1$ replaced by $2$ and $3$ replaced by $4$.
Therefore, the same calculation yields:
\begin{equation}
K_{24} = i e m_W^2 \varepsilon_{24}.
\end{equation}
Then $\mathcal{M}_{1t\alpha}$ is just repetitive occurrence of
$K_{13}$ and $K_{24}$:
\begin{align}
\mathcal{M}_{1t\alpha} =&
\frac{ie\tilde\alpha}{\xi_W(t - m_W^2)}
\left[
K_{13}\varepsilon_{24} + K_{24}\varepsilon_{13} -
\frac{1-\xi_W}{t-\xi_W m_W^2}
\left( K_{13}\varepsilon_{24} q_t^2 + K_{24}\varepsilon_{13} q_t^2
\right)
\right] \nonumber \\
=&
\frac{ie\tilde\alpha}{\xi_W(t - m_W^2)}
2 i e m_W^2 \varepsilon_{13} \varepsilon_{24}
\left[ 1 - \frac{t(1-\xi_W)}{t-\xi_W m_W^2} \right] \nonumber \\
=&
-2e^2 \tilde\alpha \varepsilon_{13} \varepsilon_{24}
\frac{m_W^2}{t-\xi_W m_W^2},
\label{eq:M1ta}
\end{align}
where the last line is again just some algebra.

Finally, the last term in $\mathcal{M}_{1t}$, independent of $\tilde\alpha$, is
with both vertices taken as $C_0$:
\begin{align}
\label{eq:M1t0first}
\mathcal{M}_{1t0} =&
\varepsilon_{1\lambda}\varepsilon_{2\sigma}\varepsilon^*_{3\nu}\varepsilon^*_{
4\kappa } C^ { \mu\nu\lambda } _ {0 \rm { WWA } } (-q_t , -p_3 , p_1)
C^{\rho\kappa\sigma}_{0\rm{WWA}}(q_t,-p_4,p_2) \\
& \times \frac{1}{t - m_W^2}\left( g_{\mu\rho} - (1-\xi_W)\frac{q_{t\mu}
q_{t\rho}}{t - \xi_W m_W^2} \right). \nonumber
\end{align}
We would like to do one last separation and consider the piece of this amplitude:
\begin{align}
\mathcal{M}_{1t00} =&
\varepsilon_{1\lambda}\varepsilon_{2\sigma}\varepsilon^*_{3\nu}\varepsilon^*_{
4\kappa } C^ { \mu\nu\lambda } _ {0 \rm { WWA } } (-q_t , -p_3 , p_1)
C^{\rho\kappa\sigma}_{0\rm{WWA}}(q_t,-p_4,p_2)
\frac{g_{\mu\rho}}{t - m_W^2}. \nonumber \\
=&
(-ie)^2
\varepsilon_{1\lambda}\varepsilon_{2\sigma}\varepsilon^*_{3\nu}\varepsilon^*_{
4\kappa } \frac{g_{\mu\rho}}{t - m_W^2}
\nonumber
\\
&\times
\left( (2p_3-p_1)^\lambda g^{\mu\nu} -
(p_3+p_1)^\mu g^{\nu\lambda} +
(2p_1-p_3)^\nu g^{\lambda\mu} \right)
\nonumber \\
&\times
\left( (2p_4-p_2)^\sigma g^{\rho\kappa} -
(p_4+p_2)^\rho g^{\kappa\sigma} +
(2p_2-p_4)^\kappa g^{\sigma\rho} \right).
\label{eq:M1t00}
\end{align}
The reason to write it separately is that this piece is completely independent of the
gauge parameters (unlike the rest of $\mathcal{M}_{1t0}$, which has $\xi_W$).
This
is in fact the full $\mathcal{M}_{1t}$ in the case of simple gauge choice
$\xi_W=1$, $\tilde\alpha=0$, and we will see in the end that it contains most of
the full amplitude in the $t$-channel. However, in the context of this work
this term is not very interesting, because we want to analyze the dependence on
the gauge, and it is already independent. Furthermore, expanding
(\ref{eq:M1t00}) into scalar products is quite tedious and not very
illuminating, so there is not much more we can do, but to keep this term as
$\mathcal{M}_{1t00}$. The rest of $\mathcal{M}_{1t0}$ reduces to $K_{13}$ and
$K_{24}$ and we have from (\ref{eq:M1t0first}):
\begin{align}
\mathcal{M}_{1t0} =& \mathcal{M}_{1t00} + \frac{1-\xi_W}{(t-m_W^2)(t-\xi_W
m_W^2)} K_{13} K_{24} \nonumber \\
=&
\mathcal{M}_{1t00} -
e^2 \varepsilon_{13} \varepsilon_{24}
\frac{m_W^4(1-\xi_W)}{(t-m_W^2)(t-\xi_W m_W^2)}.
\label{eq:M1t0}
\end{align}

We have now calculated the full $\mathcal{M}_{1t}$, which can be built from
(\ref{eq:M1ta2}), (\ref{eq:M1ta}) and (\ref{eq:M1t0}). The next diagram on the
list is $\mathcal{M}_{1u}$. Here, however, almost no extra calculations are
needed, because this amplitude is exactly like $\mathcal{M}_{1t}$ but with the
two photons interchanged. That is, we can rewrite completely the same
calculation making everywhere the following replacements:
\begin{align}
p_1 \rightarrow p_2,& \quad\quad p_2 \rightarrow p_1, \\
\varepsilon_1 \rightarrow \varepsilon_2,& \quad\quad \varepsilon_2 \rightarrow
\varepsilon_1, \\
q_t \rightarrow q_u,& \quad\quad t \rightarrow u.
\end{align}
Therefore, the final results are analogously:
\begin{align}
\mathcal{M}_{1u\alpha 2} =&  e^2 \tilde\alpha^2 \varepsilon_{14}
  \varepsilon_{23} \frac{u}{\xi_W(u - \xi_W m_W^2)}, \\
\mathcal{M}_{1u\alpha} =&
-2e^2 \tilde\alpha \varepsilon_{14} \varepsilon_{23}
\frac{m_W^2}{u-\xi_W m_W^2}, \\
\mathcal{M}_{1u0} =&
\mathcal{M}_{1u00} -
e^2 \varepsilon_{14} \varepsilon_{23}
\frac{m_W^4(1-\xi_W)}{(u-m_W^2)(u-\xi_W m_W^2)},
\end{align}
and
\begin{align}
\mathcal{M}_{1u00} =&
(-ie)^2
\varepsilon_{2\lambda}\varepsilon_{1\sigma}\varepsilon^*_{3\nu}\varepsilon^*_{
4\kappa } \frac{g_{\mu\rho}}{u - m_W^2}
\nonumber
\\
&\times
\left( (2p_3-p_2)^\lambda g^{\mu\nu} -
(p_3+p_2)^\mu g^{\nu\lambda} +
(2p_2-p_3)^\nu g^{\lambda\mu} \right)
\nonumber \\
&\times
\left( (2p_4-p_1)^\sigma g^{\rho\kappa} -
(p_4+p_1)^\rho g^{\kappa\sigma} +
(2p_1-p_4)^\kappa g^{\sigma\rho} \right).
\label{eq:M1u00}
\end{align}

Now we move to calculation of $\mathcal{M}_{2t}$ and $\mathcal{M}_{2u}$. The
interaction vertex with a Goldstone boson and its propagator are quite simple,
so we can write the amplitude out fully as:
\begin{align}
\mathcal{M}_{2t} =& -i
\varepsilon_{1\mu}\varepsilon_{2\rho}\varepsilon^*_{3\nu}\varepsilon^*_{
4\sigma }
(-(1-\tilde\alpha)e m_W g^{\mu\nu})((1-\tilde\alpha)e m_W g^{\rho\sigma})
\frac{i}{t - \xi_W m_W^2} \nonumber \\
=&
-e^2 (1 - 2\tilde\alpha + \tilde\alpha^2) \varepsilon_{13} \varepsilon_{24}
\frac{m_W^2}{t - \xi_W m_W^2}.
\end{align}
The $u$-channel part is analogously
\begin{equation}
\mathcal{M}_{2u} = -e^2 (1 - 2\tilde\alpha + \tilde\alpha^2) \varepsilon_{14}
\varepsilon_{23}
\frac{m_W^2}{u - \xi_W m_W^2},
\end{equation}
and we can write out components with different powers of $\tilde\alpha$:
\begin{align}
\mathcal{M}_{2t0} =
-e^2 \varepsilon_{13} \varepsilon_{24}
\frac{m_W^2}{t - \xi_W m_W^2},& \quad\quad
\mathcal{M}_{2u0} =
-e^2 \varepsilon_{14} \varepsilon_{23}
\frac{m_W^2}{u - \xi_W m_W^2} \\
\mathcal{M}_{2t\alpha} =
2 e^2 \tilde\alpha \varepsilon_{13} \varepsilon_{24}
\frac{m_W^2}{t - \xi_W m_W^2},& \quad\quad
\mathcal{M}_{2u\alpha} =
2 e^2 \tilde\alpha \varepsilon_{14} \varepsilon_{23}
\frac{m_W^2}{u - \xi_W m_W^2} \\
\mathcal{M}_{2t\alpha2} =
-e^2 \tilde\alpha^2 \varepsilon_{13} \varepsilon_{24}
\frac{m_W^2}{t - \xi_W m_W^2},& \quad\quad
\mathcal{M}_{2u\alpha2} =
-e^2 \tilde\alpha^2 \varepsilon_{14} \varepsilon_{23}
\frac{m_W^2}{u - \xi_W m_W^2}
\end{align}

And, finally, the last diagram $\mathcal{M}_3$ is simply the interaction
vertex with two $W$'s and two photons:
\begin{align}
\mathcal{M}_3 =& -i
\varepsilon_{1\mu}\varepsilon_{2\nu}\varepsilon^*_{3\rho}\varepsilon^*_{
4\sigma }
ie^2 \left[ -2 g_{\mu\nu} g_{\rho\sigma} + \left( 1 -
\frac{\tilde\alpha^2}{\xi_W} \right) \left(
g_{\mu\rho}g_{\nu\sigma} + g_{\mu\sigma}g_{\nu\rho} \right) \right] \nonumber \\
=&
e^2
\left[ -2 \varepsilon_{12} \varepsilon_{34}
+ \left( 1 - \frac{\tilde\alpha^2}{\xi_W} \right)
\left(\varepsilon_{13} \varepsilon_{24} +
\varepsilon_{14} \varepsilon_{23} \right)
\right].
\end{align}
The $\tilde\alpha$-independent component is
\begin{equation}
\mathcal{M}_{30} =
e^2
\left( -2 \varepsilon_{12} \varepsilon_{34}
+ \varepsilon_{13} \varepsilon_{24} +
\varepsilon_{14} \varepsilon_{23} \right),
\end{equation}
and it is convenient to further split up the part quadratic in $\tilde\alpha$
into two terms - corresponding to $t$-channel and $u$-channel respectively:
\begin{equation}
\mathcal{M}_{3t\alpha2} = -\frac{e^2\tilde\alpha^2}{\xi_W}
\varepsilon_{13} \varepsilon_{24}, \quad\quad
\mathcal{M}_{3u\alpha2} = -\frac{e^2\tilde\alpha^2}{\xi_W}
\varepsilon_{14} \varepsilon_{23}.
\end{equation}
The full $\mathcal{M}_3$ is then
\begin{equation}
\mathcal{M}_3 = \mathcal{M}_{30} + \mathcal{M}_{3t\alpha2} +
\mathcal{M}_{3u\alpha2},
\end{equation}
and there is no term in $\mathcal{M}_3$ proportional to $\tilde\alpha$, that is,
$\mathcal{M}_{3\alpha} = 0$.

We have now all the parts to construct the full amplitude $\mathcal{M}$. We will
start with $\tilde\alpha$-dependent terms and show that they add up to 0. Then we
will analyze the $\mathcal{M}_0$ component and show that the other gauge
parameter, $\xi_W$, also drops out. Since we have split $\mathcal{M}_{3\alpha2}$
into $t$ and $u$ components, we can perform $\tilde\alpha$-dependence calculation
for $t$ and $u$ channels independently. So the quadratic $t$ component is:
\begin{align}
\mathcal{M}_{t\alpha2} =&  \mathcal{M}_{1t\alpha2} + \mathcal{M}_{2t\alpha2}
+ \mathcal{M}_{3t\alpha2} \nonumber \\
=& e^2 \tilde\alpha^2 \varepsilon_{13} \varepsilon_{24}
\frac{t}{\xi_W(t - \xi_W m_W^2)}
-e^2 \tilde\alpha^2 \varepsilon_{13} \varepsilon_{24}
\frac{m_W^2}{t - \xi_W m_W^2}
-\frac{e^2\tilde\alpha^2}{\xi_W}
\varepsilon_{13} \varepsilon_{24} \nonumber \\
=&
e^2 \tilde\alpha^2 \varepsilon_{13} \varepsilon_{24}
\left(
\frac{t}{\xi_W(t - \xi_W m_W^2)} -
\frac{m_W^2}{t - \xi_W m_W^2} -
\frac{1}{\xi_W}
\right) \nonumber \\
=&
e^2 \tilde\alpha^2 \varepsilon_{13} \varepsilon_{24}
\left(
\frac{t-\xi_W m_W^2 -t + \xi_W m_W^2}{\xi_W(t - \xi_W m_W^2)} -
\right) = 0,
\end{align}
and by complete analogy
\begin{equation}
\mathcal{M}_{u\alpha2} =  \mathcal{M}_{1u\alpha2} + \mathcal{M}_{2u\alpha2}
+ \mathcal{M}_{3u\alpha2} = 0,
\end{equation}
the quadratic $\tilde\alpha$ terms have canceled! The linear terms also add up to
0 as expected:
\begin{equation}
\mathcal{M}_{t\alpha} = \mathcal{M}_{1t\alpha} + \mathcal{M}_{2t\alpha} =
-2e^2 \tilde\alpha \varepsilon_{13} \varepsilon_{24}
\frac{m_W^2}{t-\xi_W m_W^2} + 2e^2 \tilde\alpha \varepsilon_{13}
\varepsilon_{24}
\frac{m_W^2}{t-\xi_W m_W^2}=0,
\end{equation}
and so also
\begin{equation}
\mathcal{M}_{u\alpha} = \mathcal{M}_{1u\alpha} + \mathcal{M}_{2u\alpha} = 0.
\end{equation}
So the full amplitude is $\tilde\alpha$ independent! It is explicit here how only
the sum of all diagrams gives this gauge invariance, while each separate
diagram is gauge dependent.

Now let's finish with calculating $\mathcal{M}_0$, where we still have to see
the
$\xi_W$-independence. Since $\mathcal{M}_{30}$ already has no $\xi_W$ we
evaluate the sum of the other two terms:
\begin{align}
\mathcal{M}_{1t0} + \mathcal{M}_{2t0} =&
\mathcal{M}_{1t00} -
e^2 \varepsilon_{13} \varepsilon_{24}
\frac{m_W^4(1-\xi_W)}{(t-m_W^2)(t-\xi_W m_W^2)} -
e^2 \varepsilon_{13} \varepsilon_{24}
\frac{m_W^2}{t - \xi_W m_W^2} \nonumber \\
=&
\mathcal{M}_{1t00} - e^2 \varepsilon_{13} \varepsilon_{24}
\frac{m_W^2}{t - \xi_W m_W^2}
\left(
\frac{m_W^2(1-\xi_W)}{t-m_W^2} + 1
\right) \nonumber \\
=&
\mathcal{M}_{1t00} - e^2 \varepsilon_{13} \varepsilon_{24}
\frac{m_W^2}{t - m_W^2},
\end{align}
and likewise for $u$-channel:
\begin{equation}
\mathcal{M}_{1u0} + \mathcal{M}_{2u0} =
\mathcal{M}_{1u00} - e^2 \varepsilon_{14} \varepsilon_{23}
\frac{m_W^2}{u - m_W^2}.
\end{equation}
The dependence on $\xi_W$ has dropped out! The full amplitude is
then
\begin{align}
\mathcal{M} =& \mathcal{M}_{1t0} + \mathcal{M}_{2t0} + \mathcal{M}_{1u0} +
\mathcal{M}_{2u0} + \mathcal{M}_{30} \nonumber \\
=&
\mathcal{M}_{1t00} + \mathcal{M}_{1u00}
- e^2 \varepsilon_{13} \varepsilon_{24} \frac{m_W^2}{t - m_W^2}
- e^2 \varepsilon_{14} \varepsilon_{23} \frac{m_W^2}{u - m_W^2} \nonumber \\
&+ e^2 \left( -2 \varepsilon_{12} \varepsilon_{34}
+ \varepsilon_{13} \varepsilon_{24} +
\varepsilon_{14} \varepsilon_{23} \right),
\label{eq:Mfinal}
\end{align}
which is free of any gauge-fixing parameters. As mentioned before, the largest
part of the result is contained in $\mathcal{M}_{1t00}$ and
$\mathcal{M}_{1u00}$ defined in (\ref{eq:M1t00}), (\ref{eq:M1u00}), and for our
purposes there is no need to further expand them.

In this section we have performed a full tree-level calculation of the
amplitude for $W^+W^-$ production, and showed that it is independent of both
non-linear and linear gauge-fixing parameters. On the other hand, it is clear
now how this fact can be used to simplify the calculations. Choosing, for
example,
the gauge $\tilde\alpha=\xi_W=1$ eliminates the diagrams $\mathcal{M}_{2t}$ and
$\mathcal{M}_{2u}$, and removes $t$ and $u$ channel components from
$\mathcal{M}_3$ - the only term remaining there is proportional to
$\varepsilon_{12}\varepsilon_{34}$. Then the diagram $\mathcal{M}_{1t}$ with
the modified interaction vertex will produce the \emph{full} amplitude for the
$t$ channel. Specifically, in addition to $\mathcal{M}_{1t00}$, which is the
``usual" amplitude for that diagram, corresponding to gauge $\xi_W=1$,
$\tilde\alpha=0$, it will produce the other two terms in (\ref{eq:Mfinal})
proportional to $\varepsilon_{13}\varepsilon_{24}$ that are normally seen as
coming from the diagrams $\mathcal{M}_{2t}$ and $\mathcal{M}_3$.

\section{Non-linear gauge-fixing in FeynArts}

We have implemented the full Lagrangian with non-linear gauge-fixing in FeynArts, and used it to perform a number of automatic amplitude calculations. The results, as expected, turned out to be gauge-independent, which is a good indication that there were no errors in the implementation.

In this section first we will briefly describe FeynArts/FormCalc and how it is used. Then we will explain what changes have to be made to accommodate non-linear gauge-fixing, and, finally, we will discuss the calculations that we did.

\subsection{About FeynArts and FormCalc}

FeynArts/FormCalc \cite{feynarts} is a freely available open source package for \emph{Mathematica} that performs automatic calculation of Feynman diagrams up to 2-loop level. The download of the package and more information about it is available at:
$$\verb} http://www.feynarts.de }$$
In this work we have worked with the version FeynArts~3.2.

The calculation of an amplitude in FeynArts is performed as follows. The process is defined simply as a set of incoming and outgoing particles. The package then provides functions to do all the required steps: first, it generates all the possible Feynman diagrams at the given loop level, including the counterterm diagrams, second, it constructs the algebraic expressions for the diagrams using the definitions for vertices and propagators, and, finally, it can algebraically evaluate these expressions by applying various rules resulting in a greatly simplified symbolic result for the total amplitude. During this last step the loop integrals are reduced to the Passarino-Veltman integrals, also the counterterms are evaluated using the defined renormalization scheme, so the resulting amplitude should be already free of ultraviolet divergences. 

The final symbolic expression for the amplitude can be used then to calculate numerical results. For that purpose the package can generate a full FORTRAN program, which, when compiled and run, will loop through the phase space of a process producing numeric values for the cross-section, which can be compared with experimental results. In this work, however, we have not used numeric calculations.

The first steps -- generating diagrams, and translating them into expressions using propagators and vertices -- are done using the functionality of Mathematica and this part of the package is called \emph{FeynArts}, while the last steps -- algebraic evaluation and FORTRAN code generation -- internally uses another program \emph{FORM}, and so this part is called the \emph{FormCalc} package. Nevertheless, these two are closely related and can be really thought of as one complete package.

The really attractive feature of FeynArts is that the physical model for calculations is not built-in, but provided by external input files called \emph{model files}, which contain all the information about the theory: the definitions of the fields, their properties, propagators, interaction vertices, counterterms and renormalization conditions. The package comes already with some predefined models, including QED, Standard Model and MSSM, that can be used immediately for calculations. In order to implement the non-linear gauge-fixing we had to modify the existing Standard Model file, and we give now the details of the needed modifications.

\subsection{Implementation in FeynArts}

The existing definition of the Standard Model in FeynArts matches the one we have discussed in this work in the case of \emph{linear} gauge-fixing. That is, all the fields, propagators and interaction vertices are the same as listed in Appendix \ref{sec:vertices} with the choice $\tilde\alpha=\tilde\beta=\tilde\delta=\tilde\kappa=\tilde\varepsilon=0$. Therefore, the only change needed for non-linear gauge-fixing is to modify the interaction vertices with the additional terms proportional to the non-linear gauge-fixing parameters.

First, in order to compare our vertices with the ones defined in FeynArts, we  note the following differences in sign conventions. FeynArts defines the Fourier transform so that $\partial_\mu\rightarrow ip_\mu$ as opposed to $\partial_\mu\rightarrow -ip_\mu$, so there is a relative minus sign in all vertices that include particle momenta. Other differences relate to the additional factors in the definitions of the fields (which are, of course, allowed), as a result of that vertices get multiplied by a factor for each occurrence of the fields with differing definitions. 
We give in Table~1 these extra factors that should be applied to a vertex given in Appendix~\ref{sec:vertices}, in order to get the corresponding FeynArts vertex. 
\begin{table}
\label{tab:factors}
\caption{Extra factors to get a FeynArts vertex from the one in Appendix~\ref{sec:vertices}}
\begin{center}
\begin{tabular}{|c|c|}
\hline
Field & Factor \\
\hline
$Z$ & -1  \\
$\chi^-$ & $+i$ \\
$\chi^+$ & $-i$ \\
$c^Z, \bar{c}^Z$ & $-1$ \\
$\bar{c}^A, \bar{c}^Z, \bar{c}^\pm$ & $\xi_A^{-1/2}, \xi_Z^{-1/2},  \xi_W^{-1/2}$
respectively \\
\hline
\end{tabular}
\end{center}
\end{table}
In order not to be forced to change all the existing definitions for interaction vertices in FeynArts, we adopted the FeynArts conventions for the fields when implementing the non-linear gauge-fixing modifications, thus converting the interaction vertices using these rules.

Consider now the new terms in the interaction vertices that the non-linear gauge-fixing yields. Note that some of them just change the overall factor of the already existing vertices, like for Vector-Vector-Scalar-Scalar or Vector-Vector-Scalar couplings, while in other cases the non-linear gauge-fixing gives a whole new kinematic term, e.g. for Vector-Vector-Vector or Vector-Scalar-Scalar couplings. The latter case is especially notable in ghost interactions, where you can see, that the whole two new classes of interactions -- Ghost-Ghost-Vector-Vector and Ghost-Ghost-Scalar-Scalar -- were generated by the non-linear gauge-fixing. Now the reason it is important to note the new kinematic terms is that they require some additional work in order to be implemented. That comes from the fact that FeynArts defines a physical model in two steps: the \emph{generic model file} -- called {\tt Lorentz.gen} in the case of the Standard Model -- and the \emph{classes model file} -- called {\tt SM.mod}. The generic model file does not refer to the specific particles, but only to \emph{generic} fields, which can be vector, scalar, fermion or ghost. The file then defines generic couplings saying that, for example, the Vector-Vector-Vector vertex is of the form:
\begin{equation}
C^{\mu\nu\rho}_{VVV}(p_1,p_2,p_3) =
C^1_{VVV} \cdot \left[g^{\mu\nu}(p_1-p_2)^\rho + g^{\nu\rho}(p_2-p_3)^\mu + 
	g^{\rho\mu}(p_3-p_1)^\nu  \right],
\end{equation}
and Vector-Vector-Vector-Vector is:
\begin{equation}
C^{\mu\nu\rho\sigma}_{VVVV} =
C^1_{VVVV} \cdot g^{\mu\nu}g^{\rho\sigma} +
C^2_{VVVV} \cdot g^{\mu\rho}g^{\nu\sigma} +
C^3_{VVVV} \cdot g^{\mu\sigma}g^{\nu\rho},
\end{equation}
so that just the coefficients $C^1, C^2, C^3$ depend on the fields, but not the structure of the coupling. The classes model file then defines exactly what kinds of fields there are in the theory, and specifies these coefficients for all specific vertices.

This separation into generic and classes models is, of course, very useful because it allows to define common properties for the vectors, scalars, fermions, ghosts, and thus the same {\tt Lorentz.gen} file can be used, for example, for both SM and MSSM. The complication with the additional kinematic terms arising from the non-linear gauge-fixing is that they do not ``fit'' into the standard {\tt Lorentz.gen} definitions. For example, the generic Vector-Vector-Vector vertex has to be defined as
\begin{align}
C^{\mu\nu\rho}_{VVV}(p_1,p_2,p_3) = &
C^1_{VVV} \cdot \left[g^{\mu\nu}(p_1-p_2)^\rho + g^{\nu\rho}(p_2-p_3)^\mu + 
	g^{\rho\mu}(p_3-p_1)^\nu  \right] 
\\ \nonumber &
+
C^2_{VVV} \cdot p_1^\mu g^{\nu\rho} +
C^3_{VVV} \cdot p_2^\nu g^{\rho\mu} +
C^4_{VVV} \cdot p_3^\rho g^{\mu\nu},
\end{align}
in order to accommodate the extra terms proportional to 
$p_1^\mu g^{\nu\rho}-p_2^\nu g^{\rho\mu}$. 
Note that $C^4_{VVV}$ is not really used in the vertices that we have, but it has to be included in the generic coupling definition, because FeynArts requires that the generic coupling keeps the same structure under the permutation of the fields. The other generic couplings we had to extend are:
\begin{align}
C^\mu_{SSV}(p_1,p_2,p_3) =& \, C^1\cdot (p_1-p_2)^\mu + C^2\cdot p_3^\mu,
\\
C^{\mu\nu}_{GGVV} =& \, C^1_{GGVV} \cdot g^{\mu\nu},
\\
C_{GGSS} =& \, C^1_{GGSS},
\end{align}
where in the first one we had to allow for the new extra term proportional to $p_3^\mu$, and the following two are the four-vertices with ghosts, that were not present at all. The parts of {\tt Lorentz.gen} containing these changes can be found in Appendix~\ref{sec:modelfiles}.

Once the changes to the generic model file are made, the modifications of the class model file are quite straightforward: we just have to go through all the vertices, and include the terms proportional to $\tilde\alpha, \tilde\beta, \tilde\delta, \tilde\kappa, \tilde\varepsilon$ into the existing or newly defined coupling constants. Some examples of these alterations are also given in Appendix~\ref{sec:modelfiles}.

\subsection{Calculation tests}

After we have implemented the non-linear gauge-fixing by modifying the generic and class model files, we have used the new model to perform calculations of a number of amplitudes in order to test the implementation. The condition for testing is, of course, the same fact that this whole work relies on: the final amplitude for a process has to be gauge-independent, which means that in the calculation of any physical process the newly introduced gauge-fixing parameters $\tilde\alpha, \tilde\beta, \tilde\delta, \tilde\kappa, \tilde\varepsilon$ have to cancel out. Here we describe the tree-level and loop-level calculations for which we have checked this cancellation. 

\subsubsection{Tree-level calculations}

For the tree-level amplitude test we have chosen to do the same  $\gamma\gamma\rightarrow W^+W^-$ calculation that we already did by hand in Section~\ref{sec:explicitcalc}, in order to be able to analyze it better. The immediate result for the amplitude after writing out the diagrams in terms of vertices and propagators looked highly dependent on gauge-fixing parameters $\xi_W$ and $\tilde\alpha$, which is to be expected, since, as we saw, all the vertices involved do depend on $\tilde\alpha$. However, after running FormCalc, which is able to simplify the result by using the fact that the polarizations are transverse, and then using
simplifying functions of Mathematica the gauge-fixing parameters did cancel
out and the final result corresponded to (\ref{eq:Mfinal}).

By using the same simplifying procedures we were also able to confirm the gauge-invariance of other tree-level processes including $ZZ\rightarrow W^-W^+$, $e^-e^+\rightarrow W^-W^+$, $e^-e^+\rightarrow\mu^-\mu^+$.

\subsubsection{One-loop calculations}

The one-loop calculations are substantially more difficult than the tree-level, and we have not been able to show algebraically that the full one-loop amplitude is independent of the gauge parameters. Note that a single one-loop amplitude can contain almost all the vertices of the theory in its internal loop, therefore, the cancellation of all the different gauge-fixing terms is far more complicated than in the tree-level (which already was non-trivial), and the gauge-invariance check is much more strict.
Such a check would likely require a numeric calculation using the generated FORTRAN program, which can be run using different choices of the gauge-fixing parameters, and the results should not vary. However, before a numerical calculation can be carried out, it should be checked that the amplitude is free of ultraviolet divergences, otherwise the result of the numerical calculation (which doesn't include the divergent terms) could not be trusted. The cancellation of the divergent terms has to be ensured by the renormalization scheme, and for the definitions in the given {\tt SM.mod}, which has been used extensively in one-loop calculations \cite{feynartsWWWW}, it has been tested to work. But now, after modifying the model with the non-linear gauge-fixing, we have to first check again that the renormalization works and the UV divergences cancel before doing any further one-loop calculations.

FeynArts provides a way to check for UV divergences using the function {\tt UVDivergentPart}, which returns the coefficient of the divergence\footnote{FeynArts uses dimensional regularization, so the divergence is $2/(4-D)$} 
of the final amplitude. If the renormalization is successful, this coefficient should be zero. We have performed this check on the one-loop amplitudes calculated with the non-linear gauge-fixing, and, using the simplifying functions of Mathematica, showed that all the tested amplitudes are indeed free of UV divergence. The one-loop amplitudes that were tested include all the two-point functions and the processes $e^-e^+\rightarrow W^-W^+$, $e^-e^+\rightarrow\mu^-\mu^+$, $\gamma\gamma\rightarrow W^+W^-$. Thus we have shown that the renormalization scheme of the Standard Model works when the non-linear gauge-fixing is included, and the numerical results of one-loop amplitudes can be considered.


\section{Conclusions}

In this work we have presented a full derivation of the Feynman rules of the Standard Model with the non-linear gauge-fixing. We have implemented these rules in the Feyn\-Arts/FormCalc package by modifying the existing definitions for the Standard Model. We have confirmed that the new definitions give gauge invariant results in the tree-level amplitudes, and we have also checked that the one-loop  calculations are free of ultraviolet divergences. 

The new definitions have not been tested yet to give fully gauge invariant results in the one-loop level; such a check would probably have to be done by numeric calculations of the cross-sections, since the expressions for one-loop amplitudes seem to be too complicated to be analyzed algebraically. Such full check of one-loop gauge invariance would be the most reasonable next step in the continuation of this work.

Another direction in which this work could be extended is the implementation of the non-linear gauge-fixing in the Minimal Supersymmetric Standard Model (MSSM). The automatic calculations in the MSSM are as important as in the SM, because the predictions of both models have to be checked against the precision measurements, and the non-linear gauge invariance check would be a very useful tool. 

\newpage

\appendix

\section*{Appendixes}

\section{Full Lagrangian}

We give the full Lagrangian of the Standard Model grouped in terms based on species of the fields (vector, scalar, fermion, ghost) involved.
\begin{align}
\mathcal{L} =& \mathcal{L}_v + \mathcal{L}_s + \mathcal{L}_f + \mathcal{L}_g +
\\ \nonumber
&  + \mathcal{L}_{vvvv} + \mathcal{L}_{vvv} +
	\mathcal{L}_{vvss} + \mathcal{L}_{vvs} + \mathcal{L}_{vss} +
	\mathcal{L}_{ssss} + \mathcal{L}_{sss} +
\\ \nonumber
& + \mathcal{L}_{ffv} + \mathcal{L}_{ffs} + 
\\ \nonumber
& + \mathcal{L}_{ggv} + \mathcal{L}_{ggs} + \mathcal{L}_{ggvv} + \mathcal{L}_{ggss}.
\end{align}
\begin{align}
\mathcal{L}_v =& 
-W^-_\mu (-g^{\mu\nu}\partial^2 + (1-\xi_W)\partial^\mu\partial^\nu
	- m_W^2 g^{\mu\nu}) W^+_\nu 
\\ \nonumber &
-\frac{1}{2} Z_\mu (-g^{\mu\nu}\partial^2 + (1-\xi_Z)\partial^\mu\partial^\nu
	- m_Z^2 g^{\mu\nu}) Z_\nu 
-\frac{1}{2} A_\mu (-g^{\mu\nu}\partial^2 + (1-\xi_A)\partial^\mu\partial^\nu
	) A_\nu 
%
\\  \nonumber
\mathcal{L}_s =&
\chi^- (-\partial^2 - \xi_Wm_W^2) \chi^+
+ \frac{1}{2} \chi_3 (-\partial^2 - \xi_Zm_Z^2) \chi_3
+ \frac{1}{2} h (-\partial^2 - m_H^2) h
%
\\ \nonumber
\mathcal{L}_g =&
\bar{c}^-(-\partial^2 - \xi_Wm_W^2) c^+
+ \bar{c}^+(-\partial^2 - \xi_Wm_W^2) c^-
+ \bar{c}^Z(-\partial^2 - \xi_Wm_W^2) c^Z
+ \bar{c}^A(-\partial^2) c^A
%
\\ \nonumber
\mathcal{L}_f =&
\sum_{i=1}^3
	\bar{e_i} (i\slashpartial - m_{ei}) e_i
	+ \bar{\nu_{i}} (i \slashpartial) \nu_i 
	+ \bar{u_i} (i\slashpartial - m_{ui}) u_i 
	+ \bar{d_i} (i\slashpartial - m_{di}) d_i 
\end{align}
%
%
%
\begin{align}
\mathcal{L}_{vvvv} = -\frac{e^2}{s_W^2} & \left\{ 
\tfrac{1}{2}\left(W^-_\mu W^{+\mu} W^-_\nu W^{+\nu} -
	W^-_\mu W^{-\mu} W^+_\nu W^{+\nu}\right) \right.
\\ \nonumber
&+c_W^2\left(W^-_\mu W^{+\mu} Z_\nu Z^{\nu} -
	(1-\tilde{\beta}^2/\xi_W)W^-_\mu Z^{\mu} W^+_\nu Z^{\nu}\right)
\\ \nonumber
&+s_W^2\left(W^-_\mu W^{+\mu} A_\nu A^{\nu} -
	(1-\tilde{\alpha}^2/\xi_W)W^-_\mu A^{\mu} W^+_\nu A^{\nu}\right)
\\ \nonumber
&\left.+s_Wc_W\left(2W^-_\mu W^{+\mu} A_\nu Z^{\nu} 
	- (1-\tilde{\alpha}\tilde{\beta}/\xi_W)(W^-_\mu A^{\mu} W^+_\nu Z^{\nu}
	+ W^-_\mu Z^{\mu} W^+_\nu A^{\nu})\right)\right\}.
\end{align}
%
\begin{align}
\mathcal{L}_{vvv} =& -ie\frac{c_W}{s_W} \left[
Z^\nu\left(W^{-\mu}\partial_{\nu}W^+_\mu-W^{+\mu}\partial_{\nu}W^-_\mu\right)
+W^{+\nu}\left(Z^\mu\partial_{\nu}W^-_\mu-W^{-\mu}\partial_{\nu}Z_\mu\right)
\right. \\ \nonumber & \qquad  \left.
+W^{-\nu}\left(W^{+\mu}\partial_{\nu}Z_\mu-Z^\mu\partial_{\nu}W^+_\mu\right)
-\tilde{\beta}/\xi_W (
	W_\mu^+ Z^\mu \partial_\nu W^{-\nu} 
	- W_\mu^- Z^\mu \partial_\nu W^{+\nu} )
\right]
\\ \nonumber &
-ie \left[
A^\nu\left(W^{-\mu}\partial_{\nu}W^+_\mu-W^{+\mu}\partial_{\nu}W^-_\mu\right)
+W^{+\nu}\left(A^\mu\partial_{\nu}W^-_\mu-W^{-\mu}\partial_{\nu}A_\mu\right)
\right. \\ \nonumber & \qquad  \left.
+W^{-\nu}\left(W^{+\mu}\partial_{\nu}A_\mu-A^\mu\partial_{\nu}W^+_\mu\right)
-\tilde{\alpha}/\xi_W (
	W_\mu^+ A^\mu \partial_\nu W^{-\nu} 
	- W_\mu^- A^\mu \partial_\nu W^{+\nu} )
\right]
\end{align}
%
\begin{align}
\mathcal{L}_{vvss} = &
\frac{e^2}{2s_W}\left(
	i (1-\tilde{\alpha}\tilde{\delta})( A_\mu W^{+\mu} h \chi^-
	-  A_\mu W^{-\mu} h \chi^+)
\right. \\ \nonumber & \qquad\qquad\qquad\qquad \left.
	- (1-\tilde{\alpha}\tilde{\kappa})( A_\mu W^{+\mu} \chi_3 \chi^-
	+A_\mu W^{-\mu} \chi_3 \chi^+)
\right) 
\\ \nonumber &
+\frac{e^2}{2c_W}\left(
	-i (1+ \tilde{\beta}\tilde{\delta}c_W^2/s_W^2)( Z_\mu W^{+\mu} h \chi^-
	- Z_\mu W^{-\mu} h \chi^+)
\right. \\ \nonumber & \qquad\qquad\qquad\qquad \left.
	+ (1+ \tilde{\beta}\tilde{\kappa}c_W^2/s_W^2)(Z_\mu W^{+\mu} \chi_3 \chi^-
	+Z_\mu W^{-\mu} \chi_3 \chi^+ )
\right) 
\\ \nonumber &
+ e^2 A_\mu A^\mu \chi^+ \chi^-
+ \frac{e^2(c_W^2-s_W^2)}{s_W c_W} A_\mu Z^\mu \chi^+ \chi^-
+ \frac{e^2(c_W^2-s_W^2)^2}{4 s_W^2 c_W^2} Z_\mu Z^\mu \chi^+ \chi^-
\\ \nonumber &
+ \frac{e^2}{4s_W^2} \left(
	W^+_\mu W^{-\mu} h h
	+ W^+_\mu W^{-\mu} \chi_3 \chi_3
	+ 2 W^+_\mu W^{-\mu} \chi^+ \chi^-
\right)
\\ \nonumber &
+ \frac{e^2}{8s_W^2c_W^2} \left(
	Z_\mu Z^{\mu} h h
	+ Z_\mu Z^{\mu} \chi_3 \chi_3
\right).
\end{align}
%
\begin{align}
\mathcal{L}_{vvs} = &
e m_W (1-\tilde{\alpha}) 
\left( i W^+_\mu A^\mu \chi^- - i W^-_\mu A^\mu \chi^+ \right)
\\ \nonumber &
+ \frac{e m_W s_W}{c_W} (1+\tilde{\beta} c_W^2/s_W^2)
\left( -i W^+_\mu Z^\mu \chi^- + i W^-_\mu Z^\mu \chi^+ \right)
\\ \nonumber &
+ \frac{e m_W}{s_W} W^+_\mu W^{-\mu} h
+ \frac{e m_W}{2s_Wc_W^2} Z_\mu Z^{\mu} h
\end{align}
%
\begin{align}
\mathcal{L}_{vss} = &
\frac{e}{2s_W} \left[
	W^{+\mu}(h\partial_\mu\chi^- - \chi^-\partial_\mu h)
	+\tilde{\delta}( h\chi^-\partial_\mu W^{+\mu})
\right.\\&\nonumber\qquad\qquad\qquad\left.
	+W^{-\mu}(h\partial_\mu\chi^+ - \chi^+\partial_\mu h)
	+\tilde{\delta}( h\chi^+\partial_\mu W^{-\mu})
\right] 
\\ \nonumber
+& \frac{ie}{2s_W} \left[
	W^{+\mu}(\chi_3\partial_\mu\chi^- - \chi^-\partial_\mu \chi_3)
	+\tilde{\kappa} (\chi_3\chi^-\partial_\mu W^{+\mu})
\right.\\&\nonumber\qquad\qquad\qquad\left.
	-W^{-\mu}(\chi_3\partial_\mu\chi^+ - \chi^+\partial_\mu \chi_3)
	-\tilde{\kappa}(\chi_3\chi^+\partial_\mu W^{-\mu})
\right]
\\ \nonumber
	+& i e A^{\mu}(\chi^-\partial_\mu\chi^+ - \chi^+\partial_\mu \chi^-)
	+ i e\frac{c_W^2-s_W^2}{2s_Wc_W} 
	Z^{\mu}(\chi^-\partial_\mu\chi^+ - \chi^+\partial_\mu \chi^-)
\\ \nonumber
+ &\frac{e}{2s_Wc_W} \left[
	Z^\mu (h\partial_\mu\chi_3 - \chi_3\partial_\mu h)
	+\tilde{\varepsilon}(h\chi_3\partial_\mu Z^\mu) \right].
\end{align}
%
\begin{align}
\mathcal{L}_{ssss} = & -\frac{e^2m_H^2}{32s_W^2m_W^2} \left(
hhhh + \chi_3\chi_3\chi_3\chi_3 + 4\chi^+\chi^+\chi^-\chi^-
\right.\\&\nonumber\qquad\left.
 + 2(1+2\tilde{\varepsilon}^2\xi_Zm_Z^2/m_H^2) hh\chi_3\chi_3 
	+ 4(1+2\tilde{\delta}^2\xi_Wm_W^2/m_H^2) hh\chi^+\chi^- 
\right.\\&\nonumber\qquad\left.
	+ 4(1+2\tilde{\kappa}^2\xi_Wm_W^2/m_H^2)\chi^+\chi^-\chi_3\chi_3
\right)
\end{align}
%
\begin{align}
\mathcal{L}_{sss} = -\frac{em_H^2}{4s_Wm_W} \left(
hhh + (1+2\tilde{\varepsilon}\xi_Zm_Z^2/m_H^2)h\chi_3\chi_3 
	+ 2(1+2\tilde{\delta}\xi_Wm_W^2/m_H^2)h\chi^+\chi^- \right).
\end{align}
%
%
\begin{align}
\mathcal{L}_{ffv} = \sum_{i=1}^3 & \left\{ 
\frac{e}{\sqrt{2}s_W}W^+_\mu 
	(\bar{\nu_e} \gamma^\mu e_L + \bar{u}_L \gamma^\mu d_L)
+ \frac{e}{\sqrt{2}s_W}W^-_\mu 
	(\bar{e}_L \gamma^\mu \nu_e + \bar{d}_L \gamma^\mu u_L) \right.
\\ \nonumber &
+ \frac{e}{s_Wc_W} Z_\mu \left[
	+\tfrac12\bar{\nu_e}\gamma^\mu\nu_e
	-\tfrac12\bar{e}_L\gamma^\mu e_L
	+\tfrac12\bar{u}_L\gamma^\mu u_L
	-\tfrac12\bar{d}_L\gamma^\mu d_L\right.\\ \nonumber
	&\quad\quad\quad\left.
	+\sin^2\theta_w \bar{e}\gamma^\mu e
	-\tfrac23\sin^2\theta_w \bar{u}\gamma^\mu u
	+\tfrac13\sin^2\theta_w \bar{d}\gamma^\mu d
\right]
\\ \nonumber & \left.
+ e A_\mu \left( - \bar{e}\gamma^\mu e
	+\tfrac23 \bar{u}\gamma^\mu u
	-\tfrac13 \bar{d}\gamma^\mu d \right) \right\}
%
%
\\
\mathcal{L}_{ffs} = \sum_{i=1}^3 & \left\{ 
\frac{ie}{\sqrt{2}s_Wm_W} \chi^+
	(m_e\bar{\nu}_L e_R + m_d\bar{u}_L d_R - m_u\bar{u}_R d_L)
\right.\\&\nonumber
+ \frac{ie}{\sqrt{2}s_Wm_W} \chi^-
	(-m_e\bar{e}_R \nu_L + m_u\bar{d}_L u_R - m_d\bar{d}_R u_L) 
\\&\nonumber
- \frac{ie}{2s_Wm_W} \chi_3
	(m_e\bar{e} \gamma^5 e + m_d\bar{d} \gamma^5 d - m_u\bar{u} \gamma^5 u)
\\&\nonumber\left.
- \frac{e}{2s_Wm_W}
	(m_e\bar{e} e + m_d\bar{d} d + m_u\bar{u} u) \right\}
\end{align}
%
%
%
\begin{align}
\mathcal{L}_{ggv} =&
+ie W^{+\mu} \left( (\partial_\mu \bar{c}^-) c^A +
	\tilde{\alpha}\bar{c}^-\partial_\mu c^A \right)
+ie\frac{c_W}{s_W} W^{+\mu} \left( (\partial_\mu \bar{c}^-) c^Z + 
	\tilde{\beta}\bar{c}^-\partial_\mu c^Z \right)
\\ \nonumber &
-ie A^{\mu} \left( (\partial_\mu \bar{c}^-) c^+ -
	\tilde{\alpha}\bar{c}^-\partial_\mu c^+ \right)
-ie\frac{c_W}{s_W} Z^{\mu} \left( (\partial_\mu \bar{c}^-) c^+ -
	\tilde{\beta}\bar{c}^-\partial_\mu c^+ \right)
\\ \nonumber &
-ie W^{-\mu} \left( (\partial_\mu \bar{c}^+) c^A +
	\tilde{\alpha}\bar{c}^+\partial_\mu c^A \right)
-ie\frac{c_W}{s_W} W^{-\mu} \left( (\partial_\mu \bar{c}^+) c^Z + 
	\tilde{\beta}\bar{c}^+\partial_\mu c^Z \right)
\\ \nonumber &
+ie A^{\mu} \left( (\partial_\mu \bar{c}^+) c^- -
	\tilde{\alpha}\bar{c}^+\partial_\mu c^- \right)
+ie\frac{c_W}{s_W} Z^{\mu} \left( (\partial_\mu \bar{c}^+) c^- -
	\tilde{\beta}\bar{c}^+\partial_\mu c^- \right)
\\ \nonumber &
-ie\frac{c_W}{s_W}(\partial_\mu \bar{c}^Z)
\left(W^{+\mu}c^- - W^{-\mu}c^+\right)
\\ \nonumber &
-ie(\partial_\mu \bar{c}^A)
\left(W^{+\mu}c^- - W^{-\mu}c^+\right)
\\ 
%
%
\mathcal{L}_{ggs} =&
+\frac{ie\xi_W m_W}{2s_W}\, \bar{c}^- \left(
	\frac{c_W^2-s_W^2+\tilde{\kappa}}{c_W}\chi^+c^Z
	+2s_W\chi^+c^A - (1-\tilde{\kappa}) \chi_3 c^+ 
	+ i (1+\tilde{\delta}) h c^+ \right)
\nonumber\\  &
-\frac{ie\xi_W m_W}{2s_W}\, \bar{c}^+ \left(
	\frac{c_W^2-s_W^2+\tilde{\kappa}}{c_W}\chi^-c^Z
	+2s_W\chi^-c^A - (1-\tilde{\kappa})\chi_3 c^- 
	- i (1+\tilde{\delta}) h c^- \right)
\nonumber\\  &
-\frac{ie\xi_Z m_Z}{2s_W}\, \bar{c}^Z \left(
	\chi^+ c^- - \chi^- c^+ 
	-\frac{i}{c_W}(1+\tilde{\varepsilon})h c^Z \right).
\end{align}
%
%
\begin{align}
\mathcal{L}_{ggvv} =&
- e^2 \bar{c}^- \left(
	\tilde{\alpha} A_\mu W^{+\mu}c^A 
	+ \tilde{\alpha}\tfrac{c_W}{s_W}A_\mu W^{+\mu}c^Z 
	+ \tilde{\beta}\tfrac{c_W}{s_W}Z_\mu W^{+\mu}c^A 
	+ \tilde{\beta} \tfrac{c_W^2}{s_W^2}Z_\mu W^{+\mu}c^Z - 
\right. \nonumber \\&\qquad  \left. 
	- \tilde{\alpha}A_\mu A^\mu c^+ 
	- \tilde{\beta} \tfrac{c_W^2}{s_W^2}Z_\mu Z^\mu c^+
	- (\tilde{\alpha}+\tilde{\beta})\tfrac{c_W}{s_W}A_\mu Z^\mu c^+ -
\right. \nonumber \\&\qquad  \left. 
	-(\tilde{\alpha}+\tilde{\beta} \tfrac{c_W^2}{s_W^2})W_\mu^+W^{+\mu}c^- 
	+(\tilde{\alpha}+\tilde{\beta} \tfrac{c_W^2}{s_W^2})W^+_\mu W^{-\mu}c^+
	\right) 
 + \mr{h.c.}
\end{align}
%
%
\begin{align}
\mathcal{L}_{ggss} =&
\left[
\frac{e^2\xi_W}{4s_W^2}\,\bar{c}^- \left(
	(\tilde{\delta}-\tilde{\kappa}) \chi^+\chi^+c^- 
	+(\tilde{\delta}+\tilde{\kappa}) \chi^+\chi^-c^+  
	+ \right. \right.
\\ \nonumber & \quad 
	+ \frac{\tilde{\delta}+\tilde{\kappa}(c_W^2-s_W^2)}{c_W}\chi^+\chi_3c^Z 
	+i\frac{\tilde{\delta}(c_W^2-s_W^2)+\tilde{\kappa}}{c_W}\chi^+hc^Z 
	+ i(\tilde{\kappa}-\tilde{\delta})\chi_3hc^+  + 
\\ &\nonumber \quad
\left. \left.
	+ 2i\tilde{\delta}s_W \chi^+hc^A 
	+2\tilde{\kappa}s_W\chi^+\chi_3c^A 
	- \tilde{\delta}hhc^+
	- \tilde{\kappa}\chi_3\chi_3c^+ \right) + \mr{h.c.}
\right]
\\&\nonumber
+\tilde{\varepsilon}\frac{e^2\xi_Z}{4s_W^2c_W} \bar{c}^Z \left(
	-i\chi^+hc^- + i\chi^-hc^+  + \chi^+\chi_3c^- + \chi^-\chi_3c^+ 
+ \frac{1}{c_W}\chi_3\chi_3c^Z - \frac{1}{c_W}hhc^Z \right).
\end{align}

\section{Interaction vertices}
\label{sec:vertices}

We give the full list of the propagators and interaction vertices for the Standard Model with non-linear gauge-fixing. These results follow straightforwardly from the Lagrangian written above: the coefficient in front of the fields is multiplied by $i$ and we do a transformation to momentum space by substituting $\partial_\mu \rightarrow -ip_\mu$. Also there comes a factor from permuting identical fields.

This list can be checked with the one given in the Appendix of \cite{grace}. The non-linear gauge-fixing used there is the same, and so the resulting terms are the same. The only difference in conventions there is that the Goldstone bosons $\chi^\pm, \chi_3$ are defined with an additional minus sign, so all the interaction vertices that involve an odd number of Goldstones have a relative minus sign. Also note that \cite{grace} has the momentum of the antighosts defined with an opposite sign and does not include the overall $i$ factor for vertices.

Notes about fermions: some vertices are given  generally for any fermion $f$ in terms of its charges. The charges used are the $t^3$ charge of the left-handed component:
\begin{equation}
t^3(u_L,\nu_e) = +\frac{1}{2}, \qquad
t^3(d_L,e_L) = -\frac{1}{2},
\end{equation}
and the electric charge $q$:
\begin{equation}
q(u) = \frac{2}{3}, \quad
q(d) = -\frac{1}{3}, \quad
q(\nu_e) = 0, \quad
q(e) = -1.
\end{equation}
All fermion vertices apply to all 3 generations:
\begin{equation}
e\rightarrow(e,\mu,\tau), \quad
\nu_e\rightarrow(\nu_e,\nu_\mu,\nu_\tau), \quad
u\rightarrow(u,c,t), \quad
d\rightarrow(d,s,b), \quad
\end{equation}
with the same charges, but different masses. As it was mentioned, $SU(3)$ is not discussed in this work, but note that for each quark there are 3 states of different color. Also note that $(1-\gamma^5)/2$ and $(1+\gamma^5)/2$ are  the projectors of the left-handed and right-handed components, respectively.

\paragraph{Propagators}

\begin{align}
\langle X^{*\mu} X^\nu \rangle &=
\frac{-i}{p^2-m_X^2}\left(g^{\mu\nu}-\frac{p^\mu p^\nu}{p^2-\xi_X
m_X^2}(1-\xi_X)\right), \quad X\in\{W^\pm,Z,A\} 
\\
\langle \chi^X \chi^X \rangle &= \frac{i}{p^2-\xi^X m^2_X}, \quad
X\in\{W^\pm,Z\} 
\\
\langle hh \rangle &= \frac{i}{p^2-m^2_H}, 
\\
\langle \bar{c}^X c^X \rangle &= \frac{i}{p^2-\xi_X m^2_X}, \quad
X\in\{W^\pm,Z,A\}
\\
\langle \bar{f} f \rangle &= \frac{i}{\slashp - m_f}
\end{align}

\paragraph{Vector-Vector-Vector-Vector}

\begin{eqnarray}
\nonumber
W^+_\mu W^-_\nu A_\rho A_\sigma  \qquad &&
	ie^2\left( -2g^{\mu\nu}g^{\rho\sigma} + 
	(1-\tilde{\alpha}^2/\xi_W)(g^{\mu\rho}g^{\nu\sigma}+g^{\mu\sigma}g^{\nu\rho})
	\right) 
\\ \nonumber
W^+_\mu W^-_\nu A_\rho Z_\sigma  \qquad &&
	ie^2\frac{c_W}{s_W}\left( -2g^{\mu\nu}g^{\rho\sigma} + 
	(1-\tilde{\alpha}\tilde{\beta}/\xi_W)(g^{\mu\rho}g^{\nu\sigma}+g^{\mu\sigma}g^{\nu\rho})
	\right) 
\\ \nonumber
W^+_\mu W^-_\nu Z_\rho Z_\sigma  \qquad &&
	ie^2\frac{c_W^2}{s_W^2}\left( -2g^{\mu\nu}g^{\rho\sigma} + 
	(1-\tilde{\beta}^2/\xi_W)(g^{\mu\rho}g^{\nu\sigma}+g^{\mu\sigma}g^{\nu\rho})
	\right)
\\ \nonumber
W^+_\mu W^+_\nu W^-_\rho W^-_\sigma  \qquad &&
	-ie^2\frac{1}{s_W^2}\left( -2g^{\mu\nu}g^{\rho\sigma} + 
	(g^{\mu\rho}g^{\nu\sigma}+g^{\mu\sigma}g^{\nu\rho})
	\right) 
\end{eqnarray}

\paragraph{Vector-Vector-Vector}

\begin{eqnarray}
\nonumber
W^-_\mu(p_1)W^+_\nu(p_2)A_\rho(p_3) \qquad &&
	ie\left[g^{\mu\nu}(p_1-p_2)^\rho + g^{\nu\rho}(p_2-p_3)^\mu + 
	g^{\rho\mu}(p_3-p_1)^\nu  
	\right. \\ \nonumber && \left. \qquad
	+\tilde{\alpha}/\xi_W (p_1^\mu g^{\nu\rho} - p_2^\nu g^{\rho\mu}) \right]
\\ \nonumber
W^-_\mu(p_1)W^+_\nu(p_2)Z_\rho(p_3) \qquad &&
	ie\frac{c_W}{s_W}\left[g^{\mu\nu}(p_1-p_2)^\rho + g^{\nu\rho}(p_2-p_3)^\mu + 
	g^{\rho\mu}(p_3-p_1)^\nu  
	\right. \\ \nonumber && \left. \qquad
	+\tilde{\beta}/\xi_W (p_1^\mu g^{\nu\rho} - p_2^\nu g^{\rho\mu}) \right]
\end{eqnarray}

\paragraph{Vector-Vector-Scalar-Scalar}

\begin{eqnarray}
\nonumber
A_\mu W^\pm_\nu h \chi^\mp \qquad &&
	\mp e^2\frac{1}{2s_W} (1-\tilde\alpha\tilde\delta)g^{\mu\nu}
\\ \nonumber
A_\mu W^\pm_\nu \chi_3 \chi^\mp \qquad &&
	-ie^2\frac{1}{2s_W} (1-\tilde\alpha\tilde\kappa)g^{\mu\nu}
\\ \nonumber
Z_\mu W^\pm_\nu h \chi^\mp \qquad &&
	\pm e^2\frac{1}{2c_W} (1-\tilde\beta\tilde\delta c_W^2/s_W^2)g^{\mu\nu}
\\ \nonumber
Z_\mu W^\pm_\nu \chi_3 \chi^\mp \qquad &&
	 i e^2\frac{1}{2c_W} (1-\tilde\beta\tilde\kappa c_W^2/s_W^2)g^{\mu\nu}
\\ \nonumber
A_\mu A_\nu \chi^+ \chi^- \qquad && 
	2i e^2 g^{\mu\nu}
\\ \nonumber
Z_\mu A_\nu \chi^+ \chi^- \qquad && 
	i e^2 \frac{c_W^2-s_W^2}{c_Ws_W} g^{\mu\nu}
\\ \nonumber
Z_\mu Z_\nu \chi^+ \chi^- \qquad && 
	2i e^2 \left(\frac{c_W^2-s_W^2}{2c_Ws_W}\right)^2 g^{\mu\nu}
\\ \nonumber
W^+_\mu W^-_\nu hh \qquad &&
	ie^2\frac{1}{2s_W^2} g^{\mu\nu}
\\ \nonumber
W^+_\mu W^-_\nu \chi_3\chi_3 \qquad &&
	ie^2\frac{1}{2s_W^2} g^{\mu\nu}
\\ \nonumber
W^+_\mu W^-_\nu \chi^+\chi^- \qquad &&
	ie^2\frac{1}{2s_W^2} g^{\mu\nu}
\\ \nonumber
Z_\mu Z_\nu hh \qquad &&
	ie^2\frac{1}{2s_W^2c_W^2} g^{\mu\nu}
\\ \nonumber
Z_\mu Z_\nu \chi_3\chi_3 \qquad &&
	ie^2\frac{1}{2s_W^2c_W^2} g^{\mu\nu}
\end{eqnarray}

\paragraph{Vector-Vector-Scalar}

\begin{eqnarray}
\nonumber
W^\pm_\mu A_\nu \chi^\mp \qquad &&
	\mp e m_W (1-\tilde\alpha) g^{\mu\nu}
\\ \nonumber
W^\pm_\mu Z_\nu \chi^\mp \qquad &&
	\pm e \frac{m_Ws_W}{c_W}(1+\tilde\beta c_W^2/s_W^2) g^{\mu\nu}
\\ \nonumber
W^+_\mu W^-_\nu h \qquad &&
	ie \frac{m_W}{s_W} g^{\mu\nu}
\\ \nonumber
Z_\mu Z_\nu h \qquad &&
	ie \frac{m_W}{s_Wc_W^2} g^{\mu\nu}
\end{eqnarray}

\paragraph{Vector-Scalar-Scalar}

\begin{eqnarray}
\nonumber
h(p_1)\chi^\mp(p_2)W^\pm_\mu(p_3) \qquad &&
	e\frac{1}{2s_W} (p_2^\mu - p_1^\mu + \tilde\delta p_3^\mu)
\\ \nonumber
\chi_3(p_1)\chi^\mp(p_2)W^\pm_\mu(p_3) \qquad &&
	\pm ie\frac{1}{2s_W} (p_2^\mu - p_1^\mu + \tilde\kappa p_3^\mu)
\\ \nonumber
\chi^-(p_1)\chi^+(p_2)A_\mu(p_3) \qquad &&
	ie(p_2^\mu - p_1^\mu)
\\ \nonumber
\chi^-(p_1)\chi^+(p_2)Z_\mu(p_3) \qquad &&
	ie\frac{c_W^2-s_W^2}{2c_Ws_W}(p_2^\mu - p_1^\mu)
\\ \nonumber
h(p_1)\chi_3(p_2)Z_\mu(p_3) \qquad &&
	e\frac{1}{2s_Wc_W} (p_2^\mu - p_1^\mu + \tilde\varepsilon p_3^\mu)
\end{eqnarray}

\paragraph{Scalar-Scalar-Scalar-Scalar}

\begin{eqnarray}
\nonumber
hhhh \qquad &&
	-ie^2 \frac{3m_H^2}{4s_W^2m_W^2}
\\ \nonumber
\chi_3\chi_3\chi_3\chi_3 \qquad &&
	-ie^2 \frac{3m_H^2}{4s_W^2m_W^2}
\\ \nonumber
\chi_\pm\chi_\pm\chi_\mp\chi_\mp \qquad &&
	-ie^2 \frac{m_H^2}{2s_W^2m_W^2}
\\ \nonumber
hh\chi_3\chi_3 \qquad &&
	-ie^2 \frac{m_H^2}{4s_W^2m_W^2} (1 + 2\tilde\varepsilon^2\xi_Zm_Z^2/m_H^2)
\\ \nonumber
hh\chi^+\chi^- \qquad &&
	-ie^2 \frac{m_H^2}{4s_W^2m_W^2} (1 + 2\tilde\delta^2\xi_Wm_W^2/m_H^2)
\\ \nonumber
\chi^+\chi^-\chi_3\chi_3 \qquad &&
	-ie^2 \frac{m_H^2}{4s_W^2m_W^2} (1 + 2\tilde\kappa^2\xi_Wm_W^2/m_H^2)
\end{eqnarray}

\paragraph{Scalar-Scalar-Scalar}

\begin{eqnarray}
\nonumber
hhh \qquad &&
	-ie\frac{3m_H^2}{4s_Wm_W}
\\ \nonumber
h\chi^-\chi^+ \qquad &&
	-ie\frac{m_H^2}{2s_Wm_W}(1+2\tilde\delta\xi_W m_W^2/m_H^2)
\\ \nonumber
h\chi_3\chi_3 \qquad &&
	-ie\frac{m_H^2}{2s_Wm_W}(1+2\tilde\varepsilon\xi_Z m_Z^2/m_H^2)
\end{eqnarray}

\paragraph{Fermion-Fermion-Vector}

\begin{eqnarray}
\nonumber
\bar{f}f A_\mu \qquad &&
	ie q_f \gamma^\mu
\\ \nonumber
\bar{f}f Z_\mu \qquad &&
	ie\frac{1}{s_Wc_W}\gamma^\mu\left(t^3_f (1-\gamma^5)/2 - s_W^2 q_f  \right)
\\ \nonumber
\bar{\nu_e}eW^+_\mu, \,\, 
\bar{e}\nu_eW^-_\mu, \,\, 
\bar{u}d W^+, \,\,
\bar{d}u W^- \qquad &&
	ie\frac{1}{\sqrt{2}s_W}\gamma^\mu(1-\gamma^5)/2
\end{eqnarray}

\paragraph{Fermion-Fermion-Scalar}

\begin{eqnarray}
\nonumber
\bar{f}fh \qquad &&
	-ie\frac{m_f}{2s_Wm_W}
\\ \nonumber
\bar{e}e\chi_3,\,\, \bar{d}d\chi_3 \qquad &&
	e\frac{m_f}{2s_Wm_W}\gamma^5
\\ \nonumber
\bar{u}u\chi_3 \qquad &&
	-e\frac{m_f}{2s_Wm_W}\gamma^5
\\ \nonumber
\bar{\nu_e}e\chi^+ \qquad &&
	-e\frac{m_e}{\sqrt{2}s_Wm_W}(1+\gamma^5)/2
\\ \nonumber
\bar{e}\nu_e\chi^- \qquad &&
	+e\frac{m_e}{\sqrt{2}s_Wm_W}(1+\gamma^5)/2
\\ \nonumber
\bar{u}d\chi^+ \qquad &&
	-e\frac{1}{\sqrt{2}s_Wm_W} \left(
	m_d(1+\gamma^5) - m_u(1-\gamma^5) \right) /2
\\ \nonumber
\bar{d}u\chi^- \qquad &&
	-e\frac{1}{\sqrt{2}s_Wm_W} \left(
	m_u(1+\gamma^5) - m_d(1-\gamma^5) \right) /2
\end{eqnarray}

\paragraph{Ghost-Ghost-Vector}

\begin{eqnarray}
\nonumber
\bar{c}^A(p_1) c^\mp(p_2) W^\pm_\mu \qquad &&
	\mp ie p_1^\mu
\\ \nonumber
\bar{c}^Z(p_1) c^\mp(p_2) W^\pm_\mu \qquad &&
	\mp ie\frac{c_W}{s_W} p_1^\mu
\\ \nonumber
\bar{c}^\mp(p_1) c^A(p_2) W^\pm_\mu \qquad &&
	\pm ie (p_1^\mu + \tilde\alpha p_2^\mu)
\\ \nonumber
\bar{c}^\mp(p_1) c^Z(p_2) W^\pm_\mu \qquad &&
	\pm ie \frac{c_W}{s_W} (p_1^\mu + \tilde\beta p_2^\mu)
\\ \nonumber
\bar{c}^\mp(p_1) c^\pm(p_2) A_\mu \qquad &&
	\mp ie (p_1^\mu - \tilde\alpha p_2^\mu)
\\ \nonumber
\bar{c}^\mp(p_1) c^\pm(p_2) Z_\mu \qquad &&
	\mp ie \frac{c_W}{s_W} (p_1^\mu - \tilde\beta p_2^\mu)
\end{eqnarray}

\paragraph{Ghost-Ghost-Scalar}

\begin{eqnarray}
\nonumber
\bar{c}^Zc^Zh \qquad &&
	-ie\frac{\xi_Z m_Z}{2s_Wc_W} (1+\tilde\varepsilon)
\\ \nonumber
\bar{c}^Zc^\mp \chi^\pm \qquad &&
	\pm e\frac{\xi_Z m_Z}{2s_W} 
\\ \nonumber
\bar{c}^\mp c^A \chi^\pm \qquad &&
	\mp e m_W \xi_W
\\ \nonumber
\bar{c}^\mp c^Z \chi^\pm \qquad &&
	\mp e \frac{m_W \xi_W}{2s_Wc_W}(c_W^2 - s_W^2 + \tilde\kappa)
\\ \nonumber
\bar{c}^\mp c^\pm h \qquad &&
	- ie \frac{m_W \xi_W}{2s_W}(1 + \tilde\delta)
\\ \nonumber
\bar{c}^\mp c^\pm \chi_3 \qquad &&
	 \pm e \frac{m_W \xi_W}{2s_W}(1 - \tilde\kappa)
\end{eqnarray}

\paragraph{Ghost-Ghost-Vector-Vector}

\begin{eqnarray}
\nonumber
\bar{c}^\mp c^A A_\mu W^\pm_\nu \qquad &&
	-ie^2 \tilde\alpha g^{\mu\nu}
\\ \nonumber
\bar{c}^\mp c^A Z_\mu W^\pm_\nu \qquad &&
	-ie^2 \tilde\beta \frac{c_W}{s_W} g^{\mu\nu}
\\ \nonumber
\bar{c}^\mp c^Z A_\mu W^\pm_\nu \qquad &&
	-ie^2 \tilde\alpha \frac{c_W}{s_W} g^{\mu\nu}
\\ \nonumber
\bar{c}^\mp c^Z Z_\mu W^\pm_\nu \qquad &&
	-ie^2 \tilde\beta \frac{c_W^2}{s_W^2} g^{\mu\nu}
\\ \nonumber
\bar{c}^\mp c^\pm W^\mp_\mu W^\pm_\nu \qquad &&
	-ie^2 \left(\tilde\alpha + \tilde\beta \frac{c_W^2}{s_W^2} \right) g^{\mu\nu}
\\ \nonumber
\bar{c}^\mp c^\mp W^\pm_\mu W^\pm_\nu \qquad &&
	2ie^2 \left(\tilde\alpha + \tilde\beta \frac{c_W^2}{s_W^2} \right) g^{\mu\nu}
\\ \nonumber
\bar{c}^\mp c^\pm A_\mu A_\nu \qquad &&
	2ie^2 \tilde\alpha g^{\mu\nu}
\\ \nonumber
\bar{c}^\mp c^\pm Z_\mu A_\nu \qquad &&
	ie^2 (\tilde\alpha+\tilde\beta)\frac{c_W}{s_W} g^{\mu\nu}
\\ \nonumber
\bar{c}^\mp c^\pm Z_\mu Z_\nu \qquad &&
	2ie^2\tilde\beta \frac{c_W^2}{s_W^2} g^{\mu\nu}
\end{eqnarray}

\paragraph{Ghost-Ghost-Scalar-Scalar}

\begin{eqnarray}
\nonumber
\bar{c}^Z c^Z hh \qquad &&
	-ie^2\tilde\varepsilon\frac{\xi_Z}{2s_W^2c_W^2}
\\ \nonumber
\bar{c}^Z c^Z \chi_3\chi_3 \qquad &&
	ie^2\tilde\varepsilon\frac{\xi_Z}{2s_W^2c_W^2}
\\ \nonumber
\bar{c}^Z c^\pm \chi^\mp h \qquad &&
	\mp e^2\tilde\varepsilon\frac{\xi_Z}{4s_W^2c_W}
\\ \nonumber
\bar{c}^Z c^\pm \chi^\mp \chi_3 \qquad &&
	ie^2\tilde\varepsilon\frac{\xi_Z}{4s_W^2c_W}
\\ \nonumber
\bar{c}^\mp c^A \chi^\pm h \qquad &&
	\mp e^2\tilde\delta\frac{\xi_W}{2s_W}
\\ \nonumber
\bar{c}^\mp c^A \chi^\pm \chi_3 \qquad &&
	i e^2\tilde\kappa\frac{\xi_W}{2s_W}
\\ \nonumber
\bar{c}^\mp c^Z \chi^\pm h \qquad &&
	\mp e^2\frac{\xi_W}{4s_W^2c_W}
	\left(\tilde\kappa+\tilde\delta(c_W^2-s_W^2)\right)
\\ \nonumber
\bar{c}^\mp c^Z \chi^\pm \chi_3 \qquad &&
	i e^2\frac{\xi_W}{4s_W^2c_W}
	\left(\tilde\delta+\tilde\kappa(c_W^2-s_W^2)\right)
\\ \nonumber
\bar{c}^\mp c^\pm hh \qquad &&
	-i e^2 \tilde\delta \frac{\xi_W}{2s_W^2}
\\ \nonumber
\bar{c}^\mp c^\pm \chi_3\chi_3 \qquad &&
	-i e^2 \tilde\kappa \frac{\xi_W}{2s_W^2}
\\ \nonumber
\bar{c}^\mp c^\pm \chi_3h \qquad &&
	\pm e^2 (\tilde\delta-\tilde\kappa) \frac{\xi_W}{4s_W^2}
\\ \nonumber
\bar{c}^\mp c^\pm \chi^+ \chi^- \qquad &&
	i e^2 (\tilde\delta+\tilde\kappa) \frac{\xi_W}{4s_W^2}
\\ \nonumber
\bar{c}^\mp c^\mp \chi^\pm \chi^\pm \qquad &&
	i e^2 (\tilde\delta-\tilde\kappa) \frac{\xi_W}{2s_W^2}
\end{eqnarray}

\section{FeynArts model files}
\label{sec:modelfiles}

\subsection{Lorentz.gen}

Here we list the modified pieces of the {\tt Lorentz.gen} generic model file. These correspond to Vector-Vector-Vector, Vector-Scalar-Scalar, Ghost-Ghost-Vector-Vector and Ghost-Ghost-Scalar-Scalar vertices respectively.
\begin{verbatim}
      (* V-V-V: *)

  AnalyticalCoupling[ s1 V[j1, mom1, {li1}], s2 V[j2, mom2, {li2}], 
      s3 V[j3, mom3, {li3}] ] ==
    G[-1][s1 V[j1], s2 V[j2], s3 V[j3]] .
      { MetricTensor[li1, li2] FourVector[mom2 - mom1, li3] +
        MetricTensor[li2, li3] FourVector[mom3 - mom2, li1] +
        MetricTensor[li3, li1] FourVector[mom1 - mom3, li2],
        FourVector[mom1, li1] MetricTensor[li2, li3],
        FourVector[mom2, li2] MetricTensor[li3, li1],
        FourVector[mom3, li3] MetricTensor[li1, li2] }

      (* S-S-V: *)
   
  AnalyticalCoupling[ s1 S[j1, mom1], s2 S[j2, mom2],
      s3 V[j3, mom3, {li3}] ] == 
    G[-1][s1 S[j1], s2 S[j2], s3 V[j3]] .
      { FourVector[mom1 - mom2, li3],
        FourVector[mom3, li3] }

      (* U-U-V-V: *)

  AnalyticalCoupling[ s1 U[j1, mom1], s2 U[j2, mom2],
      s3 V[j3, mom3, {li3}], s4 V[j4, mom4, {li4}] ] ==
    G[1][s1 U[j1], s2 U[j2], s3 V[j3], s4 V[j4]] .
      { MetricTensor[li3, li4] }

      (* U-U-S-S: *)

  AnalyticalCoupling[ s1 U[j1, mom1], s2 U[j2, mom2],
      s3 S[j3, mom3], s4 S[j4, mom4] ] ==
    G[1][s1 U[j1], s2 U[j2], s3 S[j3], s4 S[j4]] .
      { 1 }
\end{verbatim}

\subsection{SM.mod}

We give some examples of the modified vertices as defined in the classes model file. Note the new terms proportional to the gauge-fixing parameters that are called {\tt Galpha, Gbeta, Gdelta, Gkappa, Gepsilon}. The ($W^+W^-AA$), ($W^+W^-A$) and ($W^+A\chi^-$) vertices that were needed in the tree-level $\gamma\gamma\rightarrow W^+W^-$ calculation are defined as:
\begin{verbatim}
  C[ -V[3], V[3], V[1], V[1] ] == -I EL^2 *
    { {2, 4 dZe1 + 2 dZW1 + 2 dZAA1 - 2 CW/SW dZZA1}, 
      {-1 + Galpha^2/GaugeXi[W], -2 dZe1 - dZW1 - dZAA1 + CW/SW dZZA1},
      {-1 + Galpha^2/GaugeXi[W], -2 dZe1 - dZW1 - dZAA1 + CW/SW dZZA1} }

  C[ V[1], -V[3], V[3] ] == -I EL *
    { {1, dZe1 + dZW1 + dZAA1/2 - CW/SW dZZA1/2},
      {0, 0},
      {- Galpha/GaugeXi[W], 0},
      {+ Galpha/GaugeXi[W], 0} }

  C[ S[3], -V[3], V[1] ] == -I EL MW * 
    { {1 - Galpha, dZe1 + dMWsq1/(2 MW^2) + dZW1/2 + dZAA1/2 + dZGp1/2 +
            SW/CW dZZA1/2} }
\end{verbatim}
The coefficients of the interaction vertices are listed in the first column, while the second column involving renormalization constants {\tt dZe1}, {\tt dZW1}, etc. lists the coefficients of the counterterm vertices, which we have not modified. Following are examples of other vertices that had new kinematic terms introduced, which are Vector-Scalar-Scalar and 4-vertices with ghosts. We give here ($h\chi^-W^+$), ($\bar{c}^-c^AZW^+$) and ($\bar{c}^Zc^Zhh$):
\begin{verbatim}
  C[ S[3], S[1], -V[3] ] == -I EL/(2 SW) *
    { {1, dZe1 - dSW1/SW + dZW1/2 + dZH1/2 + dZGp1/2},
      {Gdelta, 0} }

  C[ -U[4], U[1], V[2], -V[3] ] == I EL^2 (CW/SW)/Sqrt[GaugeXi[W]] * 
    { {Gbeta} }

  C[ -U[2], U[2], S[1], S[1] ] == -I EL^2 Sqrt[GaugeXi[Z]]/(2 SW^2 CW^2) *
    { {Gepsilon} }
\end{verbatim}

%

\end{document}